\documentclass[11pt,english]{article}
\usepackage[T1]{fontenc}
\usepackage{geometry}
\usepackage{float}
\usepackage{mathtools}
\usepackage{amsmath}
\usepackage{amssymb}
\usepackage{stmaryrd}
\usepackage{graphicx}
\graphicspath{ {./Figures/} }
\usepackage{wasysym}
\usepackage{apacite}
\usepackage{natbib}
\usepackage[colorlinks, allcolors=blue]{hyperref}
\usepackage[font={small,it}]{caption}
\usepackage{setspace}
\makeatletter
\@ifundefined{date}{}{\date{}}
\makeatother

\providecommand{\tabularnewline}{\\}

\makeatother

\usepackage{babel}
\begin{document}

\title{Semiparametric Difference-in-Differences with Potentially Many Control Variables}

\author{Neng-Chieh Chang\footnote{Department of Economics, University of California Los Angeles, 315 Portola Plaza, Los Angeles, CA 90095, USA. email: nengchiehchang@g.ucla.edu}}
\maketitle
\begin{abstract}
This paper discusses difference-in-differences (DID) estimation when
there exist many control variables, potentially more than the sample size. In this case, traditional
estimation methods, which require a limited number of variables, do not work. One may consider using statistical or machine learning
(ML) methods.  However, by the well-known theory of inference of ML methods proposed in \citet*{chernozhukov2018double}, directly applying ML methods to
the conventional semiparametric DID estimators will cause significant bias and make these
DID estimators fail to be $\sqrt{N}$-consistent. This article proposes three new DID estimators for three different data structures,
which are able to shrink the bias and achieve $\sqrt{N}$-consistency and asymptotic normality with mean zero when applying ML methods. This leads to straightforward inferential procedures.
In addition, I show that these new estimators have the small bias
property (SBP), meaning that their
bias will converge to zero faster than the pointwise bias of the nonparametric
estimator on which it is based. 
\end{abstract}

\begin{flushleft}
\end{flushleft}

\begin{flushleft}
\end{flushleft}

\section{Introduction}

The difference-in-differences (DID) estimator has been widely used in empirical economics to evaluate causal effects when there exists a natural experiment with a treated group and an untreated group. By comparing the variation over time in an outcome variable between  the treated group and the untreated group, the DID estimator can be used to calculate the effect of treatment on the outcome variable. Applications of DID include but are not limited to studies of the effects
of immigration on labor markets \citep*{card1990impact}, the effects of minimum
wage law on wages \citep*{david1994minimum}, the effect of tariffs liberalization
on corruption \citep*{sequeira2016corruption}, the effect of household income on children's
personalities \citep*{akee2018does}, and the effect of corporate tax on
wages \citep*{fuest2018higher}. 

The traditional linear DID estimator depends on a parallel trend
assumption that in the absence of treatment, the difference of outcomes between treated and untreated groups remains constant over time. In many situations, however, this assumption
may not hold because there are other individual characteristics that
may be associated with the variations of the outcomes. The treatment
may be taken as exogenous only after controlling these characteristics.
To address this problem, \citet*{abadie2005semiparametric} proposed the semiparametric
DID estimators. Compared to the traditional linear DID estimators, the advantages of Abadie's estimators are threefold.
First, the characteristics are treated nonparametrically so that any estimation
error caused by functional specification is avoided. Second,
the effect of treatment is allowed to vary among individuals,
while the traditional linear DID estimator does not allow this heterogeneity.
Third, the estimation framework proposed in \citet*{abadie2005semiparametric} allows
researchers to estimate how the effect of treatment varies with changes in the 
characteristics. 

This paper is an extension of \citet*{abadie2005semiparametric}.  \citet*{abadie2005semiparametric} considered
the case where the number of control variables has to be limited. A practical difficulty empirical researchers encounter is choosing what variables to
include when there is a rich data set. Although economic intuition can help us narrow down the choice set, it will not completely select all the important variables. This variable selection problem may lead to the chance of omitted variables in practice. 
In this paper, I consider the DID
estimation with many control variables, potentially more
than the sample size.  The classical estimation methods which
require a fixed number of variables do not work in this situation. One has to consider
using ML methods such as Lasso, Logit Lasso, random forests, boosted
trees, or various hybrids. However, by the well-known theory of inference of ML methods developed in \citet*{chernozhukov2018double},  if one directly applies ML methods to the conventional semiparametric DID estimators proposed in \citet*{abadie2005semiparametric}, the result will lead to significant bias and invalid inference. In particular,
the regularization bias embedded in ML methods will result in the
conventional semiparametric DID estimators failing to be $\sqrt{N}$-consistent. 

I contribute to the literature by proposing three new DID estimators
for three different data structures: repeated outcomes, repeated cross-sections,
and multilevel treatment. These new estimators can relieve the impact
of the regularization bias of ML methods and achieve $\sqrt{N}$-consistency. The key is to find the so-called Neyman-orthogonal
scores \citep*{chernozhukov2018double} of \citet*{abadie2005semiparametric}'s estimands. The
Neyman-orthogonal score is a function that identifies the parameter
of interest, and its derivatives with respect to the nuisance parameters
are zero. This property helps us remove the first-order bias caused
by ML methods so that only the second-order bias remains, which
is much smaller and easier to control than the first-order bias as
in the conventional semiparametric DID estimators. Using the cross-fitting algorithm in 
\citet*{chernozhukov2018double}, I show that the new DID estimators can
be $\sqrt{N}$-consistent and asymptotically normal when using ML
methods. Figure 1 presents a Monte Carlo simulation that illustrates the negative effect of directly combining ML methods with Abadie's estimator and the benefit of using the newly proposed DID estimator.  

\begin{figure}[H]
\begin{centering}
\includegraphics[scale=0.5]{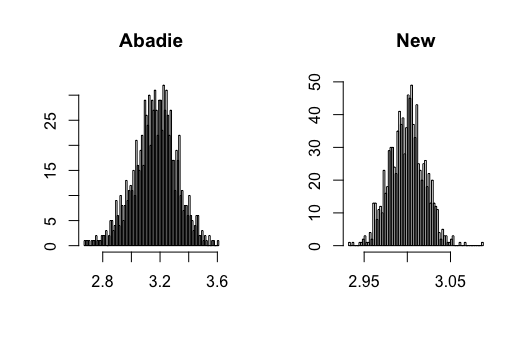}
\par\end{centering}
\caption{The true value is $\theta_{0}=3$ with sample size $N=200$ and the
number of control variables $p=300$. \emph{The left
panel} is the behavior of the conventional semiparametric DID estimator proposed in \citet*{abadie2005semiparametric}, where I estimate the propensity
score using Logit Lasso. The histogram shows that the simulated
distribution of the conventional semiparametric DID estimator is biased. \emph{The
right panel} is the behavior of the new DID estimator proposed in this paper, which is constructed by the Neyman-orthogonal score and
cross-fitting. The nuisance parameters are estimated by Logit Lasso
and random forests. The simulated distribution of the new estimator
is centered at the true value and normally distributed. Note that the simulated data
are exactly the same for both panels, and the simulation setting is
presented in Section 4.}

\end{figure}

The second contribution is concerned with the conventional semiparametric DID estimators
with a limited number of control variables considered in \citet*{abadie2005semiparametric}. In
this case, the conventional semiparametric DID estimators are able to achieve $\sqrt{N}$-consistency using kernel estimators, but they will require undersmoothing. Undersmoothing is a condition
that requires the pointwise bias of the kernel estimators to converge
to zero faster than the pointwise standard deviation. This condition will be violated if researchers use standard data-driven methods, such as cross-validation (CV), to choose the bandwidths of kernel estimators because those methods do not undersmooth. 

In this paper, I show that the new estimators do not require undersmoothing
to achieve $\sqrt{N}$-consistency. Specifically, I will show that the new estimators
have the small bias property (SBP), in terms of \citet*{newey2004twicing}, meaning that the bias of the new estimators will
converge to zero faster than the pointwise bias of the nonparametric
estimator on which it is based. The SBP, as shown in \citet*{chernozhukov2016locally}, is a sufficient condition to remove the undersmoothing requirement. Figure 2 shows the Monte Carlo simulation results of Abadie's estimator and the new estimator with bandwidths chosen by CV. We can observe that Abadie's estimator is biased since CV does not undersmooth, and the newly proposed estimator can correct this bias.  

\begin{figure}[H]
\begin{centering}
\includegraphics[scale=0.5]{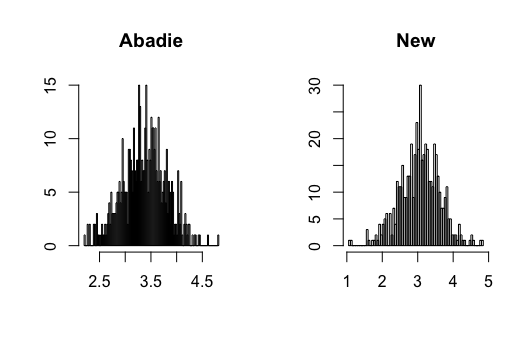}
\par\end{centering}
\begin{center}\caption{The true value is $\theta_0=3$. The first-stage kernel estimators are constructed using standard Gaussian kernel with bandwidths chosen by CV. The simulated data are exactly the same for both estimators, and the simulation setting is presented in Section 4.}\end{center}
\end{figure}
As an empirical example, I study the effect of tariff reduction on corruption behavior using the trade data between South Africa and Mozambique during 2006 and 2014. The treatment is the large tariff reduction on certain commodities occurring in 2008. This natural experiment was previously studied by \citet*{sequeira2016corruption} using the traditional linear DID estimator. I apply my proposed semiparametric DID estimator and  \citet*{abadie2005semiparametric}'s semiparemetric DID estimator on the same data set (Table 9 of \citet*{sequeira2016corruption}). In comparison to \citet*{sequeira2016corruption} that a decrease in tariff rate will decrease corruption behavior, the two semiparametric estimators consistently suggest that the effect is actually substantially larger than previously reported by \citet*{sequeira2016corruption}. A potential explanation for this difference is that the true data generating process violates the linear specification assumed in the traditional linear DID estimator. In addition, when compared to \citet*{abadie2005semiparametric}'s estimator, my proposed estimator shows that the effect is even larger.

The new estimators proposed in this paper heavily rely on the recent high-dimensional and
ML literature: \citet*{belloni2012sparse}, \citet*{Belloni14restud}, \citet*{chernozhukov2015valid}, \citet*{belloni2017program}, and \citet*{chernozhukov2018double}; and the literature of the SBP in semiparametric estimation:   \citet*{newey1998undersmoothing, newey2004twicing} and \citet*{chernozhukov2016locally}.
 
\textbf{Plan of the paper.} Section 2 describes the conventional semiparametric
DID estimators and discusses their limitations when applying ML methods.
Section 3 presents the new DID estimators and discusses their theoretical
properties. Section 4 conducts Monte Carlo simulation to shed some light on the finite sample performance of the proposed estimators. Section
5 provides an application, and Section 6 concludes the paper. 

\section{The Conventional Semiparametric DID Estimators}

Let $Y_{i}\left(t\right)$ be the outcome of interest for individual
$i$ at time $t$ and $D_{i}\left(t\right)\in\left\{ 0,1\right\} $
the treatment status. The population is observed in a pre-treatment period
$t=0$, and in a post-treatment period $t=1$. With potential outcome
notations \citep*{rubin1974estimating}, we have $Y_{i}\left(t\right)=Y_{i}^{0}\left(t\right)+\left(Y_{i}^{1}\left(t\right)-Y_{i}^{0}\left(t\right)\right)D_{i}\left(t\right)$,
where $Y_{i}^{0}\left(t\right)$ is the outcome that individual $i$
would attain at time $t$ in the absence of the treatment, and $Y_{i}^{1}\left(t\right)$
represents the outcome that individual $i$ would attain at time $t$
if exposed to the treatment. Since individuals are only exposed to
treatment at $t=1$, we have $D_{i}\left(0\right)=0$ for all $i$. To reduce
notation, I define $D_{i}\coloneqq D_{i}\left(1\right)$. Also, let  $X_{i}\in\mathbb{R}^{d}$
be a vector of control variables with dimension $d$ potentially larger than the
sample size $N$. 

The traditional linear DID estimator is the parameter $\alpha$ in the following linear model
\[
Y_{i}\left(t\right)=\mu+X_{i}'\pi\left(t\right)+\tau\cdot D_{i}+\delta\cdot t+\alpha\cdot D_{i}\left(t\right)+\varepsilon_{i}\left(t\right),
\]
where $\varepsilon_{i}\left(t\right)$ is an exogenous shock that has mean
zero and $\left(\mu, \pi\left(t\right), \tau, \delta\right)$ are the corresponding parameters. Clearly, the linear specification assumed here
is a strong assumption since the true data generating process may be nonlinear. In addition, \citet*{meyer1995workers} noticed that including control variables in this linear form may not be appropriate if the treatment has different effects for different groups in the population. To deal with these problems, \citet*{abadie2005semiparametric} proposed the semiparametric DID estimators which can
identify average treatment effect on the treated (ATT)
\[
\theta_{0}\coloneqq E\left[Y_{i}^{1}\left(1\right)-Y_{i}^{0}\left(1\right)\mid D_{i}=1\right].
\]
According to the data, there are three particular cases. 

\subsubsection*{\emph{Case 1: Random sample with repeated outcomes}}

Consider the case that researchers can observe both pre-treatment
and post-treatment outcomes for each individual of interest. That
is, researchers observe $\left\{ Y_{i}\left(0\right),Y_{i}\left(1\right),D_{i},X_{i}\right\} _{i=1}^{N}$. In this case, the ATT can be identified under the following assumptions \citep*{abadie2005semiparametric}:
\begin{description}
\item [{Assumption~2.1.}] \emph{$E\left[Y_{i}^{0}\left(1\right)-Y_{i}^{0}\left(0\right)\mid X_{i},D_{i}=1\right]=E\left[Y_{i}^{0}\left(1\right)-Y_{i}^{0}\left(0\right)\mid X_{i},D_{i}=0\right]$.}
\item [{Assumption~2.2.}] \emph{$P\left(D_{i}=1\right)>0$ and with probability
one $P\left(D_{i}=1\mid X_{i}\right)<1$.}
\end{description}
Assumption (2.1) is the conditional parallel trend assumption. It states
that conditional on individual's characteristics, the average outcomes for treated
and untreated groups would have followed parallel paths in the absence of
treatment. With these two assumptions, the ATT is identified
\citep*{abadie2005semiparametric} as 
\[
\theta_{0}=E\left[\frac{Y_{i}\left(1\right)-Y_{i}\left(0\right)}{P\left(D_{i}=1\right)}\frac{D_{i}-P\left(D_{i}=1\mid X_{i}\right)}{1-P\left(D_{i}=1\mid X_{i}\right)}\right].\tag{2.1}
\]

\subsubsection*{\emph{Case 2: Random sample with repeated cross sections}}

Often times, researchers may not be able to observe both pre-treatment
and post-treatment outcomes of the same individual. Instead, they
observe repeated cross-section data sets. Let $T_{i}$ be a time indicator
that takes value one if the observation belongs to the post-treatment
sample. Researchers observe $\left\{ Y_{i},D_{i},T_{i},X_{i}\right\} _{i=1}^{N}$,
where $Y_{i}=Y_{i}\left(0\right)+T_{i}\left(Y_{i}\left(1\right)-Y_{i}\left(0\right)\right)$.

\begin{description}
\item [{Assumption~2.3.}] Conditional on $T=0$, the data are i.i.d. from
the distribution of $\left(Y\left(0\right),D,X\right)$; conditional
on $T=1$, the data are i.i.d. from the distribution of $\left(Y\left(1\right),D,X\right)$.
\end{description}
Suppose Assumptions (2.1)-(2.3)
hold, the ATT is identified \citep*{abadie2005semiparametric} as
\[
\theta_{0}=E\left[\frac{T_{i}-\lambda_{0}}{\lambda_{0}\left(1-\lambda_{0}\right)}\frac{Y_{i}}{P\left(D_{i}=1\right)}\frac{D_{i}-P\left(D_{i}=1\mid X_{i}\right)}{1-P\left(D_{i}=1\mid X_{i}\right)}\right],\tag{2.2}
\]
where $\lambda_{0}\coloneqq P\left(T_{i}=1\right).$
\subsubsection*{\emph{Case 3: Multilevel treatments}}

In many cases, individuals can be exposed to different levels of 
treatment. Let $W\in\left\{0, w_{1},...,w_{J}\right\} $ be the level
of treatment, where $W=0$ denotes the untreated individuals. Researchers observe  $\left\{ Y_{i}\left(0\right),Y_{i}\left(1\right),W_{i},X_{i}\right\} _{i=1}^{N}$.

For $w\in\left\{ 0,w_{1},...,w_{J}\right\} $ and $t\in\left\{ 0,1\right\} $,
let $Y^{w}\left(t\right)$ be the potential outcome for treatment
level $w$ at period $t$. Denote the ATT for each level of treatment $w$ by 
\[
\theta_{0}^{w}\coloneqq E\left[Y^{w}\left(1\right)-Y^{0}\left(1\right)\mid W=w\right].
\]
Suppose that Assumptions (2.1) and (2.2) hold for each level of treatment:
\[
E\left[Y_{i}^{0}\left(1\right)-Y_{i}^{0}\left(0\right)\mid X_{i},W_{i}=w\right]=E\left[Y_{i}^{0}\left(1\right)-Y_{i}^{0}\left(0\right)\mid X_{i},W_{i}=0\right]
\]
 for $w\in\left\{ w_{1},...,w_{J}\right\} $ and \emph{$P\left(W_{i}=w\right)>0$
and with probability one $P\left(W_{i}=w\mid X_{i}\right)<1$ }for
$w\in\left\{ w_{1},...,w_{J}\right\} $. Then we have \citep*{abadie2005semiparametric}
\[
\theta_{0}^{w}=E\left[\frac{Y\left(1\right)-Y\left(0\right)}{P\left(W=w\right)}\frac{I\left(W=w\right)\cdot P\left(W=0\mid X\right)-I\left(W=0\right)\cdot P\left(W=w\mid X\right)}{P\left(W=0\mid X\right)}\right],\tag{2.3}
\]
where $I\left(\cdot\right)$ is an indicator function. 
\bigskip

Let us focus on Case 1 in which researchers confront repeated
outcomes data. To use the identification result (2.1), the first step
is to estimate the two nuisance parameters: $P\left(D_{i}=1\right)\eqqcolon p_{0}$
and $P\left(D_{i}=1\mid X_{i}\right)\eqqcolon g_{0}\left(X_{i}\right)$.
The estimator of $p_{0}$ is just a sample average $\hat{p}=N^{-1}\sum_{i=1}^{N}D_{i}$,
while the propensity score  $g_{0}$ is infinite-dimensional and needs to be estimated nonparametrically.
Denote by $\hat{g}$ the estimator of  $g_{0}$,
then the plug-in estimator based on equation (2.1) is 
\[
\hat{\theta}=\frac{1}{N}\sum_{i=1}^{N}\frac{Y_{i}\left(1\right)-Y_{i}\left(0\right)}{\hat{p}}\frac{D_{i}-\hat{g}\left(X_{i}\right)}{1-\hat{g}\left(X_{i}\right)}.
\]
When  $\hat{g}$ is estimated using  classical nonparametric
methods such as kernel or series estimators, the estimator $\hat{\theta}$ can
be $\sqrt{N}$-consistent and asymptotically normal under certain
conditions provided in the semiparametric estimation literature \citep*{newey1994asymptotic, newey1994large}. 

When $\hat{g}$ is an ML estimator, however, the estimator $\hat{\theta}$
will fail to be $\sqrt{N}$-consistent in general. By the general theory of inference of ML methods developed in \citet*{chernozhukov2018double}, the reason is twofold : 
(1) the score function based on (2.1), $\varphi\left(W,\theta_{0},p_{0},g_{0}\right)\coloneqq\frac{Y\left(1\right)-Y\left(0\right)}{P\left(D=1\right)}\frac{D-g_{0}\left(X\right)}{1-g_{0}\left(X\right)}-\theta_{0}$,
has a non-zero directional (Gateaux) derivative with respect to the propensity score $g_{0}$:
$$\partial_{g} E\left[\varphi\left(W,\theta_{0},p_{0},g_{0}\right)\right]\left[g-g_{0}\right]\neq0,$$ where the directional (Gateaux) derivative is formally defined in Section 3; (2) ML estimators usually have a convergence
rate slower than $N^{-1/2}$ due to regularization bias. Similarly, the estimators obtained by directly plugging ML estimators into (2.2) and (2.3) will not be $\sqrt{N}$-consistent in general. The Monte Carlo simulation in Section 4 supports this theoretical insight and reveals significant bias on the estimators based on (2.1)-(2.3) when using ML estimators in the first-stage nonparametric estimation. 

The next section proposes three new score functions to relieve the regularization bias of the first-stage ML estimators. These three new score functions are derived under the same
identification assumptions as those in \citet*{abadie2005semiparametric}, so that no extra assumption
is made. Heuristically, a distinctive feature of the new score functions
is that their derivatives with respect to their infinite-dimensional
nuisance parameters are zero. This property can help us  remove
the first-order bias of the first-stage estimation so that the bias
of the estimators based on these new score functions will be much smaller. In addition, I will use
the cross-fitting algorithm to improve the over-fitting phenomena
that frequently arise when using highly adaptive ML methods
\citep*{chernozhukov2018double}. 

\section{The New DID Estimators}

\subsection{The Main Algorithm}

Supposing Assumptions (2.1)-(2.3) hold, consider the following three
new score functions.

\subsubsection*{\emph{Case 1: Random sample with repeated outcomes}}

The new score function for repeated outcomes is

\begin{align*}
\psi_{1}\left(W,\theta_{0},p_{0},\eta_{10}\right) & =\frac{Y\left(1\right)-Y\left(0\right)}{P\left(D=1\right)}\frac{D-P\left(D=1\mid X\right)}{1-P\left(D=1\mid X\right)}-\theta_{0}\\
 & -\underbrace{\frac{D-P\left(D=1\mid X\right)}{P\left(D=1\right)\left(1-P\left(D=1\mid X\right)\right)}E\left[Y\left(1\right)-Y\left(0\right)\mid X,D=0\right]}_{c_{1}},\tag{3.1}
\end{align*}
with the unknown constant $p_{0}$ and the infinite-dimensional nuisance
parameter 
\[
\eta_{10}=\left(P\left(D=1\mid X\right),E\left[Y\left(1\right)-Y\left(0\right)\mid X,D=0\right]\right)\eqqcolon\left(g_{0},\ell_{10}\right).
\]

\subsubsection*{\emph{Case 2: Random sample with repeated cross sections}}

The new score function for repeated cross sections is
\[
\psi_{2}\left(W,\theta_{0},p_{0},\lambda_{0},\eta_{20}\right)=\frac{T-\lambda_{0}}{\lambda_{0}\left(1-\lambda_{0}\right)}\frac{Y}{P\left(D=1\right)}\frac{D-P\left(D=1\mid X\right)}{1-P\left(D=1\mid X\right)}-\theta_{0}-c_{2},\tag{3.2}
\]
where the adjustment term is 
\[
c_{2}=\frac{D-P\left(D=1\mid X\right)}{\lambda_{0}\left(1-\lambda_{0}\right)\cdot P\left(D=1\right)\cdot\left(1-P\left(D=1\mid X\right)\right)}\times E\left[\left(T-\lambda_{0}\right)Y\mid X,D=0\right].
\]
The nuisance parameters are the unknown constants $p_{0}$ and $\lambda_{0}$,
and the infinite-dimensional parameter
\[
\eta_{20}=\left(P\left(D=1\mid X\right),E\left[\left(T-\lambda\right)Y\mid X,D=0\right]\right)\eqqcolon\left(g_{0},\ell_{20}\right).
\]

\subsubsection*{\emph{Case 3: Multilevel treatment}}

For each $w\in\left\{ w_{1},...,w_{J}\right\} $, the new score function
for multilevel treatment is
\begin{align*}
\psi_{w}\left(W,\theta_{w0},p_{w0},\eta_{w0}\right)= & \frac{Y\left(1\right)-Y\left(0\right)}{P\left(W=w\right)}\frac{I\left(W=w\right)\cdot P\left(W=0\mid X\right)-I\left(W=0\right)\cdot P\left(W=w\mid X\right)}{P\left(W=0\mid X\right)}\\
 & -\theta_{w0}-c_{w},\tag{3.3}
\end{align*}
where the adjustment term is 
\begin{align*}
c_{w}= & \left(\frac{I\left(W=w\right)\cdot P\left(W=0\mid X\right)-I\left(W=0\right)\cdot P\left(W=w\mid X\right)}{P\left(W=w\right)\cdot P\left(W=0\mid X\right)}\right)\times\\
 & E\left[Y\left(1\right)-Y\left(0\right)\mid X,I\left(W=0\right)=1\right].
\end{align*}
The nuisance parameters are the unknown constant $p_{w0}\coloneqq P\left(W=w\right)$
and the infinite-dimensional parameter
\[
\eta_{w0}=\left(P\left(W=w\mid X\right),P\left(W=0\mid X\right),E\left[Y\left(1\right)-Y\left(0\right)\mid X,I\left(W=0\right)=1\right]\right)\eqqcolon\left(g_{0w},g_{0z},\ell_{30}\right).
\]

Notice that the above three new functions are equal to the original score
functions (2.1)-(2.3) plus the adjustment terms, $\left(c_{1},c_{2},c_{w}\right)$,
which have zero expectations. Thus, the new score functions (3.1)-(3.3) still identify the ATT in each case. I will use these new scores to construct new DID estimators. 

To avoid repetition, I will focus on the estimation of ATT when data
belongs to repeated outcomes and repeated cross sections. The estimation
of multilevel treatment is provided in appendix. Now I combine the
score functions described above with the cross-fitting estimation
algorithm of \citet*{chernozhukov2018double}.
\begin{description}
\item [{Algorithm~1~}]~
\end{description}
\begin{enumerate}
\item \emph{Take a $K$-fold random partition $\left(I_{k}\right)_{k=1}^{K}$
of observation indices $\left[N\right]=\left\{ 1,...,N\right\}$. For simplicity, assume that each fold $I_{k}$ has the same size  $n=N/K$. For each $k\in\left[K\right]=\left\{ 1,...,K\right\} $,
define the auxiliary sample $I_{k}^{c}\coloneqq\left\{ 1,...,N\right\} \setminus I_{k}$. }

\item \emph{For each $k$, construct the intermediate ATT estimators 
\[
\tilde{\theta}_{k}=\frac{1}{n}\sum_{i\in I_{k}}\frac{D_{i}-\hat{g}_{k}\left(X_{i}\right)}{\hat{p}_{k}\left(1-\hat{g}_{k}\left(X_{i}\right)\right)}\times\left(Y_{i}\left(1\right)-Y_{i}\left(0\right)-\hat{\ell}_{1k}\left(X_{i}\right)\right)~\text{(repeated outcomes) }
\]
\[
\tilde{\theta}_{k}=\frac{1}{n}\sum_{i\in I_{k}}\frac{D_{i}-\hat{g}_{k}\left(X_{i}\right)}{\hat{p}_{k}\hat{\lambda}_{k}\left(1-\hat{\lambda}_{k}\right)\left(1-\hat{g}_{k}\left(X_{i}\right)\right)}\times\left(\left(T_{i}-\hat{\lambda}_{k}\right)Y_{i}-\hat{\ell}_{2k}\left(X_{i}\right)\right)~\text{(repeated cross sections)}
\]
}where $\hat{p}_{k}=\frac{1}{n}\sum_{i\in I_{k}^{c}}D_{i}$
, $\hat{\lambda}_{k}=\frac{1}{n}\sum_{i\in I_{k}^{c}}T_{i}$, and $\left(\hat{g}_{k},\hat{\ell}_{1k},\hat{\ell}_{2k}\right)$ are the estimators of $\left(g_{0},\ell_{10},\ell_{20}\right)$ constructed using the auxiliary sample $I_{k}^{c}$. 
\item \emph{Construct the final ATT estimator $\tilde{\theta}=\frac{1}{K}\sum_{k=1}^{K}\tilde{\theta}_{k}.$}
\end{enumerate}
The estimators $\left(\hat{g}_{k},\hat{\ell}_{1k},\hat{\ell}_{2k}\right)$
can be constructed using any ML methods or classical estimators such
as kernel or series estimators. For completeness, I present the Logit
Lasso and Lasso estimators here. 

Consider a class of approximating functions of $X_{i}$,
\[
q_{i}\coloneqq\left(q_{i1}\left(X_{i}\right),...,q_{ip}\left(X_{i}\right)\right)'.
\]
For example, $q_{i}$ can be polynomials or
B-splines. Let $\Lambda\left(u\right)\coloneqq1/\left(1+\exp\left(-u\right)\right)$ be
the cumulative distribution function of the standard Logistic distribution, construct
the estimator of the propensity score $g_{0}$ by 
\[
\hat{g}_{k}\left(x_{i}\right)\coloneqq\Lambda\left(q_{i}'\hat{\beta}_{k}\right),\tag{3.4}
\]
where 
\[
\hat{\beta}_{k}\coloneqq\arg\min_{\beta\in\mathbb{R}^{p}}\frac{1}{M}\sum_{i\in I_{k}^{c}}\left\{ -D_{i}(q_{i}'\beta)+\log\left(1+\exp\left(q_{i}'\beta\right)\right)\right\} +\lambda_{k}\parallel\beta\parallel_{1}
\]
is the Logit Lasso estimator and $M=N-n$ is the sample size of the auxiliary sample  $I_{k}^{c}$.
Next, define $I_{kz}^{c}\coloneqq I_{k}^{c}\cap\left\{ i:D_{i}=0\right\} $,
$M_{k}$ the sample size of $I_{kz}^{c}$. Construct
the estimators of $\ell_{10}$ and $\ell_{20}$ by 
\[
\hat{\ell}_{1k}\left(x_{i}\right)\coloneqq q_{i}'\hat{\beta}_{1k},
\]
\[
\hat{\ell}_{2k}\left(x_{i}\right)\coloneqq q_{i}'\hat{\beta}_{2k},
\]
where 
\[
\hat{\beta}_{1k}\in\arg\min_{\beta\in\mathbb{R}^{p}}\left[\frac{1}{M_{k}}\sum_{i\in I_{kz}^{c}}\left(Y_{i}\left(1\right)-Y_{i}\left(0\right)-q_{i}'\beta\right)^{2}\right]+\frac{\lambda_{1k}}{M_{k}}\parallel\hat{\Upsilon}_{1k}\beta\parallel_{1}
\]
and 
\[
\hat{\beta}_{2k}\in\arg\min_{\beta\in\mathbb{R}^{p}}\left[\frac{1}{M_{k}}\sum_{i\in I_{kz}^{c}}\left(\left(T_{i}-\hat{\lambda}_{k}\right)Y_{i}-q_{i}'\beta\right)^{2}\right]+\frac{\lambda_{2k}}{M_{k}}\parallel\hat{\Upsilon}_{2k}\beta\parallel_{1}
\]
 are the modified Lasso estimators proposed in \citet*{belloni2012sparse}.
The choices of the penalty levels and loadings $\left(\lambda_{1k},\lambda_{2k},\hat{\Upsilon}_{1k},\hat{\Upsilon}_{2k}\right)$
suggested by \citet*{belloni2012sparse} are provided in appendix.  

\subsection{Theoretical Properties}

In this section, I discuss the theoretical properties of the new DID
estimator $\tilde{\theta}$. In particular, I will show that the estimator
$\tilde{\theta}$ can achieve $\sqrt{N}$-consistency and asymptotic
normality as long as the first-stage estimators converge at rates
faster than $N^{-1/4}$. This rate of convergence can be achieved
by many ML methods, including Lasso and Logit Lasso. Further, I will
show that when using kernel estimators in the first-stage estimation,
the estimator $\tilde{\theta}$ has the SBP while the conventional semiparametric
DID estimators do not. 

\subsubsection{The Neyman Orthogonality }

The differences between the new DID estimators and the conventional semiparametric DID estimators in \citet*{abadie2005semiparametric} are the score functions on which they are based. The key property of the new score functions (3.1)-(3.3) is that their directional (or the Gateaux) derivatives with
respect to their infinite-dimensional nuisance parameters are zero, while the scores based on (2.1)-(2.3) do not have this property.
This property is the so-called Neyman orthogonality in \citet*{chernozhukov2018double}.
The Neyman orthogonality enables us to remove the first-order bias of the first-stage estimation so that the estimators based on these
Neyman-orthogonal scores can achieve $\sqrt{N}$-consistency under
less restrictive conditions. 

The definition of the Neyman-orthogonal score provided here is slightly
different from \citet*{chernozhukov2018double} that instead of being orthogonal
against all nuisance parameters, the Neyman-orthogonal score defined
here is orthogonal against only those infinite-dimensional nuisance
parameters. Formally, let $\theta_{0}\in\Theta$ be the low-dimensional
parameter of interest, $\rho_{0}$ be the true value of the finite-dimensional
nuisance parameter $\rho$, and $\eta_{0}$ the true value of the
infinite-dimensional nuisance parameter $\eta\in\mathcal{T}$ . Suppose
that $W$ is a random element taking values in a measurable space
$\left(\mathcal{W},\mathcal{A}_{\mathcal{W}}\right)$ with probability
measure $P$. Define the directional (or the Gateaux) derivative against
the infinite-dimensional nuisance parameter $D_{r}:\tilde{\mathcal{T}}\rightarrow\mathbb{R}$,
where $\tilde{\mathcal{T}}=\left\{ \eta-\eta_{0}:\eta\in\mathcal{T}\right\} $,
\[
D_{r}\left[\eta-\eta_{0}\right]\coloneqq\partial_{r}\left\{ E_{P}\left[\psi\left(W,\theta_{0},\rho_{0},\eta_{0}+r\left(\eta-\eta_{0}\right)\right)\right]\right\} ,\eta\in\mathcal{T},
\]
for all $r\in[0,1)$. For convenience,  denote
\[
\partial_{\eta}E_{P}\psi\left(W,\theta_{0},\rho_{0},\eta_{0}\right)\left[\eta-\eta_{0}\right]\coloneqq D_{0}\left[\eta-\eta_{0}\right],\eta\in\mathcal{T}.
\]
In addition, let $\mathcal{T}_{N}\subset\mathcal{T}$ be a nuisance
realization set such that the estimator of $\eta_{0}$ take values
in this set with high probability.
\begin{description}
\item [{Definition}] (The Neyman Orthogonality)
\end{description}
\emph{The score $\psi$ obeys the Neyman orthogonality condition at
$\left(\theta_{0},\rho_{0},\eta_{0}\right)$ with respect to the nuisance
parameter realization set $\mathcal{T}_{N}\subset\mathcal{T}$ if
the directional derivative map $D_{r}\left[\eta-\eta_{0}\right]$
exists for all $r\in[0,1)$ and $\eta\in\mathcal{T}_{N}$ and vanishes
at $r=0$:
\[
\partial_{\eta}E_{P}\psi\left(W,\theta_{0},\rho_{0},\eta_{0}\right)\left[\eta-\eta_{0}\right]=0,\text{for all }\eta\in\mathcal{T}_{N}.
\]
}
\begin{description}
\item [{Lemma~1}] \emph{The new score functions (3.1)-(3.3) obey the Neyman
orthogonality.}
\end{description}

This property embedded in (3.1)-(3.3) will play the key role to make less restrictive assumptions in the following proofs of asymptotic distribution and the SBP. 

\subsubsection{Asymptotic Distribution}

In the following, I will discuss the theoretical properties of the
new estimator $\tilde{\theta}$ when data belongs to repeated outcomes
and repeated cross sections. The results of multilevel treatment can be proven using the same arguments. Let $\kappa$ and $C$ be
strictly positive constants, $K\geq2$ be a fixed integer, and $\varepsilon_{N}$
be a sequence of positive constants approaching zero. Denote by $\parallel \cdot \parallel_{P,q}$ the $L^{q}$ norm of some probability measure $P$:  $\parallel f \parallel_{P,q} \coloneqq \left(\int \mid f\left(w\right) \mid^{q}dP\left(w\right)\right)^{1/q}$ and $\parallel f \parallel_{P,\infty} \coloneqq \sup_{w} \mid f\left(w\right) \mid $.
\begin{description}
\item [{Assumption~3.1}] (Regularity Conditions for Repeated Outcomes)
\end{description}
Let $P$ be the probability law for $\left(Y\left(0\right),Y\left(1\right),D,X\right)$.
Let $D=g_{0}\left(X\right)+U$ and $Y\left(1\right)-Y\left(0\right)=\ell_{10}\left(X\right)+V_{1}$
with $E_{P}\left[U\mid X\right]=0$ and $E_{P}\left[V_{1}\mid X, D=0\right]=0$.
Define $G_{1p0}\coloneqq E_{P}\left[\partial_{p}\psi_{1}\left(W,\theta_{0},p_{0},\eta_{10}\right)\right]$
and $\Sigma_{10}\coloneqq E_{P}\left[\left(\psi_{1}\left(W,\theta_{0},p_{0},\eta_{10}\right)+G_{1p0}\left(D-p_{0}\right)\right)^{2}\right]$.
Suppose the following conditions hold: (a) $Pr\left(\kappa\leq g_{0}\left(X\right)\leq1-\kappa\right)=1$;
(b) $\parallel UV_{1}\parallel_{P,4}\leq C$; (c) $E\left[U^{2}\mid X\right]\leq C$;
(d) $E\left[V_{1}^{2}\mid X\right]\leq C$; (e) $\Sigma_{10}>0$;
and (f) given the auxiliary sample $I_{k}^{c}$, the estimator $\hat{\eta}_{1k}=\left(\hat{g}_{k},\hat{\ell}_{1k}\right)$
obeys the following conditions. With probability $1-o\left(1\right)$,
\emph{$\parallel\hat{\eta}_{1k}-\eta_{10}\parallel_{P,2}\leq\varepsilon_{N}$,
$\parallel\hat{g}_{k}-1/2\parallel_{P,\infty}\leq1/2-\kappa$,
and $\parallel\hat{g}_{k}-g_{0}\parallel_{P,2}^{2}+\parallel\hat{g}_{k}-g_{0}\parallel_{P,2}\times\parallel\hat{\ell}_{1k}-\ell_{10}\parallel_{P,2}\leq\left(\varepsilon_{N}\right)^{2}$. }
\begin{description}
\item [{Assumption~3.2}] (Regularity Conditions for Repeated Cross Sections)
\end{description}
Let $P$ be the probability law for $\left(Y,T,D,X\right)$. Let
$D=g_{0}\left(X\right)+U$ and $\left(T-\lambda_{0}\right)Y=\ell_{20}\left(X\right)+V_{2}$
with $E_{p}\left[U\mid X\right]=0$ and $E_{p}\left[V_{2}\mid X, D=0\right]=0$.
Define $G_{2p0}\coloneqq E_{P}\left[\partial_{p}\psi_{2}\left(W,\theta_{0},p_{0},\lambda_{0},\eta_{20}\right)\right]$,
$G_{2\lambda0}\coloneqq E_{P}\left[\partial_{\lambda}\psi_{2}\left(W,\theta_{0},p_{0},\lambda_{0},\eta_{20}\right)\right]$,
and $\Sigma_{20}\coloneqq E_{P}\left[\left(\psi_{1}\left(W,\theta_{0},p_{0},\eta_{10}\right)+G_{2p0}\left(D-p_{0}\right)+G_{2\lambda0}\left(T-\lambda_{0}\right)\right)^{2}\right]$.
Suppose the following conditions hold: (a) $Pr\left(\kappa\leq g_{0}\left(X\right)\leq1-\kappa\right)=1$;
(b) $\parallel UV_{2}\parallel_{P,4}\leq C$; (c) $E\left[U^{2}\mid X\right]\leq C$;
(d) $E\left[V_{2}^{2}\mid X\right]\leq C$; (e) $E_{P}\left[Y^{2}\mid X\right]\leq C$;
(f) $\mid E_{P}\left[YU\right]\mid\leq C$; (g) $\Sigma_{20}>0$;
and (h) given the auxiliary sample $I_{k}^{c}$, the estimators $\hat{\eta}_{2k}=\left(\hat{g}_{k},\hat{\ell}_{2k}\right)$
obeys the following conditions. With probability $1-o\left(1\right)$,
\emph{$\parallel\hat{\eta}_{2k}-\eta_{20}\parallel_{P,2}\leq\varepsilon_{N}$,
$\parallel\hat{g}_{k}-1/2\parallel_{P,\infty}\leq1/2-\kappa$,
and $\parallel\hat{g}_{k}-g_{0}\parallel_{P,2}^{2}+\parallel\hat{g}_{k}-g_{0}\parallel_{P,2}\times\parallel\hat{\ell}_{2k}-\ell_{20}\parallel_{P,2}\leq\left(\varepsilon_{N}\right)^{2}$. }
\begin{description}
\item [{Theorem~1~}]~
\end{description}
\emph{For repeated outcomes, suppose Assumptions (2.1), (2.2) and (3.1) hold. For repeated cross sections, suppose Assumptions  (2.1)-(2.3) and (3.2) hold. If $\varepsilon_{N}=o\left(N^{-1/4}\right)$,
then the new ATT estimator $\tilde{\theta}$ satisfies
\[
\sqrt{N}\left(\tilde{\theta}-\theta_{0}\right)\rightarrow N\left(0,\Sigma\right)
\]
with $\Sigma=\Sigma_{10}$ for repeated outcomes and $\Sigma=\Sigma_{20}$
for repeated cross sections. }
\begin{description}
\item [{Theorem~2~(Variance~Estimator)}]~
\end{description}
\emph{Construct the estimators of the asymptotic variances as
\[
\hat{\Sigma}_{1}=\frac{1}{K}\sum_{k=1}^{K}\mathbb{E}_{n,k}\left[\left(\psi_{1}\left(W,\tilde{\theta},\hat{p}_{k},\hat{\eta}_{1k}\right)+\hat{G}_{1p}\left(D-\hat{p}_{k}\right)\right)^{2}\right]\tag{repeated outcomes}
\]
\[
\hat{\Sigma}_{2}=\frac{1}{K}\sum_{k=1}^{K}\mathbb{E}_{n,k}\left[\left(\psi_{2}\left(W,\tilde{\theta},\hat{p}_{k},\hat{\lambda}_{k},\hat{\eta}_{2k}\right)+\hat{G}_{2p}\left(D-\hat{p}_{k}\right)+\hat{G}_{2\lambda}\left(T-\hat{\lambda}_{k}\right)\right)^{2}\right]\tag{repeated cross sections}
\]
where $\mathbb{E}_{n,k}\left[f\left(W\right)\right]=n^{-1}\sum_{i\in I_{k}}f\left(W_{i}\right)$,
$\hat{G}_{1p}=\hat{G}_{2p}=-\tilde{\theta}/\hat{p}_{k}$, and $\hat{G}_{2\lambda}$
is a consistent estimator of $G_{2\lambda0}$. If the assumptions
of Theorem 1 hold, then $\hat{\Sigma}_{1}=\Sigma_{10}+o_{P}\left(1\right)$
and $\hat{\Sigma}_{2}=\Sigma_{20}+o_{P}\left(1\right).$}
\bigskip

The interpretation of Theorem 1 and 2 is that the new DID  estimator $\tilde{\theta}$
can achieve $\sqrt{N}$-consistency and asymptotic normality provided that
the first-stage estimators of the infinite dimensional nuisance parameters
converge at a rate faster than $N^{-1/4}$. This rate of convergence
can be achieved by many ML methods. In particular, \citet*{van2008high}
and \citet*{belloni2012sparse} provided detail conditions for Logit Lasso
and the modified Lasso estimators to satisfy this rate of convergence. It is also
worth noting that even when the first-stage estimators do not converge
as fast as $N^{-1/4}$, the new estimator $\tilde{\theta}$ still
has smaller bias than the original estimator because the Neyman orthogonality
removes the first-order bias of the first-stage estimators. 

\subsubsection{The Small Bias Property}

Consider the conventional semiparametric DID estimators with a limited number
of control variables studied in \citet*{abadie2005semiparametric}. Let $\widehat{g}_{h}$
be the kernel estimator of $g_{0}$ with bandwidth $h\rightarrow0$ in (2.1) and (2.2). Under the standard assumptions of kernel estimation (Assumption (3.3) below), one can show that the pointwise bias of $\hat{g}_{h}$ is of order $O\left(h^{m}\right)$, where $m$ can be interpreted as the minimum number of derivatives of $g_{0}$; and the pointwise standard deviation is $sd\left(\hat{g}_{h}\left(x\right)\right)=O(\left(Nh^{d+2s}\right)^{-1/2})$. By Theorem 8.11 of \citet*{newey1994large}, one can show that the $\sqrt{N}$-consistency of the plug-in estimators based on (2.1) and (2.2) requires $\sqrt{N}h^{m}\rightarrow 0$. That is, the pointwise bias of the kernel estimator has to converge to zero faster than $N^{-1/2}$.   Since the pointwise standard deviation converges to zero slower than $N^{-1/2}$, undersmoothing is required. In this case, standard data-driven bandwidth selection methods which do not undersmooth, such as cross-validation, are invalid.

To avoid undersmoothing, by the analysis of SBP in \citet*{newey1998undersmoothing, newey2004twicing},  the estimator of the parameter of interest needs to have smaller bias than the pointwise bias of the first-stage nonparametric estimators. That is, the SBP requires that the bias of the estimator of $\theta_{0}$ converges to zero faster than $h^{m}$.

In the following, I will show that the new DID estimator $\tilde{\theta}$
has the SBP. Let $\left(\hat{g}_{kh},\hat{\ell}_{1kh},\hat{\ell}_{2kh}\right)$
be the kernel estimators of $\left(g_{0},\ell_{10},\ell_{20}\right)$
using auxiliary sample $I_{k}^{c}$. I assume here that they have
the same bandwidth $h$ and kernel $K\left(u\right)$ for convenience.

\begin{description}
\item [{Assumption~3.3}] \citep*{newey1994large}
\end{description}
\begin{enumerate}
\item $K\left(u\right)$ is differentiable of order $s$, the derivatives
of order $s$ are bounded, $K\left(u\right)$ is zero outside a bounded
set, $\int K\left(u\right)du=1$, there is a positive $m$ such
that for all $j<m$, $\int K\left(u\right)\left[\bigotimes_{\ell=1}^{j}u\right]du=0.$
\item Define $\gamma_{0}\left(x\right)=f_{0}\left(x\right)E\left(z\mid x\right)$,
where $z\in \left(1,D,Y\left(1\right)-Y\left(0\right)\mid D=0,\left(T-\lambda_{0}\right)Y\mid D=0\right)$
and $f_{0}\left(x\right)$ is the true density of $x$. Assume that
$\gamma_{0}\left(x\right)$ is continuously differentiable to order
$s$ with bounded derivatives on an open set containing  $\mathcal{X}$, where $\mathcal{X}$ is the support of $x$.
\item There is $\alpha\geq4$ such that $E\left[\mid z\mid^{\alpha}\right]<\infty$
and $E\left[\mid z\mid^{\alpha}\mid x\right]f_{0}\left(x\right)$
is bounded.
\end{enumerate}
\begin{description}
\item [{Theorem~3}]~
\end{description}
\emph{For repeated outcomes, suppose Assumptions (2.1), (2.2),  (3.1), and (3.3) hold. For repeated cross sections, suppose Assumptions (2.1)-(2.3), (3.2), and (3.3) hold. Suppose
that  $\inf_{x\in\mathcal{X}}f_{0}\left(x\right)\neq 0$, $h=h\left(N\right)$ with  $\log N/\left(\sqrt{N}h^{d+2s}\right)\rightarrow 0$. If $\sqrt{N}h^{2m}\rightarrow0$,
then
\[
\sqrt{N}\left(\tilde{\theta}-\theta_{0}\right)\rightarrow N\left(0,\Sigma\right)
\]
with $\Sigma=\Sigma_{10}$ for repeated outcomes and $\Sigma=\Sigma_{20}$
for repeated cross sections. }
\bigskip

The interpretation of Theorem 3 is that the new estimator $\tilde{\theta}$
only requires $\sqrt{N}h^{2m}\rightarrow0$ to achieve $\sqrt{N}$-consistency,
while the conventional semiparametric DID estimators require $\sqrt{N}h^{m}\rightarrow0$
under the same assumptions. With the Neyman orthogonality,
the bias of $\tilde{\theta}$ is only of the second-order of the pointwise
bias of the first-stage kernel estimators. The bias of $\tilde{\theta}$
is $h^{2m}$ instead of $h^{m}$. Hence, $\tilde{\theta}$ satisfies  the SBP. In particular, the bandwidth $h$ such that  $\log N/\left(\sqrt{N}h^{d+2s}\right)\rightarrow 0$ and $\sqrt{N}h^{2m}\rightarrow0$ exists only if $2m>d+2s$. Under this condition, the optimal bandwidth selected by  minimizing  mean-square errors (CV),  $h=N^{-1/\left(d+2s+2m\right)}$,  satisfies the conditions for $\sqrt{N}$-consistency. 
\begin{description}
\item [{Theorem~4~}]~
\end{description}
\emph{Construct the estimators of the asymptotic variances as
\[
\hat{\Sigma}_{1}=\frac{1}{K}\sum_{k=1}^{K}\mathbb{E}_{n,k}\left[\left(\psi_{1}\left(W,\tilde{\theta},\hat{p}_{k},\hat{\eta}_{1kh}\right)+\hat{G}_{1p}\left(D-\hat{p}_{k}\right)\right)^{2}\right] \tag{repeated outcomes}
\]
\[
\hat{\Sigma}_{2}=\frac{1}{K}\sum_{k=1}^{K}\mathbb{E}_{n,k}\left[\left(\psi_{2}\left(W,\tilde{\theta},\hat{p}_{k},\hat{\lambda}_{k},\hat{\eta}_{2kh}\right)+\hat{G}_{2p}\left(D-\hat{p}_{k}\right)+\hat{G}_{2\lambda}\left(T-\hat{\lambda}_{k}\right)\right)^{2}\right] \tag{repeated cross sections}
\]
where $\hat{G}_{1p}=\hat{G}_{2p}=-\tilde{\theta}/\hat{p}_{k}$ and
$\hat{G}_{2\lambda}$ is a consistent estimator of $G_{2\lambda0}$.
If the assumptions of Theorem 3 hold, then $\hat{\Sigma}_{1}=\Sigma_{10}+o_{P}\left(1\right)$
and $\hat{\Sigma}_{2}=\Sigma_{20}+o_{P}\left(1\right).$}

\section{Simulation}

In this section, I present Monte Carlo simulation results of the conventional semiparametric DID estimators and the new DID estimator
$\tilde{\theta}$ in three different data structures: repeated outcomes, repeated cross
sections, and multilevel treatment. I use both ML methods and kernel estimators in the first-stage estimation. For ML estimation, I generate high-dimensional (HD) data and estimate the propensity score by Logit Lasso (Multi-Logit Lasso for multilevel treatment). To choose the penalty parameter for Logit Lasso (Multi-Logit Lasso), I use $K$-fold CV (as recommended by \citet*{van2008high}) with $K=10$. Alternatively, one could use a method developed in \citet*{belloni2018uniformly}. The other infinite-dimensional nuisance parameters are estimated by random forests with 500 regression trees. For kernel estimation, all the infinite-dimensional nuisance parameters are estimated using the standard Gaussian kernel. 

Figure 3-20 in appendix show the simulation results. I find that the conventional semiparametric DID estimators are biased when using ML methods, while the new DID estimator $\tilde{\theta}$ can correct the bias. For kernel estimation, the conventional DID estimator with bandwidth selected by CV is biased, while the new DID estimator $\tilde{\theta}$ is centered at the true value. The data generating processes are presented in the following. 

\subsection{Repeated Outcomes}

\subsubsection{ML Estimation}

Let $N\in \left\{200,500\right\}$ be the sample size and $p\in \left\{100,300\right\}$ the dimension of control variables,
$X_{i}\sim N\left(0,I_{p\times p}\right)$. Also, let $\gamma_{0}=\left(1,1/2,1/3,1/4,1/5,0,...,0\right)\in\mathbb{R}^{p}$
and $D_{i}$ is generated by the propensity score 
\[
P\left(D=1\mid X\right)=\frac{1}{1+\exp\left(-X'\gamma_{0}\right)}\text{(Logistic)}.
\]
At $t=0$, the potential outcome is generated 
\[
Y_{i}^{0}\left(0\right)=X_{i}'\beta_{0}+\varepsilon_{1},
\]
and at $t=1$,
\[
Y_{i}^{0}\left(1\right)=Y_{i}^{0}\left(0\right)+1+\varepsilon_{2},
\]
\[
Y_{i}^{1}\left(1\right)=\theta_{0}+Y_{i}^{0}\left(1\right)+\varepsilon_{3},
\]
where $\beta_{0}=\gamma_{0}+0.5$ and $\theta_{0}=3$, and all error
terms follow $N\left(0,0.1\right)$. Researchers observe $\left\{ Y_{i}\left(0\right),Y_{i}\left(1\right),D_{i},X_{i}\right\}$ for $i=1,...,N$,
where $Y_{i}\left(0\right)=Y_{i}^{0}\left(0\right)$ and $Y_{i}\left(1\right)=Y_{i}^{0}\left(1\right)\left(1-D_{i}\right)+Y_{i}^{1}\left(1\right)D_{i}$. Figure 3-6 present the results.  

\subsubsection{Kernel Estimation}

Let $N\in \left\{200,500\right\}$ be the sample size, $D_{i}\sim Bernoulli(0.5)$, and $X_{i}\mid D_{i}\sim N\left(D_{i},1\right)$. At $t=0$, the potential outcome is generated 
\[
Y_{i}^{0}\left(0\right)=\varepsilon_{1},
\]
and at $t=1$,
\[
Y_{i}^{0}\left(1\right)=Y_{i}^{0}\left(0\right)+X_{i}+\varepsilon_{2},
\]
\[
Y_{i}^{1}\left(1\right)=\theta_{0}+Y_{i}^{0}\left(1\right)+\varepsilon_{3},
\]
where $\theta_{0}=3$ and all error
terms follow $N\left(0,0.1\right)$. Researchers observe $\left\{ Y_{i}\left(0\right),Y_{i}\left(1\right),D_{i},X_{i}\right\}$ for $i=1,...,N$,
where $Y_{i}\left(0\right)=Y_{i}^{0}\left(0\right)$ and $Y_{i}\left(1\right)=Y_{i}^{0}\left(1\right)\left(1-D_{i}\right)+Y_{i}^{1}\left(1\right)D_{i}$. Figure 7-8 present the results.

\subsection{Repeated Cross Sections}

\subsubsection{ML Estimation}

Let $N\in \left\{200,500\right\}$ be the sample size and $p\in \left\{100,300\right\}$ the dimension of control variables,
$X_{i}\sim N\left(0.3,I_{p\times p}\right)$. Also, let $\gamma_{0}=\left(1,1/2,1/3,1/4,1/5,0,...,0\right)\in\mathbb{R}^{p}$
and $D$ is generated by the propensity score 
\[
P\left(D=1\mid X\right)=\frac{1}{1+\exp\left(-X'\gamma_{0}\right)}\text{(Logistic)}.
\]
At $t=0$, the potential outcome is generated 
\[
Y_{i}^{0}\left(0\right)=1+\varepsilon_{1},
\]
and at $t=1$,
\[
Y_{i}^{0}\left(1\right)=Y_{i}^{0}\left(0\right)+1+\varepsilon_{2},
\]
\[
Y_{i}^{1}\left(1\right)=\theta_{0}+Y_{i}^{0}\left(1\right)+\varepsilon_{3},
\]
where $\beta_{0}=\gamma_{0}+0.5$ and $\theta_{0}=3$, and all error
terms follow $N\left(0,0.1\right)$. Define $Y_{i}\left(0\right)=Y_{i}^{0}\left(0\right)$
and $Y_{i}\left(1\right)=Y_{i}^{0}\left(1\right)\left(1-D_{i}\right)+Y_{i}^{1}\left(1\right)D_{i}$.
Let $T_{i}$ follow a Bernoulli distribution with parameter $0.5$.
Researchers observe $\left\{ Y_{i},T_{i},D_{i},X_{i}\right\}$ for $i=1,...,N$,
where $Y_{i}=Y_{i}\left(0\right)+T_{i}\left(Y_{i}\left(1\right)-Y_{i}\left(0\right)\right)$. Figure 9-12 present the results. 

\subsubsection{Kernel Estimation}

Let $N\in \left\{200,500\right\}$ be the sample size, $D_{i}\sim Bernoulli(0.5)$, and $X_{i}\mid D_{i}\sim N\left(D_{i},1\right)$. At $t=0$, the potential outcome is generated 
\[
Y_{i}^{0}\left(0\right)=\varepsilon_{1},
\]
and at $t=1$,
\[
Y_{i}^{0}\left(1\right)=Y_{i}^{0}\left(0\right)+X_{i}+\varepsilon_{2},
\]
\[
Y_{i}^{1}\left(1\right)=\theta_{0}+Y_{i}^{0}\left(1\right)+\varepsilon_{3},
\]
where $\theta_{0}=3$ and all error
terms follow $N\left(0,0.1\right)$. Let $Y_{i}\left(0\right)=Y_{i}^{0}\left(0\right)$ and $Y_{i}\left(1\right)=Y_{i}^{0}\left(1\right)\left(1-D_{i}\right)+Y_{i}^{1}\left(1\right)D_{i}$.
Let $T_{i}\sim Bernoulli(0.5)$.
Researchers observe $\left\{ Y_{i},T_{i},D_{i},X_{i}\right\}$ for $i=1,...,N$,
where $Y_{i}=Y_{i}\left(0\right)+T_{i}\left(Y_{i}\left(1\right)-Y_{i}\left(0\right)\right)$. Figure 13-14 present the results.

\subsection{Multilevel Treatment}

\subsubsection{ML Estimation}

Suppose there are two levels of treatment so that $W\in\left\{ 0,1,2\right\}$. Let $N\in \left\{200,500\right\}$ be the sample size and $p\in \left\{100,300\right\}$ the dimension of control variables, $X_{i}\sim N\left(0,I_{p\times p}\right)$.
Also, let $\gamma_{0}=\left(1,1/2,1/3,1/4,1/5,0,...,0\right)\in\mathbb{R}^{p}$ and
$$\left(P\left(W=0\right),P\left(W=1\right),P\left(W=2\right)\right)=\left(0.3,0.3,0.4\right)$$
At $t=0$, the potential outcome is generated 
\[
Y_{i}^{0}\left(0\right)=X'\beta_{0}+\varepsilon_{1},
\]
and at $t=1$,
\[
Y_{i}^{0}\left(1\right)=Y_{i}^{0}\left(0\right)+1+\varepsilon_{2},
\]
\[
Y_{i}^{1}\left(1\right)=\theta_{10}+Y_{i}^{0}\left(1\right)+\varepsilon_{3},
\]
\[
Y_{i}^{2}\left(1\right)=\theta_{20}+Y_{i}^{0}\left(1\right)+\varepsilon_{4},
\]
where $\beta_{0}=\gamma_{0}+0.5$ and $\theta_{10}=3$ and $\theta_{20}=6$,
and all error terms follow $N\left(0,0.1\right)$. Researchers observe
$\left\{ Y_{i}\left(0\right),Y_{i}\left(1\right),W_{i},X_{i}\right\}$ for $i=1,...,N$,
where $Y_{i}\left(0\right)=Y_{i}^{0}\left(0\right)$ and $Y_{i}\left(1\right)=Y_{i}^{0}\left(1\right)I\left(W_{i}=0\right)+Y_{i}^{1}\left(1\right)I\left(W_{i}=1\right)+Y_{i}^{2}\left(1\right)I\left(W_{i}=2\right)$.
I focus on the estimation of the second level ATT $\theta_{20}$. Figure 15-18 present the results
\subsubsection{Kernel Estimation}

Suppose there are two levels of treatment so that $W\in\left\{ 0,1,2\right\}$. Let $N$ be the sample
size, $X_{i}\mid W_{i}\sim N\left(W_{i},1\right)$, and

\[
P\left(W_{i}=0\right)=P\left(W_{i}=1\right)=P\left(W_{i}=2\right)=\frac{1}{3}.
\]
At $t=0$, the potential outcome is generated 
\[
Y_{i}^{0}\left(0\right)=\varepsilon_{1},
\]
and at $t=1$,
\[
Y_{i}^{0}\left(1\right)=Y_{i}^{0}\left(0\right)+X_{i}+\varepsilon_{2},
\]
\[
Y_{i}^{1}\left(1\right)=\theta_{10}+Y_{i}^{0}\left(1\right)+\varepsilon_{3},
\]
\[
Y_{i}^{2}\left(1\right)=\theta_{20}+Y_{i}^{0}\left(1\right)+\varepsilon_{4},
\]
where  $\theta_{10}=3$, $\theta_{20}=6$,
and all error terms follow $N\left(0,0.1\right)$. 
Let $Y_{i}\left(0\right)=Y_{i}^{0}\left(0\right)$ and $Y_{i}\left(1\right)=Y_{i}^{0}\left(1\right)I\left(W_{i}=0\right)+Y_{i}^{1}\left(1\right)I\left(W_{i}=1\right)+Y_{i}^{2}\left(1\right)I\left(W_{i}=2\right)$. Researchers observe
$\{ Y_{i}\left(0\right),Y_{i}\left(1\right),W_{i},X_{i}\}$ for $i=1,...,N$.
I focus on the estimation of the second level ATT $\theta_{20}$. Figure 19-20 present the results. 

\section{Empirical Example}

In this example, I analyze the effect of tariffs reduction on corruption behaviors using the bribe payment data collected by \citet*{sequeira2016corruption} between South Africa and Mozambique. There have been theoretical and empirical debates on whether higher tariff rates increase incentives for corruption to occur \citep*{clotfelter1983tax,sequeira2014corruption} or lower tariffs encourage agents to pay higher bribes through an income effect \citep*{feinstein1991econometric,slemrod2002tax}. The former argues that an increase in the tariff rate makes it more profitable to evade taxes on the margin. The latter argues that an increased tariff rate makes the tax payer less wealthy and this, under the decreasing risk aversion of being penalized, tend to reduce evasion \citep*{allingham1972income}. 

\citet*{sequeira2016corruption} collected primary data on the bribed payments between the ports in Mozambique and South Africa from 2007 to 2013. The treatment is the large reduction in the average nominal tariff rate (of 5 percent) occurring in 2008.    Since not all products were on the tariff reduction list, a credible control group of products is available. This allows for a DID estimation.  

This natural experiment between South Africa and Mozambique was previously studied by \citet*{sequeira2016corruption} by pooling the cross section data between 2007 and 2013, with sample size $N=1084$, and estimating the effect of treatment using the traditional linear DID. Here I focus on the specification of one of the main results (Table 9 of  \citet*{sequeira2016corruption}):
\begin{align*}
y_{it}= &\gamma_{1}TariffChangeCategory_{i}\times POST\\
 & +\mu POST+\gamma_{2}TariffChangeCategory_{i}\\
 & +\beta_{2}BaselineTariff_{i}+\Gamma_{i}\\
 & +p_{i}+w_{t}+\delta_{i}+\epsilon_{it}
\end{align*} where $y_{it}$ is the natural log of the amount of bribe paid for shipment $i$ in period $t$, conditional on paying a bribe. $TariffChangeCategory\in\left\{ 0,1\right\}$  denotes the treatment status of commodities, $POST\in\left\{ 0,1\right\}$  is an indicator for the years following 2008, and $BaselineTariff$ is the tariff rate before the tariff reduction. The specification also includes a vector of characteristics $\Gamma_{i}$, and time and individual fixed effects $p_{i}$, $w_{t}$, and $\delta_{i}$. The parameter $\gamma_{1}$ is the parameter of interest in the traditional linear DID estimation. \citet*{sequeira2016corruption} found that the amount of bribes paid dropped after the tariff reduction ($\hat{\gamma}_{1}=-2.928^{**}$).

I use the same data set but instead of using the traditional linear DID estimation, I estimate the ATT by \citet*{abadie2005semiparametric}'s DID estimator and my proposed DID estimator $\tilde{\theta}$. Since the data is repeated cross sections, I construct the estimators based on (2.2) and (3.2), respectively. The estimators with the first-stage kernel estimation contain one individual characteristic (the natural log of shipment value per ton), which is a significant characteristic in Table 9 of \citet*{sequeira2016corruption}. The estimators with the first-stage Lasso estimation contain a list of the significant characteristics in Table 9 of \citet*{sequeira2016corruption}, which includes product, shipment, firm-level characteristics,  and their interaction terms. Table 1 below shows the results. I find that all these estimators consistently suggest that a decrease in tariff rate will lead to less bribes payment, but the effect of treatment may be actually substantially larger than previously reported by \citet*{sequeira2016corruption}.

\begin{table}[H]
\begin{centering}
\begin{tabular}{c|c|c|c|c|c}

 & Sequeira (2016) & Abadie (kernel) & $\tilde{\theta}$ (kernel)  & Abadie (Lasso) & $\tilde{\theta}$ (Lasso)\tabularnewline
\hline 
ATT & -2.928{*}{*} (0.944) & -7.986** (3.028) & -8.670** (3.643) & -7.499** (2.746)& -9.191* (4.854)\tabularnewline
\hline 

\hline 
\end{tabular}
\par\end{centering}
\caption{}
\end{table}

\section{Conclusion}
In this article, I have introduced three new DID estimators based on the newly-derived Neyman-orthogonal scores. These new scores do not require any additional conditions other than the original conditions made in \citet*{abadie2005semiparametric}. The new DID estimators will be particularly appropriate when researchers would like to use ML methods in the first-stage nonparametric estimation. When using kernel estimators in the first-stage estimation , the new DID estimators do not require undersmoothing to achieve $\sqrt{N}$-consistency. Hence, researchers can use standard data-driven methods, such as CV, to select bandwidths. 

\bibliographystyle{apacite}
\bibliography{References}

\begin{thebibliography}{}

\bibitem [\protect \citeauthoryear {%
Abadie%
}{%
Abadie%
}{%
{\protect \APACyear {2005}}%
}]{%
abadie2005semiparametric}
\APACinsertmetastar {%
abadie2005semiparametric}%
\begin{APACrefauthors}%
Abadie, A.%
\end{APACrefauthors}%
\unskip\
\newblock
\APACrefYearMonthDay{2005}{}{}.
\newblock
{\BBOQ}\APACrefatitle {Semiparametric difference-in-differences estimators}
  {Semiparametric difference-in-differences estimators}.{\BBCQ}
\newblock
\APACjournalVolNumPages{The Review of Economic Studies}{72}{1}{1--19}.
\PrintBackRefs{\CurrentBib}

\bibitem [\protect \citeauthoryear {%
Akee%
, Copeland%
, Costello%
\BCBL {}\ \BBA {} Simeonova%
}{%
Akee%
\ \protect \BOthers {.}}{%
{\protect \APACyear {2018}}%
}]{%
akee2018does}
\APACinsertmetastar {%
akee2018does}%
\begin{APACrefauthors}%
Akee, R.%
, Copeland, W.%
, Costello, E\BPBI J.%
\BCBL {}\ \BBA {} Simeonova, E.%
\end{APACrefauthors}%
\unskip\
\newblock
\APACrefYearMonthDay{2018}{}{}.
\newblock
{\BBOQ}\APACrefatitle {How does household income affect child personality
  traits and behaviors?} {How does household income affect child personality
  traits and behaviors?}{\BBCQ}
\newblock
\APACjournalVolNumPages{American Economic Review}{108}{3}{775--827}.
\PrintBackRefs{\CurrentBib}

\bibitem [\protect \citeauthoryear {%
Allingham%
\ \BBA {} Sandmo%
}{%
Allingham%
\ \BBA {} Sandmo%
}{%
{\protect \APACyear {1972}}%
}]{%
allingham1972income}
\APACinsertmetastar {%
allingham1972income}%
\begin{APACrefauthors}%
Allingham, M\BPBI G.%
\BCBT {}\ \BBA {} Sandmo, A.%
\end{APACrefauthors}%
\unskip\
\newblock
\APACrefYearMonthDay{1972}{}{}.
\newblock
{\BBOQ}\APACrefatitle {Income tax evasion: A Theoretical Analysis} {Income tax
  evasion: A theoretical analysis}.{\BBCQ}
\newblock
\APACjournalVolNumPages{Journal of public economics}{1}{}{323--338}.
\PrintBackRefs{\CurrentBib}

\bibitem [\protect \citeauthoryear {%
Belloni%
, Chen%
, Chernozhukov%
\BCBL {}\ \BBA {} Hansen%
}{%
Belloni%
\ \protect \BOthers {.}}{%
{\protect \APACyear {2012}}%
}]{%
belloni2012sparse}
\APACinsertmetastar {%
belloni2012sparse}%
\begin{APACrefauthors}%
Belloni, A.%
, Chen, D.%
, Chernozhukov, V.%
\BCBL {}\ \BBA {} Hansen, C.%
\end{APACrefauthors}%
\unskip\
\newblock
\APACrefYearMonthDay{2012}{}{}.
\newblock
{\BBOQ}\APACrefatitle {Sparse models and methods for optimal instruments with
  an application to eminent domain} {Sparse models and methods for optimal
  instruments with an application to eminent domain}.{\BBCQ}
\newblock
\APACjournalVolNumPages{Econometrica}{80}{6}{2369--2429}.
\PrintBackRefs{\CurrentBib}

\bibitem [\protect \citeauthoryear {%
Belloni%
, Chernozhukov%
, Chetverikov%
\BCBL {}\ \BBA {} Wei%
}{%
Belloni%
\ \protect \BOthers {.}}{%
{\protect \APACyear {2018}}%
}]{%
belloni2018uniformly}
\APACinsertmetastar {%
belloni2018uniformly}%
\begin{APACrefauthors}%
Belloni, A.%
, Chernozhukov, V.%
, Chetverikov, D.%
\BCBL {}\ \BBA {} Wei, Y.%
\end{APACrefauthors}%
\unskip\
\newblock
\APACrefYearMonthDay{2018}{}{}.
\newblock
{\BBOQ}\APACrefatitle {Uniformly valid post-regularization confidence regions
  for many functional parameters in z-estimation framework} {Uniformly valid
  post-regularization confidence regions for many functional parameters in
  z-estimation framework}.{\BBCQ}
\newblock
\APACjournalVolNumPages{The Annals of Statistics}{46}{6B}{3643--3675}.
\PrintBackRefs{\CurrentBib}

\bibitem [\protect \citeauthoryear {%
Belloni%
, Chernozhukov%
, Fern{\'a}ndez-Val%
\BCBL {}\ \BBA {} Hansen%
}{%
Belloni%
\ \protect \BOthers {.}}{%
{\protect \APACyear {2017}}%
}]{%
belloni2017program}
\APACinsertmetastar {%
belloni2017program}%
\begin{APACrefauthors}%
Belloni, A.%
, Chernozhukov, V.%
, Fern{\'a}ndez-Val, I.%
\BCBL {}\ \BBA {} Hansen, C.%
\end{APACrefauthors}%
\unskip\
\newblock
\APACrefYearMonthDay{2017}{}{}.
\newblock
{\BBOQ}\APACrefatitle {Program evaluation and causal inference with
  high-dimensional data} {Program evaluation and causal inference with
  high-dimensional data}.{\BBCQ}
\newblock
\APACjournalVolNumPages{Econometrica}{85}{1}{233--298}.
\PrintBackRefs{\CurrentBib}

\bibitem [\protect \citeauthoryear {%
Belloni%
, Chernozhukov%
\BCBL {}\ \BBA {} Hansen%
}{%
Belloni%
\ \protect \BOthers {.}}{%
{\protect \APACyear {2014}}%
}]{%
Belloni14restud}
\APACinsertmetastar {%
Belloni14restud}%
\begin{APACrefauthors}%
Belloni, A.%
, Chernozhukov, V.%
\BCBL {}\ \BBA {} Hansen, C.%
\end{APACrefauthors}%
\unskip\
\newblock
\APACrefYearMonthDay{2014}{}{}.
\newblock
{\BBOQ}\APACrefatitle {Inference on Treatment Effects after Selection among
  High-Dimensional Controls†} {Inference on treatment effects after selection
  among high-dimensional controls†}.{\BBCQ}
\newblock
\APACjournalVolNumPages{The Review of Economic Studies}{81}{2}{608-650}.
\PrintBackRefs{\CurrentBib}

\bibitem [\protect \citeauthoryear {%
Card%
}{%
Card%
}{%
{\protect \APACyear {1990}}%
}]{%
card1990impact}
\APACinsertmetastar {%
card1990impact}%
\begin{APACrefauthors}%
Card, D.%
\end{APACrefauthors}%
\unskip\
\newblock
\APACrefYearMonthDay{1990}{}{}.
\newblock
{\BBOQ}\APACrefatitle {The impact of the Mariel boatlift on the Miami labor
  market} {The impact of the mariel boatlift on the miami labor market}.{\BBCQ}
\newblock
\APACjournalVolNumPages{ILR Review}{43}{2}{245--257}.
\PrintBackRefs{\CurrentBib}

\bibitem [\protect \citeauthoryear {%
Card%
\ \BBA {} Krueger%
}{%
Card%
\ \BBA {} Krueger%
}{%
{\protect \APACyear {1994}}%
}]{%
david1994minimum}
\APACinsertmetastar {%
david1994minimum}%
\begin{APACrefauthors}%
Card, D.%
\BCBT {}\ \BBA {} Krueger, A.%
\end{APACrefauthors}%
\unskip\
\newblock
\APACrefYearMonthDay{1994}{}{}.
\newblock
{\BBOQ}\APACrefatitle {Minimum wages and employment: a case study of the
  fast-food industry in new jersey and pennsylvania} {Minimum wages and
  employment: a case study of the fast-food industry in new jersey and
  pennsylvania}.{\BBCQ}
\newblock
\APACjournalVolNumPages{American Economic Review}{84}{4}{772--793}.
\PrintBackRefs{\CurrentBib}

\bibitem [\protect \citeauthoryear {%
Chernozhukov%
\ \protect \BOthers {.}}{%
Chernozhukov%
\ \protect \BOthers {.}}{%
{\protect \APACyear {2018}}%
}]{%
chernozhukov2018double}
\APACinsertmetastar {%
chernozhukov2018double}%
\begin{APACrefauthors}%
Chernozhukov, V.%
, Chetverikov, D.%
, Demirer, M.%
, Duflo, E.%
, Hansen, C.%
, Newey, W.%
\BCBL {}\ \BBA {} Robins, J.%
\end{APACrefauthors}%
\unskip\
\newblock
\APACrefYearMonthDay{2018}{}{}.
\newblock
{\BBOQ}\APACrefatitle {Double/debiased machine learning for treatment and
  structural parameters} {Double/debiased machine learning for treatment and
  structural parameters}.{\BBCQ}
\newblock
\APACjournalVolNumPages{The Econometrics Journal}{21}{1}{C1--C68}.
\PrintBackRefs{\CurrentBib}

\bibitem [\protect \citeauthoryear {%
Chernozhukov%
, Escanciano%
, Ichimura%
\BCBL {}\ \BBA {} Newey%
}{%
Chernozhukov%
\ \protect \BOthers {.}}{%
{\protect \APACyear {2016}}%
}]{%
chernozhukov2016locally}
\APACinsertmetastar {%
chernozhukov2016locally}%
\begin{APACrefauthors}%
Chernozhukov, V.%
, Escanciano, J\BPBI C.%
, Ichimura, H.%
\BCBL {}\ \BBA {} Newey, W\BPBI K.%
\end{APACrefauthors}%
\unskip\
\newblock
\APACrefYearMonthDay{2016}{}{}.
\newblock
{\BBOQ}\APACrefatitle {Locally robust semiparametric estimation} {Locally
  robust semiparametric estimation}.{\BBCQ}
\newblock
\APACjournalVolNumPages{arXiv preprint arXiv:1608.00033}{}{}{}.
\PrintBackRefs{\CurrentBib}

\bibitem [\protect \citeauthoryear {%
Chernozhukov%
, Hansen%
\BCBL {}\ \BBA {} Spindler%
}{%
Chernozhukov%
\ \protect \BOthers {.}}{%
{\protect \APACyear {2015}}%
}]{%
chernozhukov2015valid}
\APACinsertmetastar {%
chernozhukov2015valid}%
\begin{APACrefauthors}%
Chernozhukov, V.%
, Hansen, C.%
\BCBL {}\ \BBA {} Spindler, M.%
\end{APACrefauthors}%
\unskip\
\newblock
\APACrefYearMonthDay{2015}{}{}.
\newblock
{\BBOQ}\APACrefatitle {Valid post-selection and post-regularization inference:
  An elementary, general approach} {Valid post-selection and
  post-regularization inference: An elementary, general approach}.{\BBCQ}
\newblock
\APACjournalVolNumPages{Annu. Rev. Econ.}{7}{1}{649--688}.
\PrintBackRefs{\CurrentBib}

\bibitem [\protect \citeauthoryear {%
Clotfelter%
}{%
Clotfelter%
}{%
{\protect \APACyear {1983}}%
}]{%
clotfelter1983tax}
\APACinsertmetastar {%
clotfelter1983tax}%
\begin{APACrefauthors}%
Clotfelter, C\BPBI T.%
\end{APACrefauthors}%
\unskip\
\newblock
\APACrefYearMonthDay{1983}{}{}.
\newblock
{\BBOQ}\APACrefatitle {Tax evasion and tax rates: An analysis of individual
  returns} {Tax evasion and tax rates: An analysis of individual
  returns}.{\BBCQ}
\newblock
\APACjournalVolNumPages{The review of economics and statistics}{}{}{363--373}.
\PrintBackRefs{\CurrentBib}

\bibitem [\protect \citeauthoryear {%
Feinstein%
}{%
Feinstein%
}{%
{\protect \APACyear {1991}}%
}]{%
feinstein1991econometric}
\APACinsertmetastar {%
feinstein1991econometric}%
\begin{APACrefauthors}%
Feinstein, J\BPBI S.%
\end{APACrefauthors}%
\unskip\
\newblock
\APACrefYearMonthDay{1991}{}{}.
\newblock
{\BBOQ}\APACrefatitle {An econometric analysis of income tax evasion and its
  detection} {An econometric analysis of income tax evasion and its
  detection}.{\BBCQ}
\newblock
\APACjournalVolNumPages{The RAND Journal of Economics}{}{}{14--35}.
\PrintBackRefs{\CurrentBib}

\bibitem [\protect \citeauthoryear {%
Fuest%
, Peichl%
\BCBL {}\ \BBA {} Siegloch%
}{%
Fuest%
\ \protect \BOthers {.}}{%
{\protect \APACyear {2018}}%
}]{%
fuest2018higher}
\APACinsertmetastar {%
fuest2018higher}%
\begin{APACrefauthors}%
Fuest, C.%
, Peichl, A.%
\BCBL {}\ \BBA {} Siegloch, S.%
\end{APACrefauthors}%
\unskip\
\newblock
\APACrefYearMonthDay{2018}{}{}.
\newblock
{\BBOQ}\APACrefatitle {Do higher corporate taxes reduce wages? Micro evidence
  from Germany} {Do higher corporate taxes reduce wages? micro evidence from
  germany}.{\BBCQ}
\newblock
\APACjournalVolNumPages{American Economic Review}{108}{2}{393--418}.
\PrintBackRefs{\CurrentBib}

\bibitem [\protect \citeauthoryear {%
Meyer%
, Viscusi%
\BCBL {}\ \BBA {} Durbin%
}{%
Meyer%
\ \protect \BOthers {.}}{%
{\protect \APACyear {1995}}%
}]{%
meyer1995workers}
\APACinsertmetastar {%
meyer1995workers}%
\begin{APACrefauthors}%
Meyer, B\BPBI D.%
, Viscusi, W\BPBI K.%
\BCBL {}\ \BBA {} Durbin, D\BPBI L.%
\end{APACrefauthors}%
\unskip\
\newblock
\APACrefYearMonthDay{1995}{}{}.
\newblock
{\BBOQ}\APACrefatitle {Workers' compensation and injury duration: evidence from
  a natural experiment} {Workers' compensation and injury duration: evidence
  from a natural experiment}.{\BBCQ}
\newblock
\APACjournalVolNumPages{The American economic review}{}{}{322--340}.
\PrintBackRefs{\CurrentBib}

\bibitem [\protect \citeauthoryear {%
Newey%
}{%
Newey%
}{%
{\protect \APACyear {1994}}%
}]{%
newey1994asymptotic}
\APACinsertmetastar {%
newey1994asymptotic}%
\begin{APACrefauthors}%
Newey, W\BPBI K.%
\end{APACrefauthors}%
\unskip\
\newblock
\APACrefYearMonthDay{1994}{}{}.
\newblock
{\BBOQ}\APACrefatitle {The asymptotic variance of semiparametric estimators}
  {The asymptotic variance of semiparametric estimators}.{\BBCQ}
\newblock
\APACjournalVolNumPages{Econometrica: Journal of the Econometric
  Society}{}{}{1349--1382}.
\PrintBackRefs{\CurrentBib}

\bibitem [\protect \citeauthoryear {%
Newey%
, Hsieh%
\BCBL {}\ \BBA {} Robins%
}{%
Newey%
\ \protect \BOthers {.}}{%
{\protect \APACyear {1998}}%
}]{%
newey1998undersmoothing}
\APACinsertmetastar {%
newey1998undersmoothing}%
\begin{APACrefauthors}%
Newey, W\BPBI K.%
, Hsieh, F.%
\BCBL {}\ \BBA {} Robins, J.%
\end{APACrefauthors}%
\unskip\
\newblock
\APACrefYearMonthDay{1998}{}{}.
\newblock
{\BBOQ}\APACrefatitle {Undersmoothing and bias corrected functional estimation}
  {Undersmoothing and bias corrected functional estimation}.{\BBCQ}
\newblock

\PrintBackRefs{\CurrentBib}

\bibitem [\protect \citeauthoryear {%
Newey%
, Hsieh%
\BCBL {}\ \BBA {} Robins%
}{%
Newey%
\ \protect \BOthers {.}}{%
{\protect \APACyear {2004}}%
}]{%
newey2004twicing}
\APACinsertmetastar {%
newey2004twicing}%
\begin{APACrefauthors}%
Newey, W\BPBI K.%
, Hsieh, F.%
\BCBL {}\ \BBA {} Robins, J\BPBI M.%
\end{APACrefauthors}%
\unskip\
\newblock
\APACrefYearMonthDay{2004}{}{}.
\newblock
{\BBOQ}\APACrefatitle {Twicing kernels and a small bias property of
  semiparametric estimators} {Twicing kernels and a small bias property of
  semiparametric estimators}.{\BBCQ}
\newblock
\APACjournalVolNumPages{Econometrica}{72}{3}{947--962}.
\PrintBackRefs{\CurrentBib}

\bibitem [\protect \citeauthoryear {%
Newey%
\ \BBA {} McFadden%
}{%
Newey%
\ \BBA {} McFadden%
}{%
{\protect \APACyear {1994}}%
}]{%
newey1994large}
\APACinsertmetastar {%
newey1994large}%
\begin{APACrefauthors}%
Newey, W\BPBI K.%
\BCBT {}\ \BBA {} McFadden, D.%
\end{APACrefauthors}%
\unskip\
\newblock
\APACrefYearMonthDay{1994}{}{}.
\newblock
{\BBOQ}\APACrefatitle {Large sample estimation and hypothesis testing} {Large
  sample estimation and hypothesis testing}.{\BBCQ}
\newblock
\APACjournalVolNumPages{Handbook of econometrics}{4}{}{2111--2245}.
\PrintBackRefs{\CurrentBib}

\bibitem [\protect \citeauthoryear {%
Rubin%
}{%
Rubin%
}{%
{\protect \APACyear {1974}}%
}]{%
rubin1974estimating}
\APACinsertmetastar {%
rubin1974estimating}%
\begin{APACrefauthors}%
Rubin, D\BPBI B.%
\end{APACrefauthors}%
\unskip\
\newblock
\APACrefYearMonthDay{1974}{}{}.
\newblock
{\BBOQ}\APACrefatitle {Estimating causal effects of treatments in randomized
  and nonrandomized studies.} {Estimating causal effects of treatments in
  randomized and nonrandomized studies.}{\BBCQ}
\newblock
\APACjournalVolNumPages{Journal of educational Psychology}{66}{5}{688}.
\PrintBackRefs{\CurrentBib}

\bibitem [\protect \citeauthoryear {%
Sequeira%
}{%
Sequeira%
}{%
{\protect \APACyear {2016}}%
}]{%
sequeira2016corruption}
\APACinsertmetastar {%
sequeira2016corruption}%
\begin{APACrefauthors}%
Sequeira, S.%
\end{APACrefauthors}%
\unskip\
\newblock
\APACrefYearMonthDay{2016}{}{}.
\newblock
{\BBOQ}\APACrefatitle {Corruption, trade costs, and gains from tariff
  liberalization: evidence from Southern Africa} {Corruption, trade costs, and
  gains from tariff liberalization: evidence from southern africa}.{\BBCQ}
\newblock
\APACjournalVolNumPages{American Economic Review}{106}{10}{3029--63}.
\PrintBackRefs{\CurrentBib}

\bibitem [\protect \citeauthoryear {%
Sequeira%
\ \BBA {} Djankov%
}{%
Sequeira%
\ \BBA {} Djankov%
}{%
{\protect \APACyear {2014}}%
}]{%
sequeira2014corruption}
\APACinsertmetastar {%
sequeira2014corruption}%
\begin{APACrefauthors}%
Sequeira, S.%
\BCBT {}\ \BBA {} Djankov, S.%
\end{APACrefauthors}%
\unskip\
\newblock
\APACrefYearMonthDay{2014}{}{}.
\newblock
{\BBOQ}\APACrefatitle {Corruption and firm behavior: Evidence from African
  ports} {Corruption and firm behavior: Evidence from african ports}.{\BBCQ}
\newblock
\APACjournalVolNumPages{Journal of International Economics}{94}{2}{277--294}.
\PrintBackRefs{\CurrentBib}

\bibitem [\protect \citeauthoryear {%
Slemrod%
\ \BBA {} Yitzhaki%
}{%
Slemrod%
\ \BBA {} Yitzhaki%
}{%
{\protect \APACyear {2002}}%
}]{%
slemrod2002tax}
\APACinsertmetastar {%
slemrod2002tax}%
\begin{APACrefauthors}%
Slemrod, J.%
\BCBT {}\ \BBA {} Yitzhaki, S.%
\end{APACrefauthors}%
\unskip\
\newblock
\APACrefYearMonthDay{2002}{}{}.
\newblock
{\BBOQ}\APACrefatitle {Tax avoidance, evasion, and administration} {Tax
  avoidance, evasion, and administration}.{\BBCQ}
\newblock
\BIn{} \APACrefbtitle {Handbook of public economics} {Handbook of public
  economics}\ (\BVOL~3, \BPGS\ 1423--1470).
\newblock
\APACaddressPublisher{}{Elsevier}.
\PrintBackRefs{\CurrentBib}

\bibitem [\protect \citeauthoryear {%
Van~de Geer%
}{%
Van~de Geer%
}{%
{\protect \APACyear {2008}}%
}]{%
van2008high}
\APACinsertmetastar {%
van2008high}%
\begin{APACrefauthors}%
Van~de Geer, S.%
\end{APACrefauthors}%
\unskip\
\newblock
\APACrefYearMonthDay{2008}{}{}.
\newblock
{\BBOQ}\APACrefatitle {High-dimensional generalized linear models and the
  lasso} {High-dimensional generalized linear models and the lasso}.{\BBCQ}
\newblock
\APACjournalVolNumPages{The Annals of Statistics}{36}{2}{614--645}.
\PrintBackRefs{\CurrentBib}

\end{thebibliography}

\begin{center}
\begin{LARGE}

APPENDIX
\end{LARGE}
\end{center}

\begin{flushleft}
\textbf{\begin{Large}
Multilevel Treatment
\end{Large}}
\end{flushleft}

Similarly, I use the cross-fitting algorithm \citep*{chernozhukov2018double}. 
\begin{enumerate}
\item \emph{Take a $K$-fold random partition $\left(I_{k}\right)_{k=1}^{K}$
of observation indices $\left[N\right]=\left\{ 1,...,N\right\} $
such that the size of each fold $I_{k}$ is $n=N/K$. For each $k\in\left[K\right]=\left\{ 1,...,K\right\} $,
define the auxiliary sample $I_{k}^{c}\coloneqq\left\{ 1,...,N\right\} \setminus I_{k}$. }
\item \emph{For each $k\in\left[K\right]$, construct the estimator of $p_{0}$
and $\lambda_{0}$ by $\hat{p}_{w}=\frac{1}{n}\sum_{i\in I_{k}^{c}}D_{i}$.
Also, construct the estimators of $g_{w}$, $g_{z}$, and $\ell_{30}$
using the auxiliary sample $I_{k}^{c}$: $\hat{g}_{wk}=\hat{g}_{w}\left(\left(W_{i}\right)_{i\in I_{k}^{c}}\right)$,
$\hat{g}_{zk}=\hat{g}_{z}\left(\left(W_{i}\right)_{i\in I_{k}^{c}}\right)$,
and $\hat{\ell}_{3k}=\hat{\ell}_{3}\left(\left(W_{i}\right)_{i\in I_{k}^{c}}\right)$.}
\item \emph{For each $k$, construct the intermediate ATT estimators 
\[
\tilde{\theta}_{wk}=\frac{1}{n}\sum_{i\in I_{k}}\frac{I\left(W_{i}=w\right)\cdot\hat{g}_{zk}\left(X_{i}\right)-I\left(W_{i}=0\right)\cdot\hat{g}_{wk}\left(X_{i}\right)}{\hat{p}_{w}\hat{g}_{zk}\left(X_{i}\right)}\times\left(Y\left(1\right)-Y\left(0\right)-\hat{\ell}_{3k}\left(X_{i}\right)\right),
\]
}
\item \emph{Construct the final ATT estimators $\tilde{\theta}=\frac{1}{K}\sum_{k=1}^{K}\tilde{\theta}_{k}.$}
\end{enumerate}
The estimator $\hat{g}_{wk}$ and $\hat{g}_{zk}$ can be constructed
by Multi-Logit Lasso. 

\begin{flushleft}
\textbf{\begin{Large}
The Lasso Penalty
\end{Large}}
\end{flushleft}

The following is suggested by \citet*{belloni2012sparse}. Let $y_{i}$ denote $Y_{i}\left(1\right)-Y_{i}\left(0\right)$ or $\left(T_{i}-\hat{\lambda}_{k}\right)$, $\lambda_{k}$ denote $\lambda_{1k}$ or $\lambda_{2k}$, and $\hat{\Upsilon}_{k}$ denote $\hat{\Upsilon}_{1k}$ or $\hat{\Upsilon}_{2k}$. For $k\in\left[K\right]$, the loading $\hat{\Upsilon}_{k}$ is a diagonal matrix with entries $\hat{\gamma}_{kj}$, $j=1,...,p$, constructed by the following steps:

\[
\text{Initial }\hat{\gamma}_{kj}=\sqrt{\frac{1}{M_k}\sum_{i\in I_{kz}^{c}}q_{ij}^{2}\left(y_{i}-\bar{y}_{k}\right)^{2}},\lambda_{k}=2c\sqrt{M_{k}}\Phi^{-1}\left(1-\gamma/2p\right),
\]
\[
\text{Refined }\hat{\gamma}_{kj}=\sqrt{\frac{1}{M_{k}}\sum_{i\in I_{kz}^{c}}q_{ij}^{2}\hat{\varepsilon}_{i}^{2}},\lambda_{k}=2c\sqrt{M_{k}}\Phi^{-1}\left(1-\gamma/2p\right),
\]
where $\bar{y}_{k}=M^{-1}\sum_{i\in I_{k}^{c}}y_{i}$, $c>1$ and $\gamma\rightarrow0$. The empirical residual $\hat{\varepsilon}_{i}$ is calculated by the modified Lasso estimator $\beta_{k}^{*}$ in the previous step: $\hat{\varepsilon}_{i}=y_{i}-q_{i}'\beta_{k}^{*}$. Repeat the
second step $B>0$ times to obtain the final loading. 

\begin{center}
\textbf{\begin{Large}
PROOFS
\end{Large}}
\end{center}

\begin{description}
\item [{Proof~of~Lemma~1}]~
\item [{\emph{Repeated~outcomes:}}]~
\end{description}
The Gateaux derivative of (3.1) in the direction $\eta_{1}-\eta_{10}=\left(g-g_{0},\ell_{1}-\ell_{10}\right)$
is 
\begin{align*}
\partial_{\eta_{1}}E_{P}\left[\psi_{1}\left(W,\theta_{0},p_{0},\eta_{10}\right)\right]\left(\eta_{1}-\eta_{10}\right)= & E_{P}\left[\frac{D-1}{p_{0}\left(1-g_{0}\left(X\right)\right)^{2}}\left(Y\left(1\right)-Y\left(0\right)-\ell_{10}\right)\left(g\left(X\right)-g_{0}\left(X\right)\right)\right]\\
 & -E_{P}\left[\frac{D-g_{0}\left(X\right)}{p_{0}\left(1-g_{0}\left(X\right)\right)}\left(\ell_{1}\left(X\right)-\ell_{10}\left(X\right)\right)\right]\\
= & -E_{P}\left[\frac{g\left(X\right)-g_{0}\left(X\right)}{p_{0}\left(1-g_{0}\left(X\right)\right)}E\left[Y\left(1\right)-Y\left(0\right)-\ell_{10}\left(X\right)\mid X, D=0\right]\right]\\
 & -E_{P}\left[\frac{\left(\ell_{1}\left(X\right)-\ell_{10}\left(X\right)\right)}{p_{0}\left(1-g_{0}\left(X\right)\right)}E_{P}\left[D-g_{0}\left(X\right)\mid X\right]\right]\\
= & -E_{P}\left[\frac{g\left(X\right)-g_{0}\left(X\right)}{p_{0}\left(1-g_{0}\left(X\right)\right)}\left(\ell_{10}\left(X\right)-\ell_{10}\left(X\right)\right)\right]-0\\
= & 0,
\end{align*}
where the second inequality follows from the law of iterated expectations, the third from the definition of $\ell_{10}\left(X\right)$ and $E_{P}\left[D-g_{0}\left(X\right)\mid X\right]=0$.

\begin{description}
\item [{\emph{Repeated~cross~sections:}}]~
\end{description}
Define $\partial_{\eta_{2}}E_{P}\left[\psi_{20}\right]\left(\eta_{2}-\eta_{20}\right)\coloneqq \partial_{\eta_{2}}E_{P}\left[\psi_{2}\left(W,\theta_{0},p_{0},\lambda_{0},\eta_{20}\right)\right]\left(\eta_{2}-\eta_{20}\right)$. Similar to the proof of repeated outcomes,  the Gateaux derivative of (3.2) in the direction $\eta_{2}-\eta_{20}=\left(g-g_{0},\ell_{2}-\ell_{20}\right)$
is
\begin{align*}
\partial_{\eta_{2}}E_{P}\left[\psi_{20}\right]\left(\eta_{2}-\eta_{20}\right)= & E_{P}\left[\frac{D-1}{p_{0}'\left(1-g_{0}\left(X\right)\right)^{2}}\left[\left(T-\lambda_{0}\right)Y-\ell_{20}\left(X\right)\right]\left(g\left(X\right)-g_{0}\left(X\right)\right)\right]\\
 & -E_{P}\left[\frac{D-g_{0}\left(X\right)}{p_{0}'\left(1-g_{0}\left(X\right)\right)}\left(\ell_{2}\left(X\right)-\ell_{20}\left(X\right)\right)\right]\\
= & -E_{P}\left[\frac{g\left(X\right)-g_{0}\left(X\right)}{p_{0}'\left(1-g_{0}\left(X\right)\right)}\left(\ell_{20}\left(X\right)-\ell_{20}\left(X\right)\right)\right]\\
 & -E_{P}\left[\frac{\ell_{2}\left(X\right)-\ell_{20}\left(X\right)}{p\lambda\left(1-\lambda\right)\left(1-g\left(X\right)\right)}E_{P}\left[D-g_{0}\left(X\right)\mid X\right]\right]\\
= & 0,
\end{align*}
where $p_{0}'\coloneqq p_{0}\lambda_{0}\left(1-\lambda_{0}\right)$.
\begin{description}
\item [{\emph{Multilevel~treatment:}}]~
\end{description}
Let $\Delta_{w}=g_{w}-g_{w0}$, $\Delta_{z}=g_{z}-g_{z0}$, and $\Delta_{\ell3}=\ell_{3}-\ell_{30}$. The Gateaux derivative of (3.3) in the direction $\eta_{w}-\eta_{w0}=\left(g_{w}-g_{w0},g_{z}-g_{z0},\ell_{3}-\ell_{30}\right)$ is

\begin{align*}
\partial_{\eta_{w}}E_{P}\left[\psi_{w}\left(W,\theta_{0},p_{w0},\eta_{w0}\right)\right]\left(\eta_{w}-\eta_{w0}\right)=&	E_{P}\left[\frac{I\left(W=0\right)g_{w0}\left(X\right)}{p_{w0}g_{z0}\left(X\right)^{2}}\left(Y\left(1\right)-Y\left(0\right)-\ell_{30}\right)\Delta_{w}\right]\\
 &	-E_{P}\left[\frac{I\left(W=0\right)}{p_{w0}g_{z0}\left(X\right)}\left(Y\left(1\right)-Y\left(0\right)-\ell_{30}\right)\Delta_{z}\right]\\
 &	+E_{P}\left[\frac{I\left(W=0\right)g_{w0}\left(X\right)-I\left(W=w\right)g_{z0}\left(X\right)}{p_{w0}g_{z0}\left(X\right)}\Delta_{\ell3}\right]\\
= &	0
\end{align*}
by the law of iterated expectation on each terms. 

The proofs of Theorem 1 and 2 follow the general framework proposed in \citet*{chernozhukov2018double}.
\begin{description}
\item [{Proof~of~Theorem~1}]~
\item [{Repeated~Outcomes:}]~
\end{description}
The proof proceeds in five steps. In Step 1, I show the main result
using the auxiliary results (A.1)-(A.4). In Step 2-5, I prove the
auxiliary results. 
\[
\sup_{\eta_{1}\in\mathcal{T}_{N}}\left(E\left[\parallel\psi_{1}\left(W,\theta_{0},p_{0},\eta_{1}\right)-\psi_{1}\left(W,\theta_{0},p_{0},\eta_{10}\right)\parallel^{2}\right]\right)^{1/2}\leq\varepsilon_{N},\tag{A.1}
\]

\[
\sup_{r\in\left(0,1\right),\eta_{1}\in\mathcal{T}_{N}}\parallel\partial_{r}^{2}E\left[\psi_{1}\left(W,\theta_{0},p_{0},\eta_{10}+r\left(\eta_{1}-\eta_{10}\right)\right)\right]\parallel\leq\left(\varepsilon_{N}\right)^{2},\tag{A.2}
\]

\[
\sup_{\eta_{1}\in\mathcal{T}_{N}}\left(E_{P}\left[\parallel\partial_{p}\psi_{1}\left(W,\theta_{0},p_{0},\eta_{1}\right)-\partial_{p}\psi_{1}\left(W,\theta_{0},p_{0},\eta_{10}\right)\parallel^{2}\right]\right)^{1/2}\leq\varepsilon_{N},\tag{A.3}
\]
\[
\sup_{p\in\mathcal{P}_{N},\eta_{1}\in\mathcal{T}_{N}}\left(E_{P}\left[\parallel\partial_{p}^{2}\psi_{1}\left(W,\theta_{0},p,\eta_{1}\right)-\partial_{p}^{2}\psi_{1}\left(W,\theta_{0},p_{0},\eta_{10}\right)\parallel^{2}\right]\right)^{1/2}\leq\varepsilon_{N},\tag{A.4}
\]
where  $\mathcal{T}_{N}$
is the set of all $\eta_{1}=\left(g,\ell_{1}\right)$ consisting of
$P$-square-integrable functions $g$ and $\ell_{1}$ such that \emph{
\[
\parallel\eta_{1}-\eta_{10}\parallel_{P,2}\leq\varepsilon_{N},
\]
\[
\parallel g-1/2\parallel_{P,\infty}\leq1/2-\kappa,
\]
\[
\parallel g-g_{0}\parallel_{P,2}^{2}+\parallel g-g_{0}\parallel_{P,2}\times\parallel\ell_{1}-\ell_{10}\parallel_{P,2}\leq\left(\varepsilon_{N}\right)^{2},
\]
}and $\mathcal{P}_{N}$ is the set of all $p>0$ such that $\mid p-p_{0}\mid\leq N^{-1/2}$. By the regularity condition (3.1) and $\mid \hat{p}_{k}-p_{0}\mid=O_{P}\left(N^{-1/2}\right)$, $\hat{\eta}_{1k}\in \mathcal{T}_{N}$ and $\hat{p}_{k}\in \mathcal{P}_{N}$ with probability $1-o\left(1\right)$.

\emph{Step 1.} Observe that we have the decomposition

\begin{align*}
\sqrt{N}\left(\tilde{\theta}-\theta_{0}\right)= & \sqrt{N}\left(\frac{1}{K}\sum_{k=1}^{K}\tilde{\theta}_{k}-\theta_{0}\right)\\
= & \sqrt{N}\frac{1}{K}\sum_{k=1}^{K}\mathbb{E}_{n,k}\left[\psi_{1}\left(W,\theta_{0},\hat{p}_{k},\hat{\eta}_{1k}\right)\right]\\
= & \sqrt{N}\frac{1}{K}\sum_{k=1}^{K}\mathbb{E}_{n,k}\left[\psi_{1}\left(W,\theta_{0},p_{0},\hat{\eta}_{1k}\right)\right]\\
 & +\underbrace{\sqrt{N}\frac{1}{K}\sum_{k=1}^{K}\mathbb{E}_{n,k}\left[\partial_{p}\psi_{1}\left(W,\theta_{0},p_{0},\hat{\eta}_{1k}\right)\right]\left(\hat{p}_{k}-p_{0}\right)}_{a}\\
 & +\underbrace{\sqrt{N}\frac{1}{K}\sum_{k=1}^{K}\mathbb{E}_{n,k}\left[\partial_{p}^{2}\psi_{1}\left(W,\theta_{0},\bar{p}_{k},\hat{\eta}_{1k}\right)\right]\left(\hat{p}_{k}-p_{0}\right)^{2}}_{b},
\end{align*}
where $\bar{p}_{k}\in\left(\hat{p}_{k},p_{0}\right)$. For term (a),
by the triangle inequality, we have 
\[
\mid\mathbb{E}_{n,k}\left[\partial_{p}\psi_{1}\left(W,\theta_{0},p_{0},\hat{\eta}_{1k}\right)\right]-E_{p}\left[\partial_{p}\psi_{1}\left(W,\theta_{0},p_{0},\eta_{10}\right)\right]\mid\leq J_{1,k}+J_{2,k},
\]
where 
\[
J_{1,k}=\mid\mathbb{E}_{n,k}\left[\partial_{p}\psi_{1}\left(W,\theta_{0},p_{0},\hat{\eta}_{1k}\right)\right]-\mathbb{E}_{n,k}\left[\partial_{p}\psi_{1}\left(W,\theta_{0},p_{0},\eta_{10}\right)\right]\mid,
\]
\[
J_{2,k}=\mid\mathbb{E}_{n,k}\left[\partial_{p}\psi_{1}\left(W,\theta_{0},p_{0},\eta_{10}\right)\right]-E_{p}\left[\partial_{p}\psi_{1}\left(W,\theta_{0},p_{0},\eta_{10}\right)\right]\mid.
\]
To bound $J_{2,k}$, we have  
\begin{align*}
E_{P}\left[J_{2,k}^{2}\right]\leq & n^{-1} E_{P}\left[\left( \partial_{p}\psi_{1}\left(W,\theta_{0},p_{0},\eta_{10}\right)^{2}\right)\right]\\
= & n^{-1}E_{P}\left[\frac{1}{p_{0}^{4}}\frac{U^{2}V_{1}^{2}}{\left(1-g_{0}\right)^{2}}\right]\\
\leq & n^{-1}\left(\frac{C^2}{p_{0}^4\kappa^2}\right),
\end{align*}
where the last inequality follows from the regularity condition (3.1). By Chebyshev's inequality, $J_{2,k}=O_{P}\left(n^{-1/2}\right)=o_{P}\left(1\right)$. Next, we bound $J_{1,k}$. Conditional on the auxiliary sample
$I_{k}^{c}$, $\hat{\eta}_{1k}$ can be treated as fixed. Under the
event that $\hat{\eta}_{1k}\in\mathcal{T}_{N}$, we have
\begin{align*}
E_{P}\left[J_{1,k}^{2}\mid\left(W_{i}\right)_{i\in I_{k}^{c}}\right]= & E_{P}\left[\parallel\partial_{p}\psi_{1}\left(W,\theta_{0},p_{0},\hat{\eta}_{1k}\right)-\partial_{p}\psi_{1}\left(W,\theta_{0},p_{0},\eta_{10}\right)\parallel^{2}\mid\left(W_{i}\right)_{i\in I_{k}^{c}}\right]\\
\leq & \sup_{\eta_{1}\in\mathcal{T}_{N}}E_{P}\left[\parallel\partial_{p}\psi_{1}\left(W,\theta_{0},p_{0},\eta_{1}\right)-\partial_{p}\psi_{1}\left(W,\theta_{0},p_{0},\eta_{10}\right)\parallel^{2}\mid\left(W_{i}\right)_{i\in I_{k}^{c}}\right]\\
= & \sup_{\eta_{1}\in\mathcal{T}_{N}}E_{P}\left[\parallel\partial_{p}\psi_{1}\left(W,\theta_{0},p_{0},\eta_{1}\right)-\partial_{p}\psi_{1}\left(W,\theta_{0},p_{0},\eta_{10}\right)\parallel^{2}\right]\\
= & \varepsilon_{N}^{2}
\end{align*}
by (A.3). By Lemma A.1, $J_{1,k}=O_{P}\left(\varepsilon_{N}\right)=o_{P}\left(1\right)$.
Together, we have 
\[
\mathbb{E}_{n,k}\left[\partial_{p}\psi_{1}\left(W,\theta_{0},p_{0},\hat{\eta}_{1k}\right)\right]\stackrel{p}{\rightarrow}E_{p}\left[\partial_{p}\psi_{1}\left(W,\theta_{0},p_{0},\eta_{10}\right)\right]=G_{1p0}.
\]
For term (b), by the triangle inequality, we have
\[
\mid\mathbb{E}_{n,k}\left[\partial_{p}^{2}\psi_{1}\left(W,\theta_{0},\bar{p}_{k},\hat{\eta}_{1k}\right)\right]-E_{p}\left[\partial_{p}^{2}\psi_{1}\left(W,\theta_{0},p_{0},\eta_{10}\right)\right]\mid\leq J_{3,k}+J_{4,k},
\]
where 
\[
J_{3,k}=\mid\mathbb{E}_{n,k}\left[\partial_{p}^{2}\psi_{1}\left(W,\theta_{0},\bar{p}_{k},\hat{\eta}_{1k}\right)\right]-\mathbb{E}_{n,k}\left[\partial_{p}^{2}\psi_{1}\left(W,\theta_{0},p_{0},\eta_{10}\right)\right]\mid,
\]
\[
J_{4,k}=\mid\mathbb{E}_{n,k}\left[\partial_{p}^{2}\psi_{1}\left(W,\theta_{0},p_{0},\eta_{10}\right)\right]-E_{p}\left[\partial_{p}^{2}\psi_{1}\left(W,\theta_{0},p_{0},\eta_{10}\right)\right]\mid.
\]
To bound $J_{4,k}$, we have  
\begin{align*}
E_{P}\left[J_{4,k}^{2}\right]\leq & n^{-1} E_{P}\left[\left( \partial_{p}^{2}\psi_{1}\left(W,\theta_{0},p_{0},\eta_{10}\right)^{2}\right)\right]\\
= & n^{-1}E_{P}\left[\frac{4}{p_{0}^{6}}\frac{U^{2}V_{1}^{2}}{\left(1-g_{0}\right)^{2}}\right]\\
\leq & n^{-1}\left(\frac{4C^2}{p_{0}^{6}\kappa^2}\right),
\end{align*}
where the last inequality follows from the regularity conditions. By Chebyshev's inequality, $J_{4,k}=O_{P}\left(n^{-1/2}\right)=o_{P}\left(1\right)$.
Conditional on $I_{k}^{c}$, both $\bar{p}_{k}$ and $\hat{\eta}_{1k}$
can be treated as fixed. Under the event that $\hat{p}_{k}\in\mathcal{P}_{N}$
(thus $\bar{p}_{k}\in\mathcal{P}_{N}$) and $\hat{\eta}_{1k}\in\mathcal{T}_{N}$
, we have 
\begin{align*}
E_{P}\left[J_{3,k}^{2}\mid\left(W_{i}\right)_{i\in I_{k}^{c}}\right]= & E_{P}\left[\parallel\partial_{p}^{2}\psi_{1}\left(W,\theta_{0},\bar{p}_{k},\hat{\eta}_{1k}\right)-\partial_{p}^{2}\psi_{1}\left(W,\theta_{0},p_{0},\eta_{10}\right)\parallel^{2}\mid\left(W_{i}\right)_{i\in I_{k}^{c}}\right]\\
\leq & \sup_{p\in\mathcal{P}_{N},\eta_{1}\in\mathcal{T}_{N}}E_{P}\left[\parallel\partial_{p}\psi_{1}\left(W,\theta_{0},p,\eta_{1}\right)-\partial_{p}\psi_{1}\left(W,\theta_{0},p_{0},\eta_{10}\right)\parallel^{2}\mid\left(W_{i}\right)_{i\in I_{k}^{c}}\right]\\
= & \sup_{p\in\mathcal{P}_{N},\eta_{1}\in\mathcal{T}_{N}}E_{P}\left[\parallel\partial_{p}\psi_{1}\left(W,\theta_{0},p,\eta_{1}\right)-\partial_{p}\psi_{1}\left(W,\theta_{0},p_{0},\eta_{10}\right)\parallel^{2}\right]\\
\leq & \varepsilon_{N}^{2}
\end{align*}
by (A.4). By Lemma A.1, $J_{3,k}=O_{P}\left(\varepsilon_{N}\right)=o_{P}\left(1\right)$.
Hence, $\mathbb{E}_{n,k}\left[\partial_{p}^{2}\psi_{1}\left(W,\theta_{0},\bar{p}_{k},\hat{\eta}_{1k}\right)\right]=O_{P}\left(1\right)$. 

Combine the above results with $\hat{p}_{k}-p_{0}=\mathbb{E}_{n,k}\left[D-p_{0}\right]$
and $\left(\hat{p}_{k}-p_{0}\right)^{2}=O_{P}\left(N^{-1}\right)$,
the decomposition of $\tilde{\theta}$ becomes
\begin{align*}
\sqrt{N}\left(\tilde{\theta}-\theta_{0}\right)= & \sqrt{N}\frac{1}{K}\sum_{k=1}^{K}\mathbb{E}_{n,k}\left[\psi_{1}\left(W,\theta_{0},p_{0},\hat{\eta}_{1k}\right)\right]\\
 & +\sqrt{N}\frac{1}{K}\sum_{k=1}^{K}\mathbb{E}_{n,k}\left[G_{1p0}\left(D-p_{0}\right)\right]+o_{P}\left(1\right)\\
= & \sqrt{N}\frac{1}{K}\sum_{k=1}^{K}\mathbb{E}_{n,k}\left[\psi_{1}\left(W,\theta_{0},p_{0},\hat{\eta}_{1k}\right)+G_{1p0}\left(D-p_{0}\right)\right]+o_{P}\left(1\right)\\
= & \frac{1}{\sqrt{N}}\sum_{i=1}^{N}\left[\psi_{1}\left(W_{i},\theta_{0},p_{0},\eta_{10}\right)+G_{1p0}\left(D_{i}-p_{0}\right)\right]+\sqrt{N}R_{N}+o_{P}\left(1\right),
\end{align*}
where 
\begin{align*}
R_{N}= & \frac{1}{K}\sum_{k=1}^{K}\mathbb{E}_{n,k}\left[\psi_{1}\left(W,\theta_{0},p_{0},\hat{\eta}_{1k}\right)+G_{1p0}\left(D-p_{0}\right)\right]-\frac{1}{N}\sum_{i=1}^{N}\left[\psi_{1}\left(W_{i},\theta_{0},p_{0},\eta_{10}\right)+G_{1p0}\left(D_{i}-p_{0}\right)\right]\\
= & \frac{1}{K}\sum_{k=1}^{K}\mathbb{E}_{n,k}\left[\psi_{1}\left(W,\theta_{0},p_{0},\hat{\eta}_{1k}\right)\right]-\frac{1}{N}\sum_{i=1}^{N}\psi_{1}\left(W_{i},\theta_{0},p_{0},\eta_{10}\right).
\end{align*}
If we can show that $\sqrt{N}R_{N}=o_{P}\left(1\right)$, then we
are done.

This part is essentially identical to Step 3 in the proof of Theorem 3.1 (DML2) in \citet*{chernozhukov2018double}. I reproduce it here for reader's convenience. Since $K$ is a fixed integer, which is independent of $N$, it suffices
to show that for any $k\in\left[K\right]$,
\[
\mathbb{E}_{n,k}\left[\psi_{1}\left(W,\theta_{0},p_{0},\hat{\eta}_{1k}\right)\right]-\frac{1}{n}\sum_{i\in I_{k}}\psi_{1}\left(W_{i},\theta_{0},p_{0},\eta_{10}\right)=o_{P}\left(N^{-1/2}\right).
\]
Define the empirical process notation:
\[
\mathbb{G}_{n,k}\left[\phi\left(W\right)\right]=\frac{1}{\sqrt{n}}\sum_{i\in I_{k}}\left(\phi\left(W_{i}\right)-\int\phi\left(w\right)dP\right),
\]
where $\phi$ is any $P$-integrable function on $\mathcal{W}$. By
the triangle inequality, we have
\[
\parallel\mathbb{E}_{n,k}\left[\psi_{1}\left(W,\theta_{0},p_{0},\hat{\eta}_{1k}\right)\right]-\frac{1}{n}\sum_{i\in I_{k}}\psi_{1}\left(W_{i},\theta_{0},p_{0},\eta_{10}\right)\parallel\leq\frac{I_{1,k}+I_{2,k}}{\sqrt{n}},
\]
where
\[
I_{1,k}\coloneqq\parallel\mathbb{G}_{n,k}\left[\psi_{1}\left(W,\theta_{0},p_{0},\hat{\eta}_{1k}\right)\right]-\mathbb{G}_{n,k}\left[\psi_{1}\left(W,\theta_{0},p_{0},\eta_{10}\right)\right]\parallel,
\]
\[
I_{2,k}\coloneqq\sqrt{n}\parallel E_{P}\left[\psi_{1}\left(W,\theta_{0},p_{0},\hat{\eta}_{1k}\right)\mid\left(W_{i}\right)_{i\in I_{k}^{c}}\right]-E_{P}\left[\psi_{1}\left(W,\theta_{0},p_{0},\eta_{10}\right)\right]\parallel.
\]
To bound $I_{1,k}$, note that conditional on $\left(W_{i}\right)_{i\in I_{k}^{c}}$
the estimator $\hat{\eta}_{1k}$ is nonstochastic. Under the event
that $\hat{\eta}_{1k}\in\mathcal{T}_{N}$, we have 
\begin{align*}
E_{P}\left[I_{1,k}^{2}\mid\left(W_{i}\right)_{i\in I_{k}^{c}}\right]= & E_{P}\left[\parallel\psi_{1}\left(W,\theta_{0},p_{0},\hat{\eta}_{1k}\right)-\psi_{1}\left(W,\theta_{0},p_{0},\eta_{10}\right)\parallel^{2}\mid\left(W_{i}\right)_{i\in I_{k}^{c}}\right]\\
\leq & \sup_{\eta_{1}\in\mathcal{T}_{N}}E_{P}\left[\parallel\psi_{1}\left(W,\theta_{0},p_{0},\eta_{1}\right)-\psi_{1}\left(W,\theta_{0},p_{0},\eta_{10}\right)\parallel^{2}\mid\left(W_{i}\right)_{i\in I_{k}^{c}}\right]\\
= & \sup_{\eta_{1}\in\mathcal{T}_{N}}E_{P}\left[\parallel\psi_{1}\left(W,\theta_{0},p_{0},\eta_{1}\right)-\psi_{1}\left(W,\theta_{0},p_{0},\eta_{10}\right)\parallel^{2}\right]\\
= & \left(\varepsilon_{N}\right)^{2}
\end{align*}
by (A.1). Hence, $I_{1,k}=O_{P}\left(\varepsilon_{N}\right)$ by Lemma
A.1. To bound $I_{2,k}$, define the following function
\[
f_{k}\left(r\right)=E_{P}\left[\psi_{1}\left(W,\theta_{0},p_{0},\eta_{10}+r\left(\hat{\eta}_{1k}-\eta_{10}\right)\right)\mid\left(W_{i}\right)_{i\in I_{k}^{c}}\right]-E\left[\psi_{1}\left(W,\theta_{0},p_{0},\eta_{10}\right)\right],r\in[0,1).
\]
By Taylor series expansion, we have
\[
f_{k}\left(1\right)=f_{k}\left(0\right)+f'_{k}\left(0\right)+f''_{k}\left(\tilde{r}\right)/2,\text{for some }\tilde{r}\in\left(0,1\right).
\]
Note that $f_{k}\left(0\right)=0$ since $E\left[\psi_{1}\left(W,\theta_{0},p_{0},\eta_{10}\right)\mid\left(W_{i}\right)_{i\in I_{k}^{c}}\right]=E\left[\psi_{1}\left(W,\theta_{0},p_{0},\eta_{10}\right)\right]$.
Further, on the event $\hat{\eta}_{1k}\in\mathcal{T}_{N}$,
\[
\parallel f_{k}'\left(0\right)\parallel=\parallel\partial_{\eta_{1}}E_{P}\psi_{1}\left(W,\theta_{0},p_{0},\eta_{10}\right)\left[\hat{\eta}_{1k}-\eta_{10}\right]\parallel=0
\]
by the orthogonality of $\psi_{1}$. Also, on the event $\hat{\eta}_{1k}\in\mathcal{T}_{N}$,
\[
\parallel f_{k}''\left(\tilde{r}\right)\parallel\leq\sup_{r\in\left(0,1\right)}\parallel f_{k}''\left(r\right)\parallel\leq\left(\varepsilon_{N}\right)^{2}
\]
by (A.2). Thus,
\[
I_{2,k}=\sqrt{n}\parallel f_{k}\left(1\right)\parallel=O_{P}\left(\sqrt{n}\left(\varepsilon_{N}\right)^{2}\right).
\]
Together with the result on $I_{1,k}$, we have 
\begin{align*}
\mathbb{E}_{n,k}\left[\psi_{1}\left(W,\theta_{0},p_{0},\hat{\eta}_{1k}\right)\right]-\frac{1}{n}\sum_{i\in I_{k}}\psi_{1}\left(W_{i},\theta_{0},p_{0},\eta_{10}\right)\leq & \frac{I_{1,k}+I_{2,k}}{\sqrt{n}}\\
= & O_{P}\left(n^{-1/2}\varepsilon_{N}+\left(\varepsilon_{N}\right)^{2}\right)\\
= & o_{P}\left(N^{-1/2}\right)
\end{align*}
by $n=O\left(N\right)$ and $\varepsilon_{N}=o\left(N^{-1/4}\right)$. Hence, $\sqrt{N}R_{N}=o_{P}\left(1\right)$.

\emph{Step 2.} In this step, I present the proof of (A.1).  We have the following decomposition: 
\begin{align*}
\psi_{1}\left(W,\theta_{0},p_{0},\eta_{1}\right)-\psi_{1}\left(W,\theta_{0},p_{0},\eta_{10}\right)= & \frac{D-g\left(X\right)}{p_{0}\left(1-g\left(X\right)\right)}\left(Y\left(1\right)-Y\left(0\right)-\ell_{1}\left(X\right)\right)\\
 & -\frac{D-g_{0}\left(X\right)}{p_{0}\left(1-g_{0}\left(X\right)\right)}\left(Y\left(1\right)-Y\left(0\right)-\ell_{10}\left(X\right)\right)\\
= & \frac{U+g_{0}\left(X\right)-g\left(X\right)}{p_{0}\left(1-g\left(X\right)\right)}\left(V_{1}+\ell_{10}\left(X\right)-\ell_{1}\left(X\right)\right)\\
 & -\frac{UV_{1}}{p_{0}\left(1-g_{0}\left(X\right)\right)}.
\end{align*}
Thus, we have
\begin{align*}
\psi_{1}\left(W,\theta_{0},p_{0},\eta_{1}\right)-\psi_{1}\left(W,\theta_{0},p_{0},\eta_{10}\right)= & \frac{UV_{1}}{p_{0}\left(1-g\left(X\right)\right)}+\frac{U\left(\ell_{10}\left(X\right)-\ell_{1}\left(X\right)\right)}{p_{0}\left(1-g\left(X\right)\right)}+\frac{\left(g_{0}\left(X\right)-g\left(X\right)\right)V_{1}}{p_{0}\left(1-g\left(X\right)\right)}\\
 & +\frac{\left(g_{0}\left(X\right)-g\left(X\right)\right)\left(\ell_{10}\left(X\right)-\ell_{1}\left(X\right)\right)}{p_{0}\left(1-g\left(X\right)\right)}-\frac{UV_{1}}{p_{0}\left(1-g_{0}\left(X\right)\right)}.
\end{align*}
Given $\kappa\leq g_{0}\left(X\right)\leq1-\kappa$
and $\kappa\leq g\left(X\right)\leq1-\kappa$, 
\begin{align*}
\parallel\psi_{1}\left(W,\theta_{0},p_{0},\eta_{1}\right)-\psi_{1}\left(W,\theta_{0},p_{0},\eta_{10}\right)\parallel_{P,2}\leq & \frac{1}{p_{0}\kappa^{2}}\parallel UV_{1}\left(1-g_{0}\left(X\right)\right)\\
 & +U\left(\ell_{10}\left(X\right)-\ell_{1}\left(X\right)\right)\left(1-g_{0}\left(X\right)\right)\\
 & +V_{1}\left(g_{0}\left(X\right)-g\left(X\right)\right)\left(1-g_{0}\left(X\right)\right)\\
 & +\left(g_{0}-g\right)\left(\ell_{10}-\ell_{1}\right)\left(1-g_{0}\left(X\right)\right)\\
 & -UV_{1}\left(1-g\left(X\right)\right)\parallel_{P,2}.
\end{align*}
By $\kappa\leq g_{0}\left(X\right)\leq1-\kappa$ and $\kappa\leq g\left(X\right)\leq1-\kappa$
again, we can obtain
\begin{align*}
\parallel\psi_{1}\left(W,\theta_{0},p_{0},\eta_{1}\right)-\psi_{1}\left(W,\theta_{0},p_{0},\eta_{10}\right)\parallel_{P,2}\leq & \frac{1-\kappa}{p_{0}\kappa^{2}}\parallel UV_{1}+U\left(\ell_{10}\left(X\right)-\ell_{1}\left(X\right)\right)\\
 & +V_{1}\left(g_{0}\left(X\right)-g\left(X\right)\right)\\
 & +\left(g_{0}\left(X\right)-g\left(X\right)\right)\left(\ell_{10}\left(X\right)-\ell_{1}\left(X\right)\right)\\
 & -UV_{1}\parallel_{P,2}.
\end{align*}
Thus, by $E_{P}\left[U^{2}\mid X\right]\leq C$ and $E_{P}\left[V_{1}^{2}\mid X\right]\leq C$,

\begin{align*}
\parallel\psi_{1}\left(W,\theta_{0},p_{0},\eta_{1}\right)-\psi_{1}\left(W,\theta_{0},p_{0},\eta_{10}\right)\parallel_{P,2}\leq & \frac{(1-\kappa)\sqrt{C}}{p_{0}\kappa^{2}}\parallel\ell_{10}-\ell_{1}\parallel_{P,2}\\
 & +\frac{(1-\kappa)\sqrt{C}}{p_{0}\kappa^{2}}\parallel g_{0}-g\parallel_{P,2}\\
 & +\frac{(1-\kappa)}{p_{0}\kappa^{2}}\parallel g_{0}-g\parallel_{P,2}\parallel\ell_{10}-\ell_{1}\parallel_{P,2}\\
\leq & O\left(\varepsilon_{N}+\varepsilon_{N}+\left(\varepsilon_{N}\right)^{2}\right)\\
= & O\left(\varepsilon_{N}\right).
\end{align*}

\emph{Step 3.} In this step, I present the proof of (A.2). Define $$f\left(r\right)=E_{P}\left[\psi_{1}\left(W,\theta_{0},p_{0},\eta_{10}+r\left(\eta_{1}-\eta_{10}\right)\right)\right].$$
Then its second-order derivative is
\begin{align*}
\partial_{r}^{2}f\left(r\right)= & \frac{2}{p_{0}}E_{P}\left[\frac{\left(D-1\right)\left(g-g_{0}\right)^{2}}{\left(1-g_{0}-r\left(g-g_{0}\right)\right)^{3}}\left(Y\left(1\right)-Y\left(0\right)-\ell_{10}-r\left(\ell_{1}-\ell_{10}\right)\right)\right]\\
 & -\frac{2}{p_{0}}E_{P}\left[\frac{D-1}{\left(1-g_{0}-r\left(g-g_{0}\right)\right)^{2}}\left(\ell_{1}-\ell_{10}\right)\left(g-g_{0}\right)\right].
\end{align*}
It follows that 
\[
\mid\partial_{r}^{2}f\left(r\right)\mid\leq O\left(\parallel\left(g-g_{0}\right)\parallel_{P,2}^{2}+\parallel\left(g-g_{0}\right)\parallel_{P,2}\times\parallel\left(\ell_{1}-\ell_{10}\right)\parallel_{P,2}\right)\leq\left(\varepsilon_{N}\right)^{2}.
\]

\emph{Step }4. Notice that 
\begin{align*}
\partial_{p}\psi_{1}\left(W,\theta,p,\eta_{1}\right)= & -\frac{1}{p}\frac{D-g\left(X\right)}{1-g\left(X\right)}\left(Y\left(1\right)-Y\left(0\right)-\ell_{1}\left(X\right)\right)\\
= & -\frac{1}{p}\left(\psi_{1}\left(W,\theta,p,\eta_{1}\right)+\theta\right),
\end{align*}
then we have
\begin{align*}
\parallel\partial_{p}\psi_{1}\left(W,\theta_{0},p_{0},\eta_{1}\right)-\partial_{p}\psi_{1}\left(W,\theta_{0},p_{0},\eta_{10}\right)\parallel_{P,2}= & \frac{1}{p_{0}}\parallel\psi_{1}\left(W,\theta_{0},p_{0},\eta_{1}\right)-\psi_{1}\left(W,\theta_{0},p_{0},\eta_{10}\right)\parallel_{P,2}\\
= & O\left(\varepsilon_{N}\right)
\end{align*}
by Step 2. 

\emph{Step 5}. Notice that 
\begin{align*}
\partial_{p}^{2}\psi_{1}\left(W,\theta,p,\eta_{1}\right)= & \frac{2}{p^{3}}\frac{D-g\left(X\right)}{1-g\left(X\right)}\left(Y\left(1\right)-Y\left(0\right)-\ell_{1}\left(X\right)\right)\\
= & \frac{2}{p^{2}}\left(\psi_{1}\left(W,\theta,p,\eta_{1}\right)+\theta\right),
\end{align*}
then we have 
\begin{align*}
\partial_{p}^{2}\psi_{1}\left(W,\theta_{0},p,\eta_{1}\right)-\partial_{p}^{2}\psi_{1}\left(W,\theta_{0},p_{0},\eta_{10}\right)= & \partial_{p}^{2}\psi_{1}\left(W,\theta_{0},p_{0},\eta_{1}\right)-\partial_{p}^{2}\psi_{1}\left(W,\theta_{0},p_{0},\eta_{10}\right)\\
 & +\partial_{p^{3}}^{3}\psi_{1}\left(W,\theta_{0},\bar{p},\eta_{1}\right)\left(p-p_{0}\right)\\
= & \frac{2}{p_{0}^{2}}\left(\psi_{1}\left(W,\theta_{0},p_{0},\eta_{1}\right)-\psi_{1}\left(W,\theta_{0},p_{0},\eta_{10}\right)\right)\\
 & -\frac{6}{\bar{p}^{4}}\frac{D-g\left(X\right)}{1-g\left(X\right)}\left(Y\left(1\right)-Y\left(0\right)-\ell_{1}\left(X\right)\right)\left(p-p_{0}\right),
\end{align*}
where $\bar{p}\in\left(p,p_{0}\right)$. Thus,
\begin{align*}
\parallel\partial_{p}^{2}\psi_{1}\left(W,\theta_{0},p,\eta_{1}\right)-\partial_{p}^{2}\psi_{1}\left(W,\theta_{0},p_{0},\eta_{10}\right)\parallel_{P,2}\leq & \frac{2}{p_{0}^{2}}\parallel\psi_{1}\left(W,\theta_{0},p_{0},\eta_{1}\right)-\psi_{1}\left(W,\theta_{0},p_{0},\eta_{10}\right)\parallel_{P,2}\\
 & +\parallel\frac{6}{\bar{p}^{4}}\frac{D-g\left(X\right)}{1-g\left(X\right)}\left(Y\left(1\right)-Y\left(0\right)-\ell_{1}\left(X\right)\right)\parallel_{P,2}\\
 & \times\mid p-p_{0}\mid.
\end{align*}
The term in the second line is bounded by 
\begin{align*}
\frac{6}{\bar{p}^{4}\kappa}\parallel\left(U+g_{0}-g\right)\left(V_{1}+\ell_{10}-\ell_{1}\right)\parallel_{P,2}\leq & \frac{6}{\bar{p}^{4}\kappa}\parallel UV_{1}\parallel_{P,2}+\frac{6}{\bar{p}^{4}\kappa}\parallel U\left(\ell_{10}-\ell_{1}\right)\parallel_{P,2}\\
 & +\frac{6}{\bar{p}^{4}\kappa}\parallel V_{1}\left(g_{0}-g\right)\parallel_{P,2}\\
 & +\frac{6}{\bar{p}^{4}\kappa}\parallel g_{0}-g\parallel_{P,2}\parallel\ell_{10}-\ell_{1}\parallel_{P,2}\\
\leq & \frac{6}{\bar{p}^{4}\kappa}\left(C+\sqrt{C}\parallel\ell_{10}-\ell_{1}\parallel_{P,2}+\sqrt{C}\parallel g_{0}-g\parallel_{P,2}\right)\\
 & +\frac{6}{\bar{p}^{4}\kappa}\parallel g_{0}-g\parallel_{P,2}\parallel\ell_{10}-\ell_{1}\parallel_{P,2}\\
= & O\left(1\right)
\end{align*}
by $\parallel UV_{1}\parallel_{P,2}\leq \parallel UV_{1}\parallel_{P,4}\leq C$ , $E_{P}\left[U^{2}\mid X\right]\leq C$,
$E_{P}\left[V_{1}^{2}\mid X\right]\leq C$, and the conditions on
the rates of convergence. Together with Step 2, we obtain
\begin{align*}
\parallel\partial_{p}^{2}\psi_{1}\left(W,\theta_{0},p,\eta_{1}\right)-\partial_{p}^{2}\psi_{1}\left(W,\theta_{0},p_{0},\eta_{10}\right)\parallel_{P,2}\leq & O\left(\varepsilon_{N}\right)+O\left(1\right)\times O\left(N^{-1/2}\right)\\
= & O\left(\varepsilon_{N}\right),
\end{align*}
where I assume that $\varepsilon_{N}$ converges to zero no faster
than $N^{-1/2}$. 
\begin{description}
\item [{Repeated~cross~sections:}]~
\end{description}
In step 1, I show the main result with the following auxiliary results:
\[
\sup_{\eta_{2}\in\mathcal{T}_{N}}\left(E\left[\parallel\psi_{2}\left(W,\theta_{0},p_{0},\lambda_{0},\eta_{2}\right)-\psi_{2}\left(W,\theta_{0},p_{0},\lambda_{0},\eta_{20}\right)\parallel^{2}\right]\right)^{1/2}\leq\varepsilon_{N},\tag{A.5}
\]

\[
\sup_{r\in\left(0,1\right),\eta_{2}\in\mathcal{T}_{N}}\parallel\partial_{r}^{2}E\left[\psi_{2}\left(W,\theta_{0},p_{0},\lambda_{0},\eta_{20}+r\left(\eta_{2}-\eta_{20}\right)\right)\right]\parallel\leq\left(\varepsilon_{N}\right)^{2}.\tag{A.6}
\]
\[
\sup_{\eta_{2}\in\mathcal{T}_{N}}\left(E_{P}\left[\parallel\partial_{p}\psi_{2}\left(W,\theta_{0},p_{0},\lambda_{0},\eta_{2}\right)-\partial_{p}\psi_{2}\left(W,\theta_{0},p_{0},\lambda_{0},\eta_{20}\right)\parallel^{2}\right]\right)^{1/2}\leq\varepsilon_{N},\tag{A.7}
\]
\[
\sup_{\eta_{2}\in\mathcal{T}_{N}}\left(E_{P}\left[\parallel\partial_{\lambda}\psi_{2}\left(W,\theta_{0},p_{0},\lambda_{0},\eta_{2}\right)-\partial_{\lambda}\psi_{2}\left(W,\theta_{0},p_{0},\lambda_{0},\eta_{20}\right)\parallel^{2}\right]\right)^{1/2}\leq\varepsilon_{N},\tag{A.8}
\]
\[
\sup_{p\in\mathcal{P}_{N},\eta_{2}\in\mathcal{T}_{N}}\left(E_{P}\left[\parallel\partial_{p}^{2}\psi_{2}\left(W,\theta_{0},p,\lambda_{0},\eta_{2}\right)-\partial_{p}^{2}\psi_{2}\left(W,\theta_{0},p_{0},\lambda_{0},\eta_{20}\right)\parallel^{2}\right]\right)^{1/2}\leq\varepsilon_{N},\tag{A.9}
\]
\[
\sup_{p\in\mathcal{P}_{N},\lambda\in\Lambda_{N},\eta_{2}\in\mathcal{T}_{N}}\left(E_{P}\left[\parallel\partial_{\lambda}^{2}\psi_{2}\left(W,\theta_{0},p,\lambda,\eta_{2}\right)-\partial_{\lambda}^{2}\psi_{2}\left(W,\theta_{0},p_{0},\lambda_{0},\eta_{20}\right)\parallel^{2}\right]\right)^{1/2}\leq\varepsilon_{N},\tag{A.10}
\]
\[
\sup_{p\in\mathcal{P}_{N},\eta_{2}\in\mathcal{T}_{N}}\left(E_{P}\left[\parallel\partial_{\lambda}\partial_{p}\psi_{2}\left(W,\theta_{0},p,\lambda_{0},\eta_{2}\right)-\partial_{\lambda}\partial_{p}\psi_{2}\left(W,\theta_{0},p_{0},\lambda_{0},\eta_{20}\right)\parallel^{2}\right]\right)^{1/2}\leq\varepsilon_{N},\tag{A.11}
\]
where $\mathcal{T}_{N}$ is the set of all $\eta_{2}=\left(g,\ell_{2}\right)$
consisting of $P$-square-integrable functions $g$ and $\ell_{2}$
such that \emph{
\[
\parallel\eta_{2}-\eta_{20}\parallel_{P,2}\leq\varepsilon_{N},
\]
\[
\parallel g-1/2\parallel_{P,\infty}\leq1/2-\kappa,
\]
\[
\parallel\left(g-g_{0}\right)\parallel_{P,2}^{2}+\parallel\left(g-g_{0}\right)\parallel_{P,2}\times\parallel\left(\ell_{2}-\ell_{20}\right)\parallel_{P,2}\leq\left(\varepsilon_{N}\right)^{2},
\]
}
$\mathcal{P}_{N}$ and $\Lambda_{N}$ are the sets consisting
all $p>0$ and $\lambda>0$ such that $\mid p-p_{0}\mid\leq N^{-1/2}$
and $\mid\lambda-\lambda_{0}\mid\leq N^{-1/2}$, respectively. By the regularity condition (3.2), $\mid\hat{p}_{k}-p_{0}\mid=O_{P}\left(N^{-1/2}\right)$, and $\mid\hat{\lambda}_{k}-\lambda_{0}\mid=O_{P}\left(N^{-1/2}\right)$,
we have $\hat{\eta}_{2k}\in \mathcal{T}_{N}$, $\hat{p}_{k}\in\mathcal{P}_{N}$, and $\hat{\lambda}_{k}\in\Lambda_{N}$
with probability $1-o\left(1\right)$.  

In Step 2-4, I show the above auxiliary results.

\emph{Step 1. }Notice that 
\begin{align*}
\sqrt{N}\left(\tilde{\theta}-\theta_{0}\right)= & \sqrt{N}\left(\frac{1}{K}\sum_{k=1}^{K}\tilde{\theta}_{k}-\theta_{0}\right)\\
= & \sqrt{N}\frac{1}{K}\sum_{k=1}^{K}\mathbb{E}_{n,k}\left[\psi_{2}\left(W,\theta_{0},\hat{p}_{k},\hat{\lambda}_{k},\hat{\eta}_{2k}\right)\right]\\
= & \sqrt{N}\frac{1}{K}\sum_{k=1}^{K}\mathbb{E}_{n,k}\left[\psi_{2}\left(W,\theta_{0},p_{0},\lambda_{0},\hat{\eta}_{2k}\right)\right]\\
 & +\sqrt{N}\frac{1}{K}\sum_{k=1}^{K}\mathbb{E}_{n,k}\left[\partial_{p}\psi_{2}\left(W,\theta_{0},p_{0},\lambda_{0},\hat{\eta}_{2k}\right)\right]\left(\hat{p}_{k}-p_{0}\right)\\
 & +\sqrt{N}\frac{1}{K}\sum_{k=1}^{K}\mathbb{E}_{n,k}\left[\partial_{\lambda}\psi_{2}\left(W,\theta_{0},p_{0},\lambda_{0},\hat{\eta}_{2k}\right)\right]\left(\hat{\lambda}_{k}-\lambda_{0}\right)+o_{P}\left(1\right),
\end{align*}
where the term $o_{P}\left(1\right)$, by the same arguments for the
term $b$ in repeated outcomes and the auxiliary results (A.9)-(A.11),
contains the second-order terms 
\[
\sqrt{N}\frac{1}{K}\sum_{k=1}^{K}\mathbb{E}_{n,k}\left[\partial_{p}^{2}\psi_{2}\left(W,\theta_{0},\bar{p}_{k},\lambda_{0},\hat{\eta}_{2k}\right)\right]\left(\hat{p}_{k}-p_{0}\right)^{2},
\]
\[
\sqrt{N}\frac{1}{K}\sum_{k=1}^{K}\mathbb{E}_{n,k}\left[\partial_{\lambda}^{2}\psi_{2}\left(W,\theta_{0},\hat{p}_{k},\bar{\lambda}_{k},\hat{\eta}_{2k}\right)\right]\left(\hat{\lambda}_{k}-\lambda_{0}\right)^{2},
\]
\[
\sqrt{N}\frac{1}{K}\sum_{k=1}^{K}\mathbb{E}_{n,k}\left[\partial_{\lambda}\partial_{p}\psi_{2}\left(W,\theta_{0},\bar{p}_{k},\lambda_{0},\hat{\eta}_{2k}\right)\right]\left(\hat{\lambda}_{k}-\lambda_{0}\right)\left(\hat{p}_{k}-p_{0}\right),
\]
where $\bar{p}_{k}\in\left(\hat{p}_{k},p_{0}\right)$ and $\bar{\lambda}_{k}\in\left(\hat{\lambda}_{k},\lambda_{0}\right)$. On the other hand, 
by the same arguments for the term $a$ in repeated outcomes and the
auxiliary results (A.7)-(A.8), we have 
\[
\mathbb{E}_{n,k}\left[\partial_{p}\psi_{2}\left(W,\theta_{0},p_{0},\lambda_{0},\hat{\eta}_{2k}\right)\right]\stackrel{p}{\rightarrow}E_{p}\left[\partial_{p}\psi_{2}\left(W,\theta_{0},p_{0},\lambda_{0},\eta_{20}\right)\right]=G_{2p0},
\]
\[
\mathbb{E}_{n,k}\left[\partial_{\lambda}\psi_{2}\left(W,\theta_{0},p_{0},\lambda_{0},\hat{\eta}_{2k}\right)\right]\stackrel{p}{\rightarrow}E_{p}\left[\partial_{\lambda}\psi_{2}\left(W,\theta_{0},p_{0},\lambda_{0},\eta_{20}\right)\right]=G_{2\lambda0}.
\]
Hence, since $\hat{p}_{k}-p_{0}=\mathbb{E}_{n,k}\left[D-p_{0}\right]$ and
$\hat{\lambda}_{k}-\lambda_{0}=\mathbb{E}_{n,k}\left[T-\lambda_{0}\right]$,
we have 
\begin{align*}
\sqrt{N}\left(\tilde{\theta}-\theta_{0}\right)= & \sqrt{N}\frac{1}{K}\sum_{k=1}^{K}\mathbb{E}_{n,k}\left[\psi_{2}\left(W,\theta_{0},p_{0},\lambda_{0},\hat{\eta}_{2k}\right)\right]\\
= & \sqrt{N}\frac{1}{K}\sum_{k=1}^{K}\mathbb{E}_{n,k}\left[\psi_{2}\left(W,\theta_{0},p_{0},\lambda_{0},\hat{\eta}_{1k}\right)+G_{2p0}\left(D-p_{0}\right)+G_{2\lambda0}\left(T-\lambda_{0}\right)\right]+o_{P}\left(1\right)\\
= & \frac{1}{\sqrt{N}}\sum_{i=1}^{N}\left[\psi_{2}\left(W_{i},\theta_{0},p_{0},\lambda_{0},\eta_{20}\right)+G_{2p0}\left(D_{i}-p_{0}\right)+G_{2\lambda0}\left(T_{i}-\lambda_{0}\right)\right]\\
 & +\sqrt{N}R_{N}'+o_{P}\left(1\right),
\end{align*}
where 
\begin{align*}
R_{N}'= & \frac{1}{K}\sum_{k=1}^{K}\mathbb{E}_{n,k}\left[\psi_{2}\left(W,\theta_{0},p_{0},\lambda_{0},\hat{\eta}_{2k}\right)+G_{2p0}\left(D-p_{0}\right)+G_{2\lambda0}\left(T-\lambda_{0}\right)\right]\\
 & -\frac{1}{N}\sum_{i=1}^{N}\left[\psi_{2}\left(W_{i},\theta_{0},p_{0},\lambda_{0},\eta_{20}\right)+G_{2p0}\left(D_{i}-p_{0}\right)+G_{2\lambda0}\left(T_{i}-\lambda_{0}\right)\right]\\
= & \frac{1}{K}\sum_{k=1}^{K}\mathbb{E}_{n,k}\left[\psi_{2}\left(W,\theta_{0},p_{0},\lambda_{0},\hat{\eta}_{2k}\right)\right]-\frac{1}{N}\sum_{i=1}^{N}\psi_{2}\left(W_{i},\theta_{0},p_{0},\lambda_{0},\eta_{10}\right).
\end{align*}
Using (A.5)-(A.6) and the same arguments as the step 1 in repeated
outcomes, one can show that $\sqrt{N}R_{N}'=o_{P}\left(1\right)$.
Hence, it remains to prove the auxiliary results (A.5)-(A.11). 

\emph{Step 2. }
Recall that $p_{0}'= p_{0}\lambda_{0}\left(1-\lambda_{0}\right)$. For (A.5), notice that
\begin{align*}
\psi_{2}\left(W,\theta_{0},p_{0},\lambda_{0},\eta_{2}\right)-\psi_{2}\left(W,\theta_{0},p_{0},\lambda_{0},\eta_{20}\right)= & \frac{D-g\left(X\right)}{p_{0}'\left(1-g\left(X\right)\right)}\left(\left(T-\lambda_{0}\right)Y-\ell_{2}\left(X\right)\right)\\
 & -\frac{D-g_{0}\left(X\right)}{p_{0}'\left(1-g_{0}\left(X\right)\right)}\left(\left(T-\lambda_{0}\right)Y-\ell_{20}\left(X\right)\right)\\
= & \frac{U+g_{0}\left(X\right)-g\left(X\right)}{p_{0}'\left(1-g\left(X\right)\right)}\left(V_{2}+\ell_{20}\left(X\right)-\ell_{2}\left(X\right)\right)\\
 & -\frac{UV_{2}}{p_{0}'\left(1-g_{0}\left(X\right)\right)}.
\end{align*}
The decomposition becomes
\begin{align*}
\psi_{2}\left(W,\theta_{0},p_{0},\lambda_{0},\eta_{2}\right)-\psi_{2}\left(W,\theta_{0},p_{0},\lambda_{0},\eta_{20}\right)= & \frac{UV_{2}}{p_{0}'\left(1-g\left(X\right)\right)}+\frac{U\left(\ell_{20}\left(X\right)-\ell_{2}\left(X\right)\right)}{p_{0}'\left(1-g\left(X\right)\right)}\\
 & +\frac{\left(g_{0}\left(X\right)-g\left(X\right)\right)V_{2}}{p_{0}'\left(1-g\left(X\right)\right)}\\
 & +\frac{\left(g_{0}\left(X\right)-g\left(X\right)\right)\left(\ell_{20}\left(X\right)-\ell_{2}\left(X\right)\right)}{p_{0}'\left(1-g\left(X\right)\right)}\\
 & -\frac{UV_{2}}{p_{0}'\left(1-g_{0}\left(X\right)\right)}.
\end{align*}
Given that $\kappa\leq g_{0}\left(X\right)\leq1-\kappa$,
$\kappa\leq g\left(X\right)\leq1-\kappa$, we have
\begin{align*}
\parallel\psi_{2}\left(W,\theta_{0},p_{0},\lambda_{0},\eta_{2}\right)-\psi_{2}\left(W,\theta_{0},p_{0},\lambda_{0},\eta_{20}\right)\parallel_{P,2}\leq & \frac{1}{p_{0}'\kappa^{2}}\parallel UV_{2}\left(1-g_{0}\left(X\right)\right)\\
 & +U\left(\ell_{20}\left(X\right)-\ell_{2}\left(X\right)\right)\left(1-g_{0}\left(X\right)\right)\\
 & +V_{2}\left(g_{0}\left(X\right)-g\left(X\right)\right)\left(1-g_{0}\left(X\right)\right)\\
 & +\left(g_{0}-g\right)\left(\ell_{20}-\ell_{2}\right)\left(1-g_{0}\left(X\right)\right)\\
 & -UV_{2}\left(1-g\left(X\right)\right)\parallel_{P,2}.
\end{align*}
By $\kappa\leq g_{0}\left(X\right)\leq1-\kappa$, $\kappa\leq g\left(X\right)\leq1-\kappa$
again, we obtain
\begin{align*}
\parallel\psi_{2}\left(W,\theta_{0},p_{0},\lambda_{0},\eta_{2}\right)-\psi_{2}\left(W,\theta_{0},p_{0},\lambda_{0},\eta_{20}\right)\parallel_{P,2}\leq & \frac{1-\kappa}{p_{0}'\kappa^{2}}\parallel UV_{2}\\
 & +U\left(\ell_{20}\left(X\right)-\ell_{2}\left(X\right)\right)\\
 & +V_{2}\left(g_{0}\left(X\right)-g\left(X\right)\right)\\
 & +\left(g_{0}-g\right)\left(\ell_{20}-\ell_{2}\right)\\
 & -UV_{2}\parallel_{P,2}.
\end{align*}
Given $E_{P}\left[U^{2}\mid X\right]\leq C$, $E_{P}\left[V_{2}^{2}\mid X\right]\leq C$,
and the conditions on the rates of convergence, 
\begin{align*}
\parallel\psi_{2}\left(W,\theta_{0},p_{0},\lambda_{0},\eta_{2}\right)-\psi_{2}\left(W,\theta_{0},p_{0},\lambda_{0},\eta_{20}\right)\parallel_{P,2}\leq & \frac{(1-\kappa)\sqrt{C}}{p_{0}'\kappa^{2}}\parallel\ell_{20}\left(X\right)-\ell_{2}\left(X\right)\parallel_{P,2}\\
 & +\frac{(1-\kappa)\sqrt{C}}{p_{0}'\kappa^{2}}\parallel g_{0}\left(X\right)-g\left(X\right)\parallel_{P,2}\\
 & +\frac{(1-\kappa)}{p_{0}'\kappa^{2}}\parallel g_{0}-g\parallel_{P,2}\parallel\ell_{20}-\ell_{2}\parallel_{P,2}\\
\leq & O\left(\varepsilon_{N}+\varepsilon_{N}+\left(\varepsilon_{N}\right)^{2}\right)\\
= & O\left(\varepsilon_{N}\right).
\end{align*}
For (A.6), let $f\left(r\right)=E_{P}\left[\psi_{2}\left(W,\theta_{0},p_{0},\lambda_{0},\eta_{20}+r\left(\eta_{2}-\eta_{20}\right)\right)\right]$.
Then the second-order derivative is
\begin{align*}
\partial_{r}^{2}f\left(r\right)= & \frac{2}{p_{0}'}E_{P}\left[\frac{\left(D-1\right)\left(g-g_{0}\right)^{2}}{\left(1-g_{0}-r\left(g-g_{0}\right)\right)^{3}}\left(\left(T-\lambda_{0}\right)Y-\ell_{20}-r\left(\ell_{2}-\ell_{20}\right)\right)\right]\\
 & -\frac{2}{p_{0}'}E_{P}\left[\frac{D-1}{\left(1-g_{0}-r\left(g-g_{0}\right)\right)^{2}}\left(\ell_{2}-\ell_{20}\right)\left(g-g_{0}\right)\right]
\end{align*}
It follows that 
\[
\mid\partial_{r}^{2}f\left(r\right)\mid\leq O\left(\parallel\left(g-g_{0}\right)\parallel_{P,2}^{2}+\parallel\left(g-g_{0}\right)\parallel_{P,2}\times\parallel\left(\ell_{2}-\ell_{20}\right)\parallel_{P,2}\right)\leq\left(\varepsilon_{N}\right)^{2}.
\]

\emph{Step 3. }For (A.7), notice that 
\begin{align*}
\partial_{p}\psi_{2}\left(W,\theta,p,\lambda,\eta_{2}\right)= & -\frac{1}{p^{2}\lambda\left(1-\lambda\right)}\frac{D-g\left(X\right)}{1-g\left(X\right)}\left(\left(T-\lambda\right)Y-\ell_{2}\left(X\right)\right)\\
= & -\frac{1}{p}\left(\psi_{2}\left(W,\theta,p,\lambda,\eta_{2}\right)+\theta\right),
\end{align*}
then we have
\begin{align*}
\parallel\partial_{p}\psi_{2}\left(W,\theta_{0},p_{0},\lambda_{0},\eta_{2}\right)-\partial_{p}\psi_{2}\left(W,\theta_{0},p_{0},\lambda_{0},\eta_{20}\right)\parallel_{P,2}= & \frac{1}{p_{0}}\parallel\psi_{2}\left(W,\theta_{0},p_{0},\lambda_{0},\eta_{2}\right)\\
 & -\psi_{2}\left(W,\theta_{0},p_{0},\lambda_{0},\eta_{20}\right)\parallel_{P,2}\\
= & O\left(\varepsilon_{N}\right)
\end{align*}
by the proof of (A.5).

For (A.8), notice that 
\begin{align*}
\partial_{\lambda}\psi_{2}\left(W,\theta,p,\lambda,\eta_{2}\right)= & -\frac{1-2\lambda}{\lambda^{2}\left(1-\lambda\right)^{2}}\frac{D-g\left(X\right)}{p\left(1-g\left(X\right)\right)}\left(\left(T-\lambda\right)Y-\ell_{2}\left(X\right)\right)\\
 & -\frac{Y}{p\lambda\left(1-\lambda\right)}\frac{D-g\left(X\right)}{1-g\left(X\right)}.
\end{align*}
Define $\partial_{\lambda}\psi_{20}\coloneqq\partial_{\lambda}\psi_{2}\left(W,\theta_{0},p_{0},\lambda_{0},\eta_{20}\right)$,
then
\begin{align*}
\parallel\partial_{\lambda}\psi_{2}\left(W,\theta_{0},p_{0},\lambda_{0},\eta_{2}\right)-\partial_{\lambda}\psi_{20}\parallel_{P,2}= & \parallel\psi_{2}\left(W,\theta_{0},p_{0},\lambda_{0},\eta_{2}\right)-\psi_{2}\left(W,\theta_{0},p_{0},\lambda_{0},\eta_{20}\right)\parallel_{P,2}\\
 & \times \frac{\mid1-2\lambda_{0}\mid}{\lambda_{0}\left(1-\lambda_{0}\right)}+\parallel\frac{Y}{p_{0}'}\left(\frac{D-g\left(X\right)}{1-g\left(X\right)}-\frac{D-g_{0}\left(X\right)}{1-g_{0}\left(X\right)}\right)\parallel_{P,2}\\
= & O\left(\varepsilon_{N}\right)+\parallel\frac{Y}{p_{0}'}\left(\frac{D-g\left(X\right)}{1-g\left(X\right)}-\frac{D-g_{0}\left(X\right)}{1-g_{0}\left(X\right)}\right)\parallel_{P,2}\\
\leq & O\left(\varepsilon_{N}\right)+\frac{1}{p_{0}'\kappa^{2}}\parallel Y\left(g-g_{0}\right)\left(D-1\right)\parallel_{P,2}\\
\leq & O\left(\varepsilon_{N}\right)+\frac{\sqrt{C}}{p_{0}'\kappa^{2}}\parallel g-g_{0}\parallel_{P,2}\\
= & O\left(\varepsilon_{N}\right),
\end{align*}
by (A.5) and $E_{P}\left[Y^{2}\mid X\right]\leq C$. 

\emph{Step 4. }For (A.9), notice that we have 
\[
\partial_{p}^{2}\psi_{2}\left(W,\theta,p,\lambda,\eta_{2}\right)=\frac{2}{p^{3}\lambda\left(1-\lambda\right)}\frac{D-g\left(X\right)}{1-g\left(X\right)}\left(\left(T-\lambda\right)Y-\ell_{2}\left(X\right)\right).
\]
Define $\partial_{p}^{2}\psi_{20}\coloneqq\partial_{p}^{2}\psi_{2}\left(W,\theta_{0},p_{0},\lambda_{0},\eta_{20}\right)$,
then we have 
\begin{align*}
\partial_{p}^{2}\psi_{2}\left(W,\theta_{0},p,\lambda_{0},\eta_{2}\right)-\partial_{p}^{2}\psi_{20}= & \partial_{p}^{2}\psi_{2}\left(W,\theta_{0},p_{0},\lambda_{0},\eta_{2}\right)-\partial_{p}^{2}\psi_{20}\\
 & +\partial_{p}^{3}\psi_{2}\left(W,\theta_{0},\bar{p},\lambda_{0},\eta_{2}\right)\left(p-p_{0}\right)\\
= & \frac{2}{p^{2}}\left(\psi_{2}\left(W,\theta_{0},p_{0},\lambda_{0},\eta_{2}\right)-\psi_{2}\left(W,\theta_{0},p_{0},\lambda_{0},\eta_{20}\right)\right)\\
 & +\partial_{p}^{3}\psi_{2}\left(W,\theta_{0},\bar{p},\lambda_{0},\eta_{2}\right)\left(p-p_{0}\right),
\end{align*}
where $\bar{p}\in\left(p,p_{0}\right)$. Hence, we have
\begin{align*}
\parallel\partial_{p}^{2}\psi_{2}\left(W,\theta_{0},p,\lambda_{0},\eta_{2}\right)-\partial_{p}^{2}\psi_{20}\parallel_{P,2}\leq & \frac{2}{p^{2}}\parallel\psi_{2}\left(W,\theta_{0},p_{0},\lambda_{0},\eta_{2}\right)-\psi_{2}\left(W,\theta_{0},p_{0},\lambda_{0},\eta_{20}\right)\parallel_{P,2}\\
 & +\parallel\frac{D-g\left(X\right)}{1-g\left(X\right)}\left(\left(T-\lambda_{0}\right)Y-\ell_{2}\left(X\right)\right)\parallel_{P,2}\\
 & \times\frac{6}{\bar{p}^{4}\lambda_{0}\left(1-\lambda_{0}\right)}\mid p-p_{0}\mid.
\end{align*}
By (A.5), we have $\parallel\psi_{2}\left(W,\theta_{0},p_{0},\lambda_{0},\eta_{2}\right)-\psi_{2}\left(W,\theta_{0},p_{0},\lambda_{0},\eta_{20}\right)\parallel_{P,2}=O\left(\varepsilon_{N}\right)$.
The term in the second line is bounded by
\begin{align*}
\frac{1}{\kappa}\parallel\left(U+g_{0}-g\right)\left(V_{2}+\ell_{20}-\ell_{2}\right)\parallel_{P,2}\leq & \frac{1}{\kappa}\parallel UV_{2}\parallel_{P,2}+\frac{1}{\kappa}\parallel U\left(\ell_{20}-\ell_{2}\right)\parallel_{P,2}\\
 & +\frac{1}{\kappa}\parallel V_{2}\left(g_{0}-g\right)\parallel_{P,2}+\frac{1}{\kappa}\parallel g_{0}-g\parallel_{P,2}\parallel\ell_{20}-\ell_{2}\parallel_{P,2}\\
\leq & \frac{1}{\kappa}\left(C+\sqrt{C}\parallel\ell_{20}-\ell_{2}\parallel_{P,2}+\sqrt{C}\parallel g_{0}-g\parallel_{P,2}\right)\\
 & +\frac{1}{\kappa}\parallel g_{0}-g\parallel_{P,2}\parallel\ell_{20}-\ell_{2}\parallel_{P,2}\\
= & O\left(1\right)
\end{align*}
by $\parallel UV_{2}\parallel_{P,2}\leq \parallel UV_{2}\parallel_{P,4} \leq C$, $E_{P}\left[U^{2}\mid X\right]\leq C$,
and $E_{P}\left[V_{2}^{2}\mid X\right]\leq C$. Thus, we obtain 
\begin{align*}
\parallel\partial_{p}^{2}\psi_{2}\left(W,\theta_{0},p,\lambda_{0},\eta_{2}\right)-\partial_{p}^{2}\psi_{20}\parallel_{P,2}\leq & O\left(\varepsilon_{N}\right)+O\left(1\right)\times O\left(N^{-1/2}\right)\\
= & O\left(\varepsilon_{N}\right),
\end{align*}
where I assume that $\varepsilon_{N}$ converges to zero no faster
than $N^{-1/2}$.

For (A.10), notice that we have
\begin{align*}
\partial_{\lambda}^{2}\psi_{2}\left(W,\theta,p,\lambda,\eta_{2}\right)= & \frac{c_{1}}{p\lambda^{3}\left(1-\lambda\right)^{3}}\frac{D-g\left(X\right)}{1-g\left(X\right)}\left(\left(T-\lambda\right)Y-\ell_{2}\left(X\right)\right)\\
 & +\frac{2-4\lambda}{p\lambda^{2}\left(1-\lambda\right)^{2}}\frac{D-g\left(X\right)}{1-g\left(X\right)}Y,
\end{align*}
where $c_{1}$ is a constant depending on $\lambda$. Define $\partial_{\lambda}^{2}\psi_{20}\coloneqq\partial_{\lambda}^{2}\psi_{2}\left(W,\theta_{0},p_{0},\lambda_{0},\eta_{20}\right)$,
we have
\begin{align*}
\partial_{\lambda}^{2}\psi_{2}\left(W,\theta_{0},p,\lambda,\eta_{2}\right)-\partial_{\lambda}^{2}\psi_{20}= & \partial_{\lambda}^{2}\psi_{2}\left(W,\theta_{0},p_{0},\lambda_{0},\eta_{2}\right)-\partial_{\lambda}^{2}\psi_{20}\\
 & +\partial_{\lambda}^{2}\partial_{p}\psi_{2}\left(W,\theta_{0},\bar{p},\lambda,\eta_{2}\right)\left(p-p_{0}\right)\\
 & +\partial_{\lambda}^{3}\psi_{2}\left(W,\theta_{0},p_{0},\bar{\lambda},\eta_{2}\right)\left(\lambda-\lambda_{0}\right)\\
= & \frac{c_{1}}{\lambda_{0}^{2}\left(1-\lambda_{0}\right)^{2}}\left(\psi_{2}\left(W,\theta_{0},p_{0},\lambda_{0},\eta_{2}\right)-\psi_{2}\left(W,\theta_{0},p_{0},\lambda_{0},\eta_{20}\right)\right)\\
 & +\frac{2-4\lambda_{0}}{p_{0}\lambda_{0}^{2}\left(1-\lambda_{0}\right)^{2}}\left(\frac{D-g\left(X\right)}{1-g\left(X\right)}-\frac{D-g_{0}\left(X\right)}{1-g_{0}\left(X\right)}\right)Y\\
 & +\partial_{\lambda}^{2}\partial_{p}\psi_{2}\left(W,\theta_{0},\bar{p},\lambda,\eta_{2}\right)\left(p-p_{0}\right)\\
 & +\partial_{\lambda}^{3}\psi_{2}\left(W,\theta_{0},p_{0},\bar{\lambda},\eta_{2}\right)\left(\lambda-\lambda_{0}\right),
\end{align*}
where $\bar{p}\in\left(p,p_{0}\right)$ and $\bar{\lambda}\in\left(\lambda,\lambda_{0}\right)$.
By the triangle inequality, we have
\begin{align*}
\parallel\partial_{\lambda}^{2}\psi_{2}\left(W,\theta_{0},p,\lambda,\eta_{2}\right)-\partial_{\lambda}^{2}\psi_{20}\parallel_{P,2}\leq & \frac{\mid c_{1}\mid}{\lambda^{2}\left(1-\lambda\right)^{2}}\parallel\psi_{2}\left(W,\theta_{0},p_{0},\lambda_{0},\eta_{2}\right)-\psi_{2}\left(W,\theta_{0},p_{0},\lambda_{0},\eta_{20}\right)\parallel_{P,2}\\
 & +\frac{\mid2-4\lambda_{0}\mid}{p_{0}\lambda_{0}^{2}\left(1-\lambda_{0}\right)^{2}}\parallel\left(\frac{D-g\left(X\right)}{1-g\left(X\right)}-\frac{D-g_{0}\left(X\right)}{1-g_{0}\left(X\right)}\right)Y\parallel_{P,2}\\
 & +\parallel\partial_{\lambda}^{2}\partial_{p}\psi_{2}\left(W,\theta_{0},\bar{p},\lambda,\eta_{2}\right)\parallel_{P,2}\mid p-p_{0}\mid\\
 & +\parallel\partial_{\lambda}^{3}\psi_{2}\left(W,\theta_{0},p_{0},\bar{\lambda},\eta_{2}\right)\parallel_{P,2}\mid\lambda-\lambda_{0}\mid.
\end{align*}
The norm term is the second line is bounded by 
\begin{align*}
\frac{1}{\kappa^{2}}\parallel Y\left(D-1\right)\left(g-g_{0}\right)\parallel_{P,2}\leq & \frac{\sqrt{C}}{\kappa^{2}}\parallel g-g_{0}\parallel_{P,2}\\
= & O\left(\varepsilon_{N}\right),
\end{align*}
by $E_{P}\left[Y^{2}\mid X\right]\leq C$ and $D\in\left\{ 0,1\right\} $.
The two high-order terms are bounded by
\begin{align*}
\parallel\partial_{\lambda}^{2}\partial_{p}\psi_{2}\left(W,\theta_{0},\bar{p},\lambda,\eta_{2}\right)\parallel_{P,2}\leq & \frac{\mid c_{1}\mid}{\bar{p}^{2}\lambda^{3}\left(1-\lambda\right)^{3}}\parallel\frac{D-g\left(X\right)}{1-g\left(X\right)}\left(\left(T-\lambda\right)Y-\ell_{2}\left(X\right)\right)\parallel_{P,2}\\
 & +\frac{\mid2-4\lambda\mid}{\bar{p}\lambda^{2}\left(1-\lambda\right)^{2}}\parallel\frac{D-g\left(X\right)}{1-g\left(X\right)}Y\parallel_{P,2}.
\end{align*}
and
\begin{align*}
\parallel\partial_{\lambda}^{3}\psi_{2}\left(W,\theta_{0},p_{0},\bar{\lambda},\eta_{2}\right)\parallel_{P,2}\leq & \frac{\mid c_{2}\mid}{p_{0}\bar{\lambda}^{4}\left(1-\bar{\lambda}\right)^{4}}\parallel\frac{D-g\left(X\right)}{1-g\left(X\right)}\left(\left(T-\bar{\lambda}\right)Y-\ell_{2}\left(X\right)\right)\parallel_{P,2}\\
 & +\frac{\mid c_{3}\mid}{p_{0}\bar{\lambda}^{3}\left(1-\bar{\lambda}\right)^{3}}\parallel\frac{D-g\left(X\right)}{1-g\left(X\right)}\times Y\parallel_{P,2},
\end{align*}
where $c_{2}$ and $c_{3}$ are constants depending on $\lambda$.
Using the same arguments in (A.9), one can show that 
\[
\parallel\frac{D-g\left(X\right)}{1-g\left(X\right)}\left(\left(T-\lambda\right)Y-\ell_{2}\left(X\right)\right)\parallel_{P,2}\leq O\left(1\right),
\]
\[
\parallel\frac{D-g\left(X\right)}{1-g\left(X\right)}\left(\left(T-\bar{\lambda}\right)Y-\ell_{2}\left(X\right)\right)\parallel_{P,2}\leq O\left(1\right).
\]
Also, we have 
\begin{align*}
\parallel\frac{D-g\left(X\right)}{1-g\left(X\right)}\times Y\parallel_{P,2}= & \parallel\frac{U+g_{0}\left(X\right)-g\left(X\right)}{1-g\left(X\right)}\times Y\parallel_{P,2}\\
\leq & \frac{1}{\kappa}\left(\parallel UY\parallel_{P,2}+\parallel\left(g_{0}-g\right)Y\parallel_{P,2}\right)\\
\leq & \frac{1}{\kappa}\left(C+\sqrt{C}\parallel g_{0}-g\parallel_{P,2}\right)\\
= & O\left(1\right)
\end{align*}
by $\parallel UY\parallel_{P,2}\leq C$ and $E_{P}\left[Y^{2}\mid X\right]\leq C$. 

Finally, we obtain
\begin{align*}
\parallel\partial_{\lambda}^{2}\psi_{2}\left(W,\theta_{0},p,\lambda,\eta_{2}\right)-\partial_{\lambda}^{2}\psi_{20}\parallel_{P,2}\leq & O\left(\varepsilon_{N}\right)+O\left(\varepsilon_{N}\right)+O\left(1\right)O\left(N^{-1/2}\right)+O\left(1\right)O\left(N^{-1/2}\right)\\
= & O\left(\varepsilon_{N}\right),
\end{align*}
where I assume that $\varepsilon_{N}$ converges to zero no faster
than $N^{-1/2}$.

For (A.11), notice that the derivative is
\begin{align*}
\partial_{\lambda}\partial_{p}\psi_{2}\left(W,\theta,p,\lambda,\eta_{2}\right)= & \frac{1-2\lambda}{p^{2}\lambda^{2}\left(1-\lambda\right)^{2}}\frac{D-g\left(X\right)}{1-g\left(X\right)}\left(\left(T-\lambda\right)Y-\ell_{2}\left(X\right)\right)\\
 & +\frac{Y}{p^{2}\lambda\left(1-\lambda\right)}\frac{D-g\left(X\right)}{1-g\left(X\right)}.
\end{align*}
Define $\partial_{\lambda}\partial_{p}\psi_{20}\coloneqq\partial_{\lambda}\partial_{p}\psi_{2}\left(W,\theta_{0},p_{0},\lambda_{0},\eta_{20}\right),$
then we have
\begin{align*}
\partial_{\lambda}\partial_{p}\psi_{2}\left(W,\theta_{0},p,\lambda_{0},\eta_{2}\right)-\partial_{\lambda}\partial_{p}\psi_{20}= & \partial_{\lambda}\partial_{p}\psi_{2}\left(W,\theta_{0},p_{0},\lambda_{0},\eta_{2}\right)-\partial_{\lambda}\partial_{p}\psi_{20}\\
 & +\partial_{\lambda}\partial_{p}^{2}\psi_{2}\left(W,\theta_{0},\bar{p},\lambda_{0},\eta_{2}\right)\left(p-p_{0}\right),
\end{align*}
where $\bar{p}\in\left(p,p_{0}\right)$. By the triangle inequality,
we obtain
\begin{align*}
\parallel\partial_{\lambda}\partial_{p}\psi_{2}\left(W,\theta_{0},p,\lambda_{0},\eta_{2}\right)-\partial_{\lambda}\partial_{p}\psi_{20}\parallel_{P,2}\leq & \frac{1}{p}\parallel\partial_{\lambda}\psi_{2}\left(W,\theta_{0},p_{0},\lambda_{0},\eta_{2}\right)-\partial_{\lambda}\psi_{2}\left(W,\theta_{0},p_{0},\lambda_{0},\eta_{20}\right)\parallel_{P,2}\\
 & +\parallel\partial_{\lambda}\partial_{p}^{2}\psi_{2}\left(W,\theta_{0},\bar{p},\lambda_{0},\eta_{2}\right)\parallel_{P,2}\mid p-p_{0}\mid.
\end{align*}
Using the same arguments in (A.9) and (A.10), one can show that the high-order
term is bounded by 
\begin{align*}
\parallel\partial_{\lambda}\partial_{p}^{2}\psi_{2}\left(W,\theta_{0},\bar{p},\lambda_{0},\eta_{2}\right)\parallel_{P,2}\leq & \parallel\frac{2-4\lambda_{0}}{\bar{p}^{3}\lambda_{0}^{2}\left(1-\lambda_{0}\right)^{2}}\frac{D-g\left(X\right)}{1-g\left(X\right)}\left(\left(T-\lambda_{0}\right)Y-\ell_{2}\left(X\right)\right)\parallel_{P,2}\\
 & +\parallel\frac{2Y}{\bar{p}^{3}\lambda_{0}\left(1-\lambda_{0}\right)}\frac{D-g\left(X\right)}{1-g\left(X\right)}\parallel_{P,2}\\
\leq & O\left(1\right).
\end{align*}
Together with (A.8), we obtain 
\begin{align*}
\parallel\partial_{\lambda}\partial_{p}\psi_{2}\left(W,\theta_{0},p,\lambda_{0},\eta_{2}\right)-\partial_{\lambda}\partial_{p}\psi_{20}\parallel_{P,2}\leq & O\left(\varepsilon_{N}\right)+O\left(1\right)O\left(N^{-1/2}\right)\\
= & O\left(\varepsilon_{N}\right),
\end{align*}
where I assume that $\varepsilon_{N}$ converges to zero no faster
than $N^{-1/2}$.
\begin{description}
\item [{Proof~of~Theorem~2}]~
\item [{Repeated~outcomes:}]~
\end{description}
In Step 1, I show the main result using the auxiliary results 
\[
\sup_{p\in\mathcal{P}_{N},\eta_{1}\in\mathcal{T}_{N}}\left(E_{P}\left[\parallel\bar{\psi}_{1}\left(W,\theta_{0},p,\eta_{1}\right)-\bar{\psi}_{1}\left(W,\theta_{0},p_{0},\eta_{10}\right)\parallel^{2}\right]\right)^{1/2}\leq\varepsilon_{N},\tag{A.12}
\]

\[
\left(E_{P}\left[\bar{\psi}_{1}\left(W,\theta_{0},p_{0},\eta_{10}\right)^4\right]\right)^{1/4}\leq C_{1}, \tag{A.13}
\]
where $\mathcal{P}_{N}$ and $\mathcal{T}_{N}$ are specified in the proof of Theorem 1, $C_{1}$ is a constant, and
\[
\bar{\psi}_{1}\left(W,\theta,p,\eta_{1}\right)\coloneqq\frac{1}{p}\frac{D-g\left(X\right)}{1-g\left(X\right)}\left(Y\left(1\right)-Y\left(0\right)-\ell_{1}\left(X\right)\right)-\frac{D\theta}{p}.
\]
In fact, we have $E_{P}\left[\left(\bar{\psi}_{1}\left(W,\theta_{0},p_{0},\eta_{10}\right)\right)^{2}\right]=\Sigma_{10}$.
In Step 2, I show the auxiliary results (A.12) and (A.13).

\emph{Step 1. }Notice that 
\begin{align*}
\hat{\Sigma}_{1}= & \frac{1}{K}\sum_{k=1}^{K}\mathbb{E}_{n,k}\left[\left(\psi_{1}\left(W,\tilde{\theta},\hat{p}_{k},\hat{\eta}_{1k}\right)+\hat{G}_{1p}\left(D-\hat{p}_{k}\right)\right)^{2}\right]\\
= & \frac{1}{K}\sum_{k=1}^{K}\mathbb{E}_{n,k}\left[\left(\frac{1}{\hat{p}_{k}}\frac{D-\hat{g}_{k}\left(X\right)}{1-\hat{g}_{k}\left(X\right)}\left(Y\left(1\right)-Y\left(0\right)-\hat{\ell}_{1k}\left(X\right)\right)-\frac{D\tilde{\theta}}{\hat{p}_{k}}\right)^{2}\right]\\
= & \frac{1}{K}\sum_{k=1}^{K}\mathbb{E}_{n,k}\left[\bar{\psi}_{1}\left(W,\tilde{\theta},\hat{p}_{k},\hat{\eta}_{1k}\right)^{2}\right],
\end{align*}
where the second equality follows from $\hat{G}_{1p}=-\tilde{\theta}/\hat{p}_{k}$.

Since $K$ is fixed, which is independent of $N$, it suffices to
show that for each $k\in \left[k\right]$,
\[
I_{k}\coloneqq\mid\mathbb{E}_{n,k}\left[\bar{\psi}_{1}\left(W,\tilde{\theta},\hat{p}_{k},\hat{\eta}_{1k}\right)^{2}\right]-E_{P}\left[\bar{\psi}_{1}\left(W,\theta_{0},p_{0},\eta_{10}\right)^{2}\right]\mid=o_{P}\left(1\right).
\]
By the triangle inequality, we have
\[
I_{k}\leq I_{3,k}+I_{4,k},
\]
where
\[
I_{3,k}\coloneqq\mid\mathbb{E}_{n,k}\left[\bar{\psi}_{1}\left(W,\tilde{\theta},\hat{p}_{k},\hat{\eta}_{1k}\right)^{2}\right]-\mathbb{E}_{n,k}\left[\bar{\psi}_{1}\left(W,\theta_{0},p_{0},\eta_{10}\right)^{2}\right]\mid,
\]
\[
I_{4,k}\coloneqq\mid\mathbb{E}_{n,k}\left[\bar{\psi}_{1}\left(W,\theta_{0},p_{0},\eta_{10}\right)^{2}\right]-E_{P}\left[\bar{\psi}_{1}\left(W,\theta_{0},p_{0},\eta_{10}\right)^{2}\right]\mid.
\]
To bound $I_{4,k}$, we have  
\begin{align*}
E_{P}\left[I_{4,k}^{2}\right]\leq & n^{-1} E_{P}\left[ \bar{\psi}_{1}\left(W,\theta_{0},p_{0},\eta_{10}\right)^{4}\right]\\
\leq & n^{-1}C_{1}^{4},
\end{align*}
where the last inequality follows from (A.13). By Chebyshev's inequality, $I_{4,k}=O_{P}\left(n^{1/2}\right)$. 

Next, we bound $I_{3,k}$. This part is essentially identical to the proof of Theorem 3.2 in \citet*{chernozhukov2018double}, I reproduce it here for reader's convenience. Observe that for any number $a$ and $\delta a$,
\[
\mid\left(a+\delta a\right)^{2}-a^{2}\mid\leq2\left(\delta a\right)\left(a+\delta a\right).
\]
Denote $\psi_{i}=\bar{\psi}_{1}\left(W_{i},\theta_{0},p_{0},\eta_{10}\right)$
and $\hat{\psi}_{i}=\bar{\psi}_{1}\left(W_{i},\tilde{\theta},\hat{p}_{k},\hat{\eta}_{1k}\right)$,
and $a\coloneqq\psi_{i}$, $a+\delta a\coloneqq\hat{\psi}_{i}$. Then
\begin{align*}
I_{3,k}= & \mid\frac{1}{n}\sum_{i\in I_{k}}\left(\hat{\psi}_{i}\right)^{2}-\left(\psi_{i}\right)^{2}\mid\leq\frac{1}{n}\sum_{i\in I_{k}}\mid\left(\hat{\psi}_{i}\right)^{2}-\left(\psi_{i}\right)^{2}\mid\\
\leq & \frac{2}{n}\sum_{i\in I_{k}}\mid\hat{\psi}_{i}-\psi_{i}\mid\times\left(\mid\psi_{i}\mid+\mid\hat{\psi}_{i}-\psi_{i}\mid\right)\\
\leq & \left(\frac{2}{n}\sum_{i\in I_{k}}\mid\hat{\psi}_{i}-\psi_{i}\mid^{2}\right)^{1/2}\left(\frac{2}{n}\sum_{i\in I_{k}}\left(\mid\psi_{i}\mid+\mid\hat{\psi}_{i}-\psi_{i}\mid\right)^{2}\right)^{1/2}\\
\leq & \left(\frac{2}{n}\sum_{i\in I_{k}}\mid\hat{\psi}_{i}-\psi_{i}\mid^{2}\right)^{1/2}\left[\left(\frac{2}{n}\sum_{i\in I_{k}}\mid\psi_{i}\mid^{2}\right)^{1/2}+\left(\frac{2}{n}\sum_{i\in I_{k}}\mid\hat{\psi}_{i}-\psi_{i}\mid^{2}\right)^{1/2}\right].
\end{align*}
Thus,
\[
I_{3,k}^{2}\lesssim S_{N}\times\left(\frac{1}{n}\sum_{i\in I_{k}}\parallel\bar{\psi}_{1}\left(W_{i},\theta_{0},p_{0},\eta_{10}\right)\parallel^{2}+S_{N}\right),
\]
where
\[
S_{N}\coloneqq\frac{1}{n}\sum_{i\in I_{k}}\parallel\bar{\psi}_{1}\left(W_{i},\tilde{\theta},\hat{p}_{k},\hat{\eta}_{1k}\right)-\bar{\psi}_{1}\left(W_{i},\theta_{0},p_{0},\eta_{10}\right)\parallel^{2}.
\]
Since $\frac{1}{n}\sum_{i\in I_{k}}\parallel\bar{\psi_{1}}\left(W_{i},\theta_{0},p_{0},\eta_{0}\right)\parallel^{2}=O_{P}\left(1\right)$,
it suffices to bound $S_{N}$. We have the decomposition
\begin{align*}
S_{N}= & \frac{1}{n}\sum_{i\in I_{k}}\parallel\bar{\psi}_{1}\left(W_{i},\theta_{0},\hat{p}_{k},\hat{\eta}_{1k}\right)+\partial_{\theta}\bar{\psi}_{1}\left(W_{i},\bar{\theta},\hat{p}_{k},\hat{\eta}_{1k}\right)\left(\tilde{\theta}-\theta_{0}\right)-\bar{\psi}_{1}\left(W_{i},\theta_{0},p_{0},\eta_{10}\right)\parallel^{2}\\
= & \frac{1}{n}\sum_{i\in I_{k}}\parallel\bar{\psi}_{1}\left(W_{i},\theta_{0},\hat{p}_{k},\hat{\eta}_{1k}\right)+\frac{D_{i}}{\hat{p}_{k}}\left(\tilde{\theta}-\theta_{0}\right)-\bar{\psi}_{1}\left(W_{i},\theta_{0},p_{0},\eta_{10}\right)\parallel^{2}\\
\leq & \frac{1}{n}\sum_{i\in I_{k}}\parallel\frac{D_{i}}{\hat{p}_{k}}\left(\tilde{\theta}-\theta_{0}\right)\parallel^{2}+\frac{1}{n}\sum_{i\in I_{k}}\parallel\bar{\psi}_{1}\left(W_{i},\theta_{0},\hat{p}_{k},\hat{\eta}_{1k}\right)-\bar{\psi}_{1}\left(W_{i},\theta_{0},p_{0},\eta_{10}\right)\parallel^{2},
\end{align*}
where $\bar{\theta}\in\left(\tilde{\theta}-\theta_{0}\right)$. The 
first term is bounded by
\begin{align*}
\frac{1}{n}\sum_{i\in I_{k}}\parallel\frac{D_{i}}{\hat{p}_{k}}\left(\tilde{\theta}-\theta_{0}\right)\parallel^{2}\leq & \left(\frac{1}{n}\sum_{i\in I_{k}}\left(\frac{D_{i}}{\hat{p}_{k}}\right)^{2}\right)\parallel\tilde{\theta}-\theta_{0}\parallel^{2}\\
= & \left(\frac{1}{n}\sum_{i\in I_{k}}\left(\frac{D_{i}}{p_{0}}\right)^{2}+o_{P}\left(1\right)\right)\parallel\tilde{\theta}-\theta_{0}\parallel^{2}\\
= & O_{P}\left(1\right)\times O_{P}\left(N^{-1}\right).
\end{align*}
Also, notice that conditional on $\left(W_{i}\right)_{i\in I_{k}^{c}}$,
both $\hat{p}_{k}$ and $\hat{\eta}_{1k}$ can be treated as fixed.
Under the event that $\hat{p}_{k}\in\mathcal{P}_{N}$ and $\hat{\eta}_{1k}\in\mathcal{T}_{N}$,
we have
\begin{align*}
E_{P}\left[\parallel\bar{\psi}_{1}\left(W_{i},\theta_{0},\hat{p}_{k},\hat{\eta}_{1k}\right)-\bar{\psi}_{1}\left(W_{i},\theta_{0},p_{0},\eta_{10}\right)\parallel^{2}\mid\left(W_{i}\right)_{i\in I_{k}^{c}}\right]\\
\leq\sup_{p\in\mathcal{P}_{N},\eta_{1}\in\mathcal{T}_{N}}E_{P}\left[\parallel\bar{\psi}_{1}\left(W_{i},\theta_{0},p,\eta_{1}\right)-\bar{\psi}_{1}\left(W_{i},\theta_{0},p_{0},\eta_{10}\right)\parallel^{2}\right] & =\left(\varepsilon_{N}\right)^{2}
\end{align*}
by (A.12). It follows that $S_{N}=O_{P}\left(N^{-1}+\left(\varepsilon_{N}\right)^{2}\right)$.
Therefore, we obtain
\[
I_{k}=O_{P}\left(N^{-1/2}\right)+O_{P}\left(N^{-1/2}+\varepsilon_{N}\right)=o_{P}\left(1\right).
\]

\emph{Step 2. }It remains to prove (A.12) and (A.13). By Taylor series expansion, 
\begin{align*}
\bar{\psi}_{1}\left(W,\theta_{0},p,\eta_{1}\right)-\bar{\psi}_{1}\left(W,\theta_{0},p_{0},\eta_{10}\right)= & \bar{\psi}_{1}\left(W,\theta_{0},p_{0},\eta_{1}\right)-\bar{\psi}_{1}\left(W,\theta_{0},p_{0},\eta_{10}\right)\\
 & +\partial_{p}\psi_{1}\left(W,\theta_{0},\bar{p},\eta_{1}\right)\left(p-p_{0}\right)\\
= & \psi_{1}\left(W,\theta_{0},p_{0},\eta_{1}\right)-\psi_{1}\left(W,\theta_{0},p_{0},\eta_{10}\right)\\
 & +\partial_{p}\psi_{1}\left(W,\theta_{0},\bar{p},\eta_{1}\right)\left(p-p_{0}\right),
\end{align*}
where $\bar{p}\in\left(p,p_{0}\right)$. Then we have 
\begin{align*}
\parallel\bar{\psi}_{1}\left(W,\theta_{0},p,\eta_{1}\right)-\bar{\psi}_{1}\left(W,\theta_{0},p_{0},\eta_{10}\right)\parallel_{P,2}\leq & \parallel\psi_{1}\left(W,\theta_{0},p_{0},\eta_{1}\right)-\psi_{1}\left(W,\theta_{0},p_{0},\eta_{10}\right)\parallel_{P,2}\\
 & +\parallel\frac{1}{\bar{p}^{2}}\frac{D-g\left(X\right)}{1-g\left(X\right)}\left(Y\left(1\right)-Y\left(0\right)-\ell_{1}\left(X\right)\right)+\frac{D\theta_{0}}{\bar{p}^{2}}\parallel_{P,2}\\
 & \times\mid p-p_{0}\mid.
\end{align*}
By (A.1), we have $\parallel\psi_{1}\left(W,\theta_{0},p_{0},\eta_{1}\right)-\psi_{1}\left(W,\theta_{0},p_{0},\eta_{10}\right)\parallel_{P,2}=O\left(\varepsilon_{N}\right)$.
The term in the second line is bounded by 
\begin{align*}
\parallel\frac{1}{\bar{p}^{2}}\frac{U+g_{0}-g}{1-g}\left(U+\ell_{10}-\ell_{1}\right)\parallel_{P,2}+\parallel\frac{D\theta_{0}}{\bar{p}^{2}}\parallel_{P,2}\leq & \frac{1}{\bar{p}^{2}\kappa}\parallel UV_{1}\parallel_{P,2}+\frac{1}{\bar{p}^{2}\kappa}\parallel U\left(\ell_{10}-\ell_{1}\right)\parallel_{P,2}\\
 & +\frac{1}{\bar{p}^{2}\kappa}\parallel V_{1}\left(g_{0}-g\right)\parallel_{P,2}+\frac{1}{\bar{p}^{2}}\mid\theta_{0}\mid\\
 & +\frac{1}{\bar{p}^{2}\kappa}\parallel g_{0}-g_{1}\parallel_{P,2}\parallel\ell_{10}-\ell_{1}\parallel_{P,2}\\
\leq & \frac{1}{\bar{p}^{2}\kappa}\left(C+\sqrt{C}\parallel\ell_{10}-\ell_{1}\parallel_{P,2}+\sqrt{C}\parallel g_{0}-g\parallel_{P,2}\right)\\
 & +\frac{C}{\bar{p}^{2}p_{0}\kappa}+\frac{1}{\bar{p}^{2}\kappa}\parallel g_{0}-g_{1}\parallel_{P,2}\parallel\ell_{10}-\ell_{1}\parallel_{P,2}\\
= & O\left(1\right),
\end{align*}
where I use $\parallel UV_{1}\parallel_{P,2}\leq \parallel UV_{1}\parallel_{P,4}\leq C$ , $E_{P}\left[U^{2}\mid X\right]\leq C$,
$E_{P}\left[V_{1}^{2}\mid X\right]\leq C$, and
\begin{align*}
\mid\theta_{0}\mid= & \mid E_{P}\left[\frac{Y\left(1\right)-Y\left(0\right)}{p_{0}}\frac{D-g_{0}\left(X\right)}{1-g_{0}\left(X\right)}\right]\mid\\
\leq & \frac{1}{p_{0}\kappa}\mid E_{P}\left[\left(Y\left(1\right)-Y\left(0\right)\right)U\right]\mid\\
= & \frac{1}{p_{0}\kappa}\mid E_{P}\left[\left(\ell_{10}\left(X\right)+V_{1}\right)U\right]\mid\\
= & \frac{1}{p_{0}\kappa}\mid E_{P}\left[UV_{1}\right]\mid\\
\leq & \frac{C}{p_{0}\kappa}
\end{align*}
by $\mid E_{P}\left[UV_{1}\right]\mid\leq\parallel UV_{1}\parallel_{P,4}\leq C$.
Thus, we obtain 
\begin{align*}
\parallel\bar{\psi}_{1}\left(W,\theta_{0},p,\eta_{1}\right)-\bar{\psi}_{1}\left(W,\theta_{0},p_{0},\eta_{10}\right)\parallel_{P,2}\leq & O\left(\varepsilon_{N}\right)+O\left(1\right)O\left(N^{-1/2}\right)\\
= & O\left(\varepsilon_{N}\right),
\end{align*}
where I assume that $\varepsilon_{N}$ converges to zero no faster
than $N^{-1/2}$. 

For (A.13), 
\begin{align*}
\parallel \bar{\psi}_{1}\left(W, \theta_{0}, p_{0}, \eta_{10}\right)\parallel_{P,4}=& \parallel\frac{1}{p_{0}}\frac{UV_{1}}{1-g_{0}}-\frac{D\theta_{0}}{p_{0}}\parallel_{P,4}\\
\leq & \parallel\frac{1}{p_{0}}\frac{UV_{1}}{1-g_{0}}\parallel_{P,4}+ \parallel\frac{D\theta_{0}}{p_{0}}\parallel_{P,4}\\
\leq & \frac{1}{p_{0}\kappa}\parallel UV_{1} \parallel_{P,4}+\frac{1}{p_{0}}\mid \theta_{0} \mid\\
\leq & \frac{C}{p_{0}\kappa}+\frac{C}{p_{0}^{2}\kappa}
\end{align*}
since $\parallel UV_{1}\parallel_{P,4}\leq C$.
\begin{description}
\item [{Repeated~cross~sections:}]~
\end{description}
In Step 1, I show the main result with the auxiliary results:
\[
\sup_{p\in\mathcal{P}_{N},\lambda\in\Lambda_{N},\eta_{2}\in\mathcal{T}_{N}}\left(E_{P}\left[\parallel\bar{\psi}_{2}\left(W,\theta_{0},p,\lambda,G_{2\lambda0},\eta_{2}\right)-\bar{\psi}_{2}\left(W,\theta_{0},p_{0},\lambda_{0},G_{2\lambda0},\eta_{20}\right)\parallel^{2}\right]\right)^{2}\leq\varepsilon_{N},\tag{A.14}
\]

\[
\left(E_{P}\left[\bar{\psi}_{2}\left(W,\theta_{0},p_{0},\lambda_{0},G_{2\lambda0},\eta_{20}\right)^{4}\right]\right)^{1/4}\leq C_{2}, \tag{A.15}
\]
where $\left({P}_{N}, \Lambda_{N}, \mathcal{T}_{N} \right)$ are specified in the proof of Theorem 1, $C_{2}$ is a constant, and
\[
\bar{\psi}_{2}\left(W,\theta,p,\lambda,G_{2\lambda},\eta_{2}\right)\coloneqq\frac{1}{\lambda\left(1-\lambda\right)p}\frac{D-g\left(X\right)}{1-g\left(X\right)}\left(\left(T-\lambda\right)Y-\ell_{2}\left(X\right)\right)-\frac{D\theta}{p}+G_{2\lambda}\left(T-\lambda\right).
\]
In fact, we have $E_{P}\left[\left(\bar{\psi}_{2}\left(W,\theta_{0},p_{0},\lambda_{0},G_{2\lambda_{0}},\eta_{20}\right)\right)^{2}\right]=\Sigma_{20}$. In Step 2, I prove (A.14) and (A.15).

\emph{Step 1. }Notice that 
\begin{align*}
\hat{\Sigma}_{2}= & \frac{1}{K}\sum_{k=1}^{K}\mathbb{E}_{n,k}\left[\left(\psi_{2}\left(W,\tilde{\theta},\hat{p}_{k},\hat{\eta}_{1k}\right)+\hat{G}_{2p}\left(D-\hat{p}_{k}\right)+\hat{G}_{2\lambda}\left(T-\hat{\lambda}_{k}\right)\right)^{2}\right]\\
= & \frac{1}{K}\sum_{k=1}^{K}\mathbb{E}_{n,k}\left[\left(\frac{1}{\hat{\lambda}_{k}\left(1-\hat{\lambda}_{k}\right)\hat{p}_{k}}\frac{D-\hat{g}_{k}\left(X\right)}{1-\hat{g}_{k}\left(X\right)}\left(\left(T-\hat{\lambda}_{k}\right)Y-\hat{\ell}_{2k}\left(X\right)\right)-\frac{D\tilde{\theta}}{\hat{p}_{k}}+\hat{G}_{2\lambda}\left(T-\hat{\lambda}_{k}\right)\right)^{2}\right]\\
= & \frac{1}{K}\sum_{k=1}^{K}\mathbb{E}_{n,k}\left[\bar{\psi}_{2}\left(W,\tilde{\theta},\hat{p}_{k},\hat{\lambda}_{k},\hat{G}_{2\lambda},\hat{\eta}_{2k}\right)^{2}\right],
\end{align*}
where the second inequality follows from $\hat{G}_{2p}=-\tilde{\theta}/\hat{p}_{k}$
.

Since $K$ is fixed, which is independent of $N$, it suffices to
show that 
\[
J_{k}\coloneqq\mid\mathbb{E}_{n,k}\left[\bar{\psi}_{2}\left(W,\tilde{\theta},\hat{p}_{k},\hat{\lambda}_{k},\hat{G}_{2\lambda},\hat{\eta}_{2k}\right)^{2}\right]-E_{P}\left[\bar{\psi}_{2}\left(W,\theta_{0},p_{0},\lambda_{0},G_{2\lambda0},\eta_{20}\right)^{2}\right]\mid=o_{P}\left(1\right).
\]
By the triangle inequality, we have
\[
J_{k}\leq J_{5,k}+J_{6,k},
\]
where 
\[
J_{5,k}\coloneqq\mid\mathbb{E}_{n,k}\left[\bar{\psi}_{2}\left(W,\tilde{\theta},\hat{p}_{k},\hat{\lambda}_{k},\hat{G}_{2\lambda},\hat{\eta}_{2k}\right)^{2}\right]-\mathbb{E}_{n,k}\left[\bar{\psi}_{2}\left(W,\theta_{0},p_{0},\lambda_{0},G_{2\lambda0},\eta_{20}\right)^{2}\right]\mid,
\]

\[
J_{6,k}\coloneqq\mid\mathbb{E}_{n,k}\left[\bar{\psi}_{2}\left(W,\theta_{0},p_{0},\lambda_{0},G_{2\lambda0},\eta_{20}\right)^{2}\right]-E_{P}\left[\bar{\psi}_{2}\left(W,\theta_{0},p_{0},\lambda_{0},G_{2\lambda0},\eta_{20}\right)^{2}\right]\mid.
\]
By the same arguments for $I_{4,k}$ in the proof of repeated outcomes and (A.15), we can show $J_{6,k}=o_{P}\left(1\right)$. Also,
by the same arguments for $I_{3,k}$ in the proof of repeated outcomes, we have 
\[
J_{5,k}^{2}\lesssim S_{N}'\times\left(\frac{1}{n}\sum_{i\in I_{k}}\parallel\bar{\psi_{2}}\left(W,\theta_{0},p_{0},\lambda_{0},G_{2\lambda0},\eta_{20}\right)\parallel^{2}+S_{N}'\right),
\]
where
\[
S_{N}'\coloneqq\frac{1}{n}\sum_{i\in I_{k}}\parallel\bar{\psi_{2}}\left(W,\tilde{\theta},\hat{p}_{k},\hat{\lambda}_{k},\hat{G}_{2\lambda},\hat{\eta}_{2k}\right)-\bar{\psi_{2}}\left(W,\theta_{0},p_{0},\lambda_{0},G_{2\lambda0},\eta_{20}\right)\parallel^{2}.
\]
Since $\frac{1}{n}\sum_{i\in I_{k}}\parallel\bar{\psi_{2}}\left(W,\theta_{0},p_{0},\lambda_{0},G_{2\lambda0},\eta_{20}\right)\parallel^{2}=O_{P}\left(1\right)$,
it remains to bound $S_{N}'$. Define $\bar{\psi}_{20}\coloneqq\bar{\psi_{2}}\left(W,\theta_{0},p_{0},\lambda_{0},G_{2\lambda0},\eta_{20}\right)$.
By the triangle inequality, we have 
\begin{align*}
S_{N}'\leq & \frac{1}{n}\sum_{i\in I_{k}}\parallel\bar{\psi}_{2}\left(W_{i},\theta_{0},\hat{p}_{k},\hat{\lambda}_{k},G_{2\lambda0},\hat{\eta}_{2k}\right)-\bar{\psi}_{20}\parallel^{2}\\
 & +\frac{1}{n}\sum_{i\in I_{k}}\parallel\partial_{\theta}\bar{\psi}_{2}\left(W_{i},\bar{\theta},\hat{p}_{k},\hat{\lambda}_{k},\hat{G}_{2\lambda},\hat{\eta}_{2k}\right)\left(\tilde{\theta}-\theta_{0}\right)\parallel^{2}\\
 & +\frac{1}{n}\sum_{i\in I_{k}}\parallel\partial_{G_{2\lambda}}\bar{\psi}_{2}\left(W_{i},\theta_{0},\hat{p}_{k},\hat{\lambda}_{k},\overline{G_{2\lambda}},\hat{\eta}_{2k}\right)\left(\hat{G}_{2\lambda}-G_{2\lambda0}\right)\parallel^{2},
\end{align*}
where $\bar{\theta}\in\left(\tilde{\theta},\theta_{0}\right)$ and
$\overline{G_{2\lambda}}\in\left(\widehat{G_{2\lambda}},G_{2\lambda0}\right)$.
Then we have 
\begin{align*}
S_{N}'\leq & \frac{1}{n}\sum_{i\in I_{k}}\parallel\bar{\psi}_{2}\left(W_{i},\theta_{0},\hat{p}_{k},\hat{\lambda}_{k},G_{2\lambda0},\hat{\eta}_{2k}\right)-\bar{\psi}_{20}\parallel^{2}\\
 & +\frac{1}{n}\sum_{i\in I_{k}}\parallel\frac{D_{i}}{\hat{p}_{k}}\left(\tilde{\theta}-\theta_{0}\right)\parallel^{2}+\frac{1}{n}\sum_{i\in I_{k}}\parallel\left(T_{i}-\hat{\lambda}_{k}\right)\left(\hat{G}_{2\lambda}-G_{2\lambda0}\right)\parallel^{2}.
\end{align*}
 The two terms in the last line are bounded by 
\[
\left(\frac{1}{n}\sum_{i\in I_{k}}\left(\frac{D_{i}}{\hat{p}_{k}}\right)^{2}\right)\times\parallel\tilde{\theta}-\theta_{0}\parallel^{2}=O_{P}\left(1\right)\times O_{P}\left(N^{-1}\right)
\]
and
\[
\left(\frac{1}{n}\sum_{i\in I_{k}}\left(T_{i}-\hat{\lambda}_{k}\right)^{2}\right)\times\parallel\hat{G}_{2\lambda}-G_{2\lambda0}\parallel^{2}=O_{P}\left(1\right)\times\left(o_{P}\left(1\right)\right)^{2}.
\]
Conditional on the auxiliary sample $I_{k}^{c}$, $\left(\hat{p}_{k},\hat{\lambda}_{k},\hat{\eta}_{2k}\right)$
can be treated as fixed. Also, under the event that $\hat{p}_{k}\in\mathcal{P}_{N}$,
$\hat{\lambda}_{k}\in\Lambda_{N}$, and $\hat{\eta}_{2k}\in\mathcal{T}_{N}$,
we have
\begin{align*}
E_{P}\left[\parallel\bar{\psi}_{2}\left(W_{i},\theta_{0},\hat{p}_{k},\hat{\lambda}_{k},G_{2\lambda0},\hat{\eta}_{2k}\right)-\bar{\psi}_{2}\left(W_{i},\theta_{0},p_{0},\lambda_{0},G_{2\lambda0},\eta_{20}\right)\parallel^{2}\mid\left(W_{i}\right)_{i\in I_{k}^{c}}\right]\\
\leq\sup_{p\in\mathcal{P}_{N},\lambda\in\Lambda_{N},\eta_{2}\in\mathcal{T}_{N}}E_{P}\left[\parallel\bar{\psi}_{2}\left(W_{i},\theta_{0},p,\lambda,G_{2\lambda0},\eta_{2}\right)-\bar{\psi}_{2}\left(W_{i},\theta_{0},p_{0},\lambda_{0},G_{2\lambda0},\eta_{20}\right)\parallel^{2}\right] & =\left(\varepsilon_{N}\right)^{2}
\end{align*}
by (A.14). It follows that $S_{N}'=O_{P}\left(N^{-1}+\varepsilon_{N}^{2}\right)+\left(o_{P}\left(1\right)\right)^{2}$
so that 
\[
J_{k}=o_{P}\left(1\right)+O_{P}\left(N^{-1/2}+\varepsilon_{N}\right)+o_{P}\left(1\right)=o_{P}\left(1\right).
\]

\emph{Step 2. }It remains to show (A.14) and (A.15). Define $\bar{\psi}_{20}\coloneqq \bar{\psi}_{2}\left(W,\theta_{0},p_{0},\lambda_{0},G_{2\lambda0},\eta_{20}\right)$. By the triangle inequality
and 
\[
\bar{\psi}_{2}\left(W,\theta_{0},p_{0},\lambda_{0},G_{2\lambda0},\eta_{2}\right)-\bar{\psi}_{20}=\psi_{2}\left(W,\theta_{0},p_{0},\lambda_{0},\eta_{2}\right)-\psi_{2}\left(W,\theta_{0},p_{0},\lambda_{0},\eta_{20}\right),
\]
we have 
\begin{align*}
\parallel\bar{\psi}_{2}\left(W,\theta_{0},p,\lambda,G_{2\lambda0},\eta_{2}\right)-\bar{\psi}_{20}\parallel_{P,2}\leq & \parallel\psi_{2}\left(W,\theta_{0},p_{0},\lambda_{0},\eta_{2}\right)-\psi_{2}\left(W,\theta_{0},p_{0},\lambda_{0},\eta_{20}\right)\parallel_{P,2}\\
 & +\parallel\partial_{\lambda}\bar{\psi}_{2}\left(W_{i},\theta_{0},p_{0},\bar{\lambda},G_{2\lambda0},\eta_{2}\right)\parallel_{P,2}\mid\lambda-\lambda_{0}\mid\\
 & +\parallel\partial_{p}\bar{\psi}_{2}\left(W_{i},\theta_{0},\bar{p},\lambda,G_{2\lambda0},\eta_{2}\right)\parallel_{P,2}\mid p-p_{0}\mid,
\end{align*}
where $\bar{p}\in\left(p,p_{0}\right)$ and $\bar{\lambda}\in\left(\lambda,\lambda_{0}\right)$.
The term in the second line is bounded by 
\begin{align*}
\parallel\partial_{\lambda}\bar{\psi}_{2}\left(W_{i},\theta_{0},p_{0},\bar{\lambda},G_{2\lambda0},\eta_{2}\right)\parallel_{P,2}\leq & \frac{\mid1-2\bar{\lambda}\mid}{p_{0}\bar{\lambda}^{2}\left(1-\bar{\lambda}\right)^{2}}\parallel\frac{D-g\left(X\right)}{1-g\left(X\right)}\left(\left(T-\bar{\lambda}\right)Y-\ell_{2}\left(X\right)\right)\parallel_{P,2}\\
 & +\frac{1}{p_{0}\bar{\lambda}\left(1-\bar{\lambda}\right)}\parallel\frac{D-g\left(X\right)}{1-g\left(X\right)}\times Y\parallel_{P,2}+\mid G_{2\lambda0}\mid\\
\leq & O\left(1\right)
\end{align*}
by the same arguments in (A.9)-(A.11) and 
\begin{align*}
\mid G_{2\lambda0}\mid= & \mid E_{P}\left[-\frac{1-2\lambda_{0}}{\lambda_{0}^{2}\left(1-\lambda_{0}\right)^{2}p_{0}}\frac{D-g_{0}}{1-g_{0}}\left(\left(T-\lambda_{0}\right)Y-\ell_{20}\right)-\frac{Y}{\lambda_{0}\left(1-\lambda_{0}\right)p_{0}}\frac{D-g_{0}}{1-g_{0}}\right]\mid\\
\leq & \frac{\mid1-2\lambda_{0}\mid}{\lambda_{0}^{2}\left(1-\lambda_{0}\right)^{2}p_{0}\kappa}\mid E_{P}\left[UV_{2}\right]\mid+\frac{1}{\lambda_{0}\left(1-\lambda_{0}\right)p_{0}\kappa}\mid E_{P}\left[YU\right]\mid\\
\leq & \frac{\mid1-2\lambda_{0}\mid}{\lambda_{0}^{2}\left(1-\lambda_{0}\right)^{2}p_{0}\kappa}C+\frac{1}{\lambda_{0}\left(1-\lambda_{0}\right)p_{0}\kappa}C\\
= & O\left(1\right)
\end{align*}
since $\mid E_{P}\left[UV_{2}\right]\mid\leq\parallel UV_{2}\parallel_{P,4}\leq C$
and $\mid E_{P}\left[YU\right]\mid\leq C$. Also, we have 
\begin{align*}
\parallel\partial_{p}\bar{\psi}_{2}\left(W_{i},\theta_{0},\bar{p},\lambda,G_{2\lambda0},\eta_{2}\right)\parallel_{P,2}\leq & \frac{1}{\lambda\left(1-\lambda\right)\bar{p}^{2}}\parallel\frac{D-g\left(X\right)}{1-g\left(X\right)}\left(\left(T-\lambda\right)Y-\ell_{2}\left(X\right)\right)\parallel_{P,2}\\
 & +\parallel\frac{D\theta_{0}}{\bar{p}^{2}}\parallel_{P,2}\\
\leq & O\left(1\right)
\end{align*}
by the same arguments in (A.9)-(A.11) and
\begin{align*}
\mid\theta_{0}\mid= & \mid E_{P}\left[\frac{D-g_{0}\left(X\right)}{p_{0}'\left(1-g_{0}\left(X\right)\right)}\left(T-\lambda_{0}\right)Y\right]\mid\\
\leq & \frac{1}{p_{0}'\kappa}\mid E_{P}\left[\left(T-\lambda_{0}\right)YU\right]\mid\\
= & \frac{1}{p_{0}\kappa}\mid E_{P}\left[\left(\ell_{20}\left(X\right)+V_{2}\right)U\right]\mid\\
= & \frac{1}{p_{0}\kappa}\mid E_{P}\left[UV_{2}\right]\mid\\
\leq & \frac{C}{p_{0}\kappa}
\end{align*}
since $\mid E_{P}\left[UV_{2}\right]\mid\leq\parallel UV_{2}\parallel_{P,4}\leq C$.
Together with (A.5), we have 
\begin{align*}
\parallel\bar{\psi}_{2}\left(W,\theta_{0},p,\lambda,G_{2\lambda0},\eta_{2}\right)-\bar{\psi}_{20}\parallel_{P,2}\leq & O\left(\varepsilon_{N}\right)+O\left(1\right)O\left(N^{-1/2}\right)+O\left(1\right)O\left(N^{-1/2}\right)\\
= & O\left(\varepsilon_{N}\right),
\end{align*}
where I assume that $\varepsilon_{N}$ converges to zero no faster
than $N^{-1/2}$. 

For (A.15), we have 
\begin{align*}
\parallel \bar{\psi}_{2}\left(W,\theta_{0}, p_{0}, \lambda_{0}, G_{2\lambda0}, \eta_{20}\right)\parallel_{P,4}=& \parallel \frac{1}{\lambda_{0}\left(1-\lambda_{0}\right)p_{0}}\frac{UV_{2}}{1-g_{0}}-\frac{D\theta_{0}}{p_{0}}+G_{2\lambda0}\left(T-\lambda_{0}\right)\parallel_{P,4}\\
\leq &  \frac{1}{\lambda_{0}\left(1-\lambda_{0}\right)p_{0}\kappa}\parallel UV_{2}\parallel_{P,4}+ \frac{1}{p_{0}}\mid \theta_{0}\mid +\mid G_{2\lambda0}\mid \\
\leq & O\left(1\right)
\end{align*}
since $\parallel UV_{2}\parallel_{P,4}\leq C$.
\begin{description}
\item [{Proof~of~Theorem~3:}]~
\end{description} 
By the same arguments in the proof of Theorem 1, we can have 
\[
\sqrt{N}\left(\tilde{\theta}-\theta_{0}\right)=\frac{1}{\sqrt{N}}\sum_{i=1}^{N}\psi_{1}\left(W_{i},\theta_{0},p_{0},\eta_{10}\right)+G_{1p0}\left(D-p_{0}\right)+O_{P}\left(\varepsilon_{N}+\sqrt{N}\left(\varepsilon_{N}\right)^{2}\right)
\]
for repeated outcomes and
\begin{align*}
\sqrt{N}\left(\tilde{\theta}-\theta_{0}\right)= & \frac{1}{\sqrt{N}}\sum_{i=1}^{N}\psi_{2}\left(W_{i},\theta_{0},p_{0},\lambda_{0},\eta_{20}\right)+G_{2p0}\left(D-p_{0}\right)+G_{2\lambda0}\left(T-\lambda_{0}\right)\\
 & +O_{P}\left(\varepsilon_{N}+\sqrt{N}\left(\varepsilon_{N}\right)^{2}\right)
\end{align*}
for repeated cross sections. The term $\varepsilon_{N}$ is the rate of convergence of the kernel
estimators $\hat{g}_{kh}$,  $\hat{\ell}_{1kh}$, and $\hat{\ell}_{2kh}$.
It remains to show $
\parallel\hat{g}_{kh}-g_{0}\parallel_{P,2}=o_{P}\left(N^{-1/4}\right) 
$,
$
\parallel\hat{\ell}_{1kh}-\ell_{10}\parallel_{P,2}=o_{P}\left(N^{-1/4}\right) 
$, and
$
\parallel\hat{\ell}_{2kh}-\ell_{20}\parallel_{P,2}=o_{P}\left(N^{-1/4}\right)
$.

Here I use the standard result of kernel estimation in \citet*{newey1994large}. Let $\hat{\gamma}_{kh}\left(x\right)$ denote the kernel estimator of $\gamma_{0}\left(x\right)=f_{0}\left(x\right)E_{P}\left[z\mid x\right]$ using the auxiliary sample $I_{k}^{c}$, where $z\in \left\{1,D,Y\left(1\right)-Y\left(0\right)\mid D=0,\left(T-\lambda_{0}\right)Y\mid D=0\right\}$. By Assumption (3.3) and Lemma 8.10 of \citet*{newey1994large}, we have 

\begin{align*}
\sup_{x\in\mathcal{X}}\mid\hat{\gamma}_{kh}\left(x\right)-\gamma_{0}\left(x\right)\mid=& O_{P}\left(\left(\log N\right)^{1/2}\left(Nh^{d+2s}\right)^{-1/2}+h^{m}\right)\\
= & o_{P}\left(N^{-1/4}\right)
\end{align*}
by the conditions on $h$. Let $\hat{f}_{kh}\left(x\right)$ denote $\hat{\gamma}_{k}\left(x\right)$ with $z=1$ and $\hat{m}_{kh}\left(x\right)$ denote $\hat{\gamma}_{k}\left(x\right)$ with $z=D$. Then
\[
\hat{g}_{kh}\left(x\right)=\frac{\hat{m}_{kh}\left(x\right)}{\hat{f}_{kh}\left(x\right)}=\frac{\hat{m}_{kh}\left(x\right)/f_{0}\left(x\right)}{\hat{f}_{kh}\left(x\right)/f_{0}\left(x\right)}.
\] 
For the denominator, we have
\[
\sup_{x\in\mathcal{X}}\mid\frac{\hat{f}_{kh}\left(x\right)}{f_{0}\left(x\right)}-1 \mid\leq \frac{\sup_{x\in\mathcal{X}}\mid\hat{f}_{kh}\left(x\right)-f_{0}\left(x\right) \mid}{\inf_{x\in\mathcal{X}}f_{0}\left(x\right)}=o_{P}\left(N^{-1/4}\right)
\]
given $\inf_{x\in\mathcal{X}}f_{0}\left(x\right)\neq 0$. For the numerator, let $m_{0}\left(x\right)$ denote $\gamma_{0}\left(x\right)$ with $z=D$, we have 
\[
\sup_{x\in\mathcal{X}}\mid\frac{\hat{m}_{kh}\left(x\right)}{f_{0}\left(x\right)}-g_{0}\left(x\right) \mid \leq \frac{\sup_{x\in\mathcal{X}}\mid\hat{m}_{kh}\left(x\right)-m_{0}\left(x\right) \mid}{\inf_{x\in\mathcal{X}}f_{0}\left(x\right)}=o_{P}\left(N^{-1/4}\right)
\]
given $\inf_{x\in\mathcal{X}}f_{0}\left(x\right)\neq 0$. The above two inequalities imply that uniformly over $x\in \mathcal{X}$,  

\[
\hat{g}_{kh}\left(x\right)=\frac{g_{0}\left(x\right)+o_{P}\left(N^{-1/4}\right)}{1+o_{P}\left(N^{-1/4}\right)}= g_{0}\left(x\right)+o_{P}\left(N^{-1/4}\right).
\]
That is, $\sup_{x\in\mathcal{X}}\mid \hat{g}_{kh}\left(x\right)-g_{0}\left(x\right)\mid=o_{P}\left(N^{-1/4}\right)$. Using the same arguments, one can also show that $\sup_{x\in\mathcal{X}}\mid \hat{\ell}_{1kh}\left(x\right)-\ell_{0}\left(x\right)\mid=o_{P}\left(N^{-1/4}\right)$ and $\sup_{x\in\mathcal{X}}\mid \hat{\ell}_{2kh}\left(x\right)-\ell_{0}\left(x\right)\mid=o_{P}\left(N^{-1/4}\right)$. Since uniform convergence implies $L^{2}$-norm convergence, we complete the proof. 
\begin{description}
\item [{Proof~of~Theorem~4:}]~
\end{description}
The proof is the same as the proof in Theorem 2 provided the assumptions in Theorem 3 hold.  
\begin{description}
\item [{Lemma~A.1}] (CONDITIONAL CONVERGENCE IMPLIES UNCONDITIONAL) 
\end{description}
\emph{Let
$\left\{ X_{m}\right\} $ and $\left\{ Y_{m}\right\} $ be sequences
of random vectors. (i) If for $\epsilon_{m}\rightarrow0$, $\text{Pr}\left(\parallel X_{m}\parallel>\epsilon_{m}\mid Y_{m}\right)\stackrel{p}{\rightarrow}0$,
then $\text{Pr}\left(\parallel X_{m}\parallel>\epsilon_{m}\right)\rightarrow0$.
This occurs if $E\left[\parallel X_{m}\parallel^{q}/\epsilon_{m}^{q}\mid Y_{m}\right]\stackrel{p}{\rightarrow}0$
for some $q\geq1$, by Markov's inequality. (ii) Let $\left\{ A_{m}\right\} $
be a sequence of positive constants. If $\parallel X_{m}\parallel=O_{P}\left(A_{m}\right)$
conditional on $Y_{m}$, namely, that for any $\ell_{m}\rightarrow\infty$,
$\text{Pr}\left(\parallel X_{m}\parallel>\ell_{m}A_{m}\mid Y_{m}\right)\stackrel{p}{\rightarrow}0$,
then $\parallel X_{m}\parallel=O_{P}\left(A_{m}\right)$ unconditionally,
namely, that for any $\ell_{m}\rightarrow\infty$, $\text{Pr}\left(\parallel X_{m}\parallel>\ell_{m}A_{m}\right)\stackrel{}{\rightarrow}0$. }

PROOF: This lemma is the Lemma 6.1 in \citet*{chernozhukov2018double}.

\pagebreak

\begin{center} \begin{Large}
SIMULATION
\end{Large} \end{center}

\begin{figure}[H]
\begin{centering}
\includegraphics[scale=0.4]{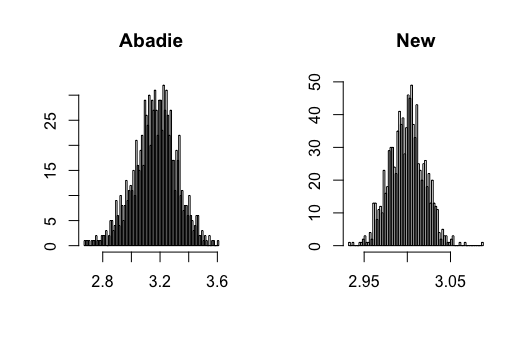}
\par\end{centering}
\begin{center}\caption{Repeated outcomes: $N=200$ and $p=300$.} \end{center}
\end{figure}

\begin{figure}[H]
\begin{centering}
\includegraphics[scale=0.4]{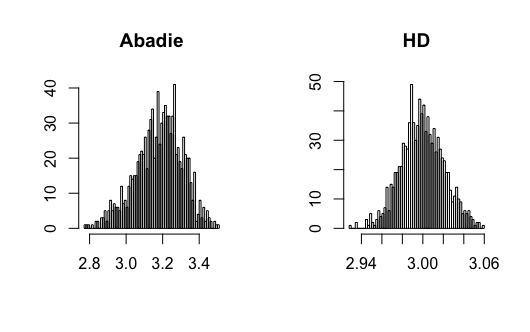}
\par\end{centering}
\begin{center}\caption{Repeated outcomes: $N=200$ and $p=100$.} \end{center}
\end{figure}

\begin{figure}[H]
\begin{centering}
\includegraphics[scale=0.4]{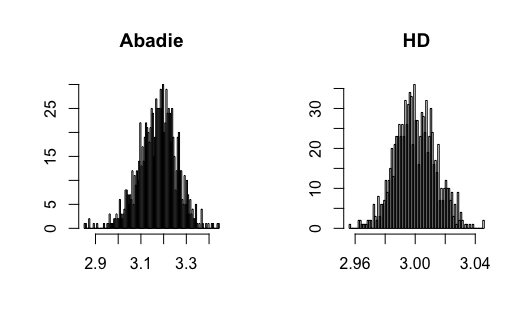}
\par\end{centering}
\begin{center}\caption{Repeated outcomes: $N=500$ and $p=300$.}\end{center}
\end{figure}

\begin{figure}[H]
\begin{centering}
\includegraphics[scale=0.4]{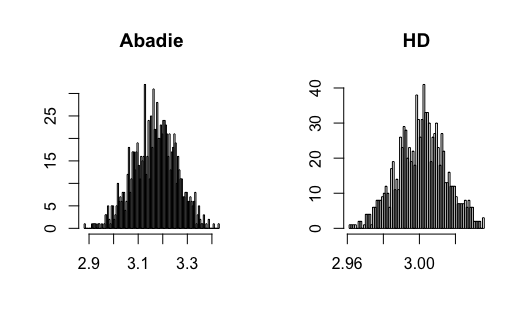}
\par\end{centering}
\begin{center}\caption{Repeated outcomes: $N=500$ and $p=100$.}\end{center}
\end{figure}

\begin{figure}[H]
\begin{centering}
\includegraphics[scale=0.4]{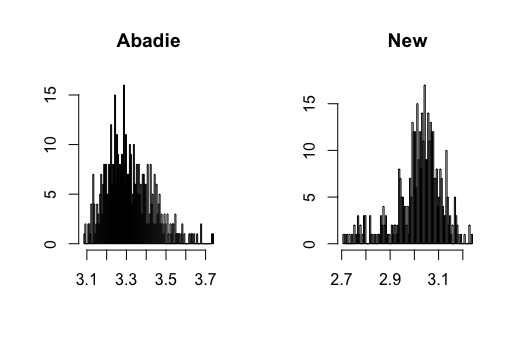}
\par\end{centering}
\begin{center}\caption{Repeated outcomes: $N=200$}\end{center}
\end{figure}

\begin{figure}[H]
\begin{centering}
\includegraphics[scale=0.4]{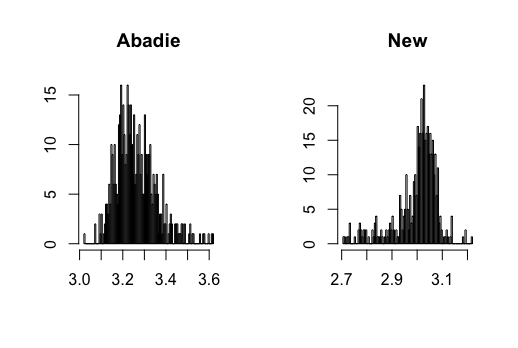}
\par\end{centering}
\begin{center}\caption{Repeated outcomes: $N=500$}\end{center}
\end{figure}

\begin{figure}[H]
\begin{centering}
\includegraphics[scale=0.4]{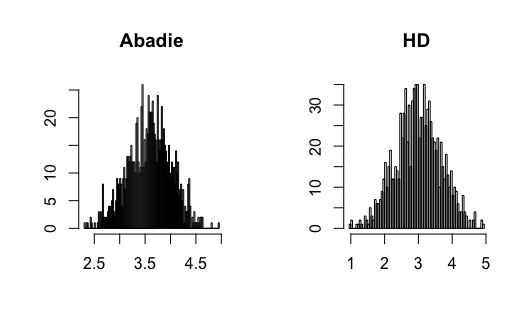}
\par\end{centering}
\begin{center}\caption{Repeated cross sections: $N=200$ and $p=300$.} \end{center}
\end{figure}

\begin{figure}[H]
\begin{centering}
\includegraphics[scale=0.4]{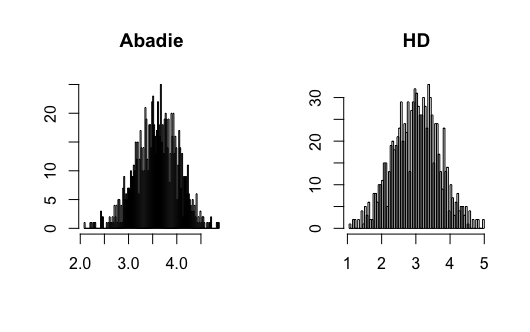}
\par\end{centering}
\begin{center}\caption{Repeated cross sections: $N=200$ and $p=100$.}\end{center}
\end{figure}

\begin{figure}[H]
\begin{centering}
\includegraphics[scale=0.4]{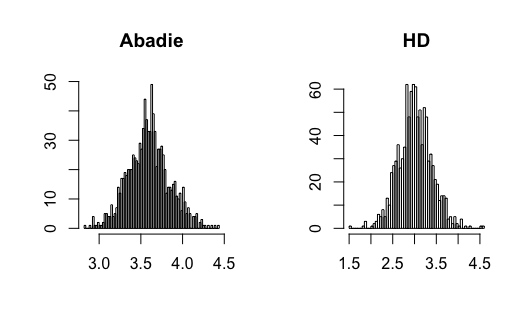}
\par\end{centering}
\begin{center}\caption{Repeated cross sections: $N=500$ and $p=300$.}\end{center}
\end{figure}

\begin{figure}[H]
\begin{centering}
\includegraphics[scale=0.4]{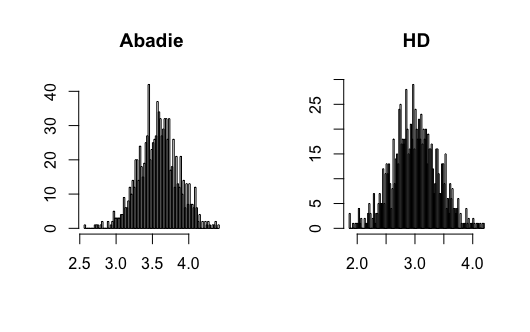}
\par\end{centering}
\begin{center}\caption{Repeated cross sections: $N=500$ and $p=100$.}\end{center}
\end{figure}

\begin{figure}[H]
\begin{centering}
\includegraphics[scale=0.4]{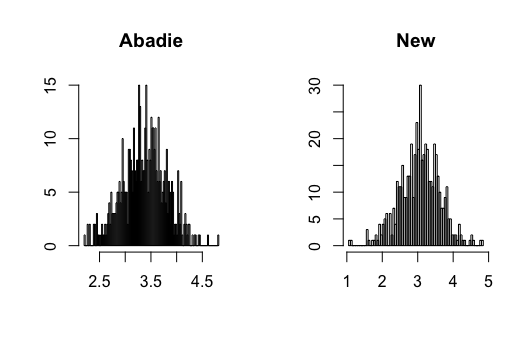}
\par\end{centering}
\begin{center}\caption{Repeated cross sections: $N=200$}\end{center}
\end{figure}

\begin{figure}[H]
\begin{centering}
\includegraphics[scale=0.4]{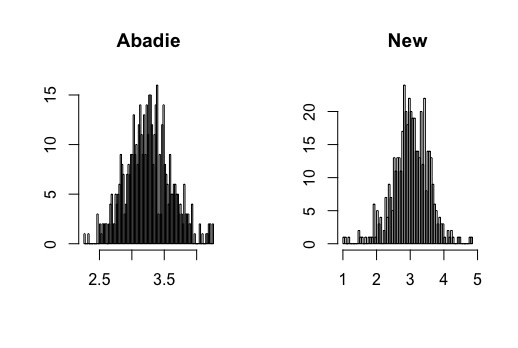}
\par\end{centering}
\begin{center}\caption{Repeated cross sections: $N=500$}\end{center}
\end{figure}

\begin{figure}[H]
\begin{centering}
\includegraphics[scale=0.4]{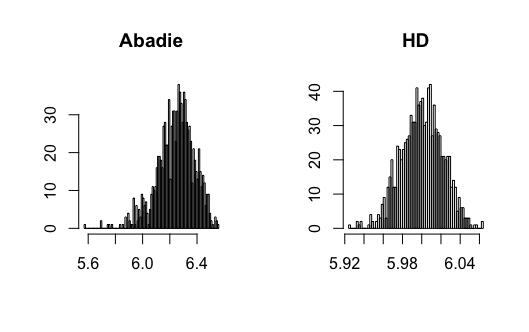}
\par\end{centering}
\begin{center}\caption{Multilevel treatment: $N=200$ and $p=300$.}\end{center}
\end{figure}

\begin{figure}[H]
\begin{centering}
\includegraphics[scale=0.4]{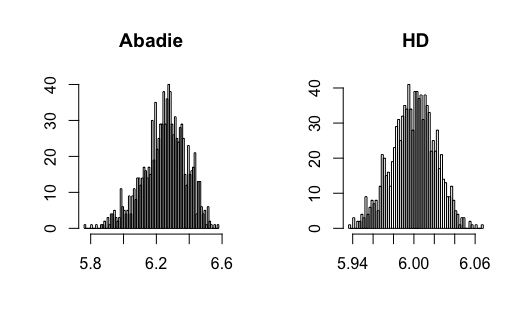}
\par\end{centering}
\begin{center}\caption{Multilevel treatment: $N=200$ and $p=100$.}\end{center}
\end{figure}

\begin{figure}[H]
\begin{centering}
\includegraphics[scale=0.4]{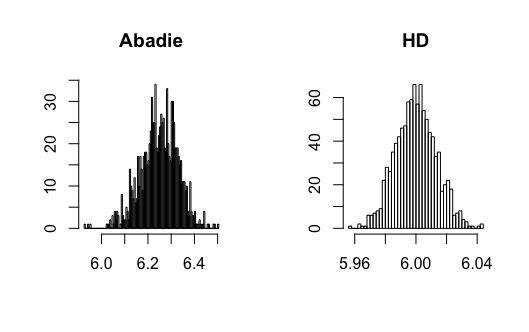}
\par\end{centering}
\begin{center}\caption{Multilevel treatment: $N=500$ and $p=300$.}\end{center}
\end{figure}

\begin{figure}[H]
\begin{centering}
\includegraphics[scale=0.4]{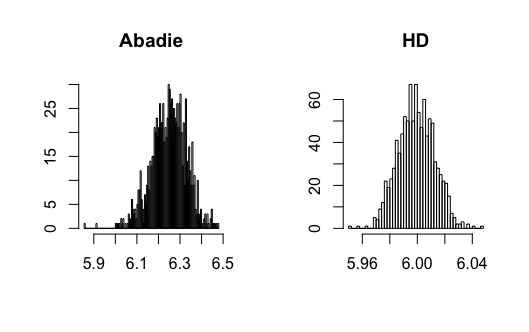}
\par\end{centering}
\begin{center}\caption{Multilevel treatment: $N=500$ and $p=100$.}\end{center}
\end{figure}

\begin{figure}[H]
\begin{centering}
\includegraphics[scale=0.4]{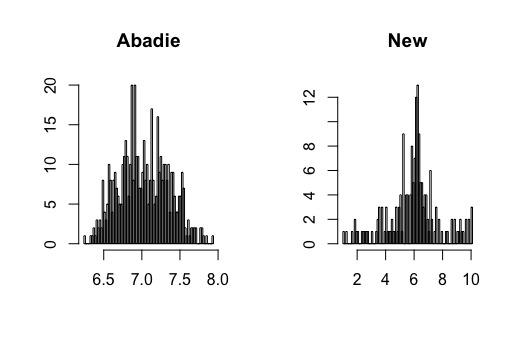}
\par\end{centering}
\begin{center}\caption{Multilevel treatment: $N=200$} \end{center}
\end{figure}

\begin{figure}[H]
\begin{centering}
\includegraphics[scale=0.4]{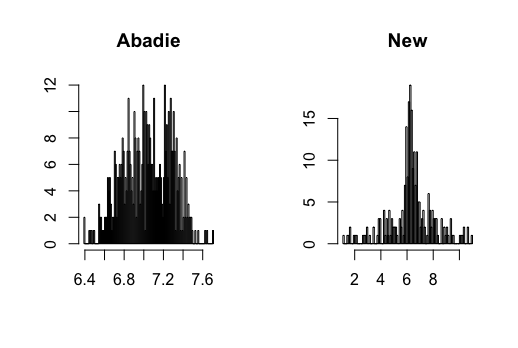}
\par\end{centering}
\begin{center}\caption{Multilevel treatment: $N=500$} \end{center}
\end{figure}

\end{document}